%% file: paper.tex
\pdfoutput=1
\documentclass[letterpaper,twocolumn,10pt]{article}
\usepackage{usenix,epsfig,endnotes}

\usepackage{hypernat}
\usepackage{graphicx}
\usepackage{epsfig}
\usepackage{textcomp}
\usepackage{epstopdf}
\usepackage{cancel}
\usepackage{soul}
\usepackage{array}
\usepackage{colortbl}
\usepackage{xspace}
\usepackage{amsmath}
\usepackage[ruled,vlined,linesnumbered]{algorithm2e}
\usepackage{pifont}
\usepackage[noend]{algpseudocode}
\usepackage{multirow}
\usepackage{titlesec}

\newcommand{\sysname}{{Shape-GD}\xspace}

\newcommand{\intbasis}{{\cal intuitive basis}\xspace} % intuitive basis ???

\newcommand{\ignore}[1]{}
\newcommand{\mohit}[1]{{\footnotesize\color{blue}[Mohit: #1]}}

\newcommand{\mikhail}[1]{{\footnotesize\color{red}[Mikhail: #1]}}
\newcommand{\todo}[1]{{\footnotesize\color{red}[TODO: #1]}}
\newcommand{\nonl}{\renewcommand{\nl}{\let\nl\oldnl}}% Remove line number for one line
\newcommand{\codecomm}[1]{{\nonl\it\small\mdseries\textcolor{blue}{#1}}}

\makeatletter
\newcommand{\removelatexerror}{\let\@latex@error\@gobble}
\makeatother
\usepackage{url}

\def\BibTeX{{\rm B\kern-.05em{\sc i\kern-.025em b}\kern-.08em
    T\kern-.1667em\lower.7ex\hbox{E}\kern-.125emX}}

\titlespacing\section{0pt}{12pt plus 4pt minus 2pt}{2pt plus 2pt minus 2pt}
\titlespacing\subsection{0pt}{12pt plus 4pt minus 2pt}{2pt plus 2pt minus 2pt}
\begin{document}
%don't want date printed
\date{}

%\title{{\em Shape GD}: Guarding Noisy Neighborhoods with Weak Detectors}
%\title{The Last Mile for ML: 
%Early and Robust Malware Detection \\ in Enterprise Networks\vspace{-0.1in}}
%\title{Exploiting Latent Attack Semantics for \\ Malware Detection\vspace{-0.43in}}

%make title bold and 14 pt font (Latex default is non-bold, 16 pt)
%\title{\Large \bf Discovering Transient Attack Correlations in the WINE dataset 
%	and Using Them For Early Malware Detection\vspace{-0.0in}}

\title{
%\vspace{-1.65in}
\Large \bf 
The Shape of Alerts: Detecting Malware Using Distributed Detectors \\ 
by Robustly Amplifying Transient Correlations
%\vspace{-1.65in}
}

\author{
{\rm Mikhail Kazdagli}\\
mikhail.kazdagli@utexas.edu
\and
{\rm Constantine Caramanis}\\
constantine@utexas.edu
\and
{\rm Sanjay Shakkottai}\\
shakkott@austin.utexas.edu
\and
{\rm Mohit Tiwari}\\
tiwari@austin.utexas.edu\\\\
The University of Texas at Austin
} % end author

\maketitle

%\section*{Abstract}
%\vspace{-0.08in}
%
%Behavioral malware detectors are deployed widely yet suffer
%great detectors that can
%analyze data with hundreds of dimensions at extremely low false false
%positive rates
%
%
%Opportunity because the attack vectors are blunt.
%
%Behavioral malware detectors, by using statistical methods, promise to expose
%previously unknown malware and are an important security primitive.  However,
%even the best behavioral detectors suffer from high false positives and
%negatives. In this paper, we address the challenge of aggregating weak
%per-device behavioral detectors (local detectors or LDs) in {\em noisy}
%communities (i.e., ones that produce alerts at unpredictable rates) into an
%accurate and robust global anomaly detector (GD).  
%\ignore{
%Our system consists of feature vector (FV) driven local detectors at individual
%nodes, and a global detector (GD) that 
%aggregates alerts (that flag the possibility of malware being present)
%from this community of local detectors.}
 
%Anomaly-based malware detectors on each machine into
%an enterprise-wide global malware detector capable of providing early and
%robust alerts.  
%
%\vspace*{-0.65in}
\subsection*{Abstract}
%\vspace{-0.05in}

We introduce a new malware detector -- \sysname -- that 
aggregates per-machine detectors into a robust global detector. 
\sysname is based on two insights:
%-- one structural and the other statistical: 
{\em 1. Structural:} actions such as visiting a website (waterhole attack)
by nodes correlate well
with malware spread, and create dynamic {\em neighborhoods} of nodes that were
exposed to the same attack vector. However, neighborhood sizes vary unpredictably
and require aggregating an unpredictable number of local detectors' outputs into
a global alert.  {\em 2. Statistical:} feature vectors corresponding to true
and false positives of local detectors have markedly different conditional
distributions -- i.e. their {\em shapes} differ. The shape of
neighborhoods can identify infected neighborhoods without having to
estimate neighborhood sizes -- on 5 years of Symantec detectors' logs,
\sysname reduces false positives from $\sim$1M down to $\sim$110K and
raises alerts 345 days (on average) before commercial anti-virus products; in a 
waterhole attack simulated using Yahoo web-service logs, \sysname detects
infected machines when only $\sim$100 of $\sim$550K are compromised.

\section{Introduction}
\label{sec:intro}
\input{intro2}

\section{Related Work}
\label{sec:related-work}

\input{related-work}

\section{\sysname Algorithm}
\label{sec:model}
\input{algorithm_3}

%\section{Shape GD Algorithm}
%\label{sec:model}
%\input{theory/model}
%
%\section{Algorithm  (Symantec)}
%\label{sec:model-symantec}
%\input{algorithm_symantec}
%\input{theory}

%\subsection{}
%\label{sec:deployment}
%\section{Computing Shape GD's parameters} 
%\label{sec:deployment}
%\input{deployment}

%\section{WINE Experiments}
%\label{sec:wine}
%\input{wine_section}

\section{Experimental Setup}
\label{sec:exp-set}
\input{experimental-setup}

%\vspace{-0.1in}
\section{Case Study 1: Symantec Wine Dataset}
\label{sec:results-symantec}
\input{results_symantec_2}

\section{Case Study 2: Waterhole Attack}
\label{sec:results-waterhole}
\input{results_v3}

\section{Discussion}
\label{sec:discussion_main}
\input{discussion_main_text}

%\vspace{-0.1in}
\section{Conclusions}
\label{sec:conclusion}
\input{conclusion}

{
%\footnotesize 
%\bibliographystyle{acm}
%\vspace{0.1in}
%\bibliographystyle{unsrt}
%\bibliographystyle{ieeetr}
%\bibliographystyle{abbrv}
%\bibliographystyle{IEEEtran}
%\bibliography{global}
}

%{\footnotesize \bibliographystyle{acm}
%\bibliography{global}}

%\bibliographystyle{IEEEtranS}
%\bibliography{global}

{\footnotesize \bibliographystyle{acm}
\bibliography{global}}

\newpage

%\bibliographystyle{abbrv}
%% \bibliographystyle{sig-alternate-05-2015}
%\bibliography{global}

%\begin{thebibliography}{1}
%\end{thebibliography}

\appendix
%\counterwithin{figure}{section}
\section*{Appendix}
%\section{Appendix - Running out of space}
\input{scratchpad}

%\label{sec:eval-clustering}
%\input{clustering_results}
%
%\section{Local Detectors}
%\label{sec:LDs}
%\input{local_detectors}
%
%\section{How many FVs does Shape GD need to make robust predictions?} 
%\label{sec:nbd-size}
%\input{robust_prediction}
%
%\section{Computation and Communication Costs of Shape-GD}
%\label{sec:overhead-appendix}
%\input{overhead-appendix}
%
%\section{Discussion}
%\label{sec:discussion}
%\input{discussion}

%\section*{B. Shape GD on an Enterprise Dataset}
%\label{sec:accenture}
%\input{accenture}

%\section*{C. Previous submissions}
%\label{sec:accenture}
%\input{previous_submissions}

\end{document}

%% file: intro2.tex
% \ignore{
% Malware causes heavy damage.
% Behavioral detectors are an important line of defense.
% Deployed in industry today.

% Local detectors.

% Global detectors. Count. We use COUNT for GD. Why - explain.

% Challenges: 
% early -- filtering. 
% robust to community size -- shape.

% Key insight: Fig 1. 
% Filtering -- early detection. 
% Shape -- robust to nbd size estimation.

% }

%----------------
\begin{figure}[t]
\includegraphics[width=0.48\textwidth]{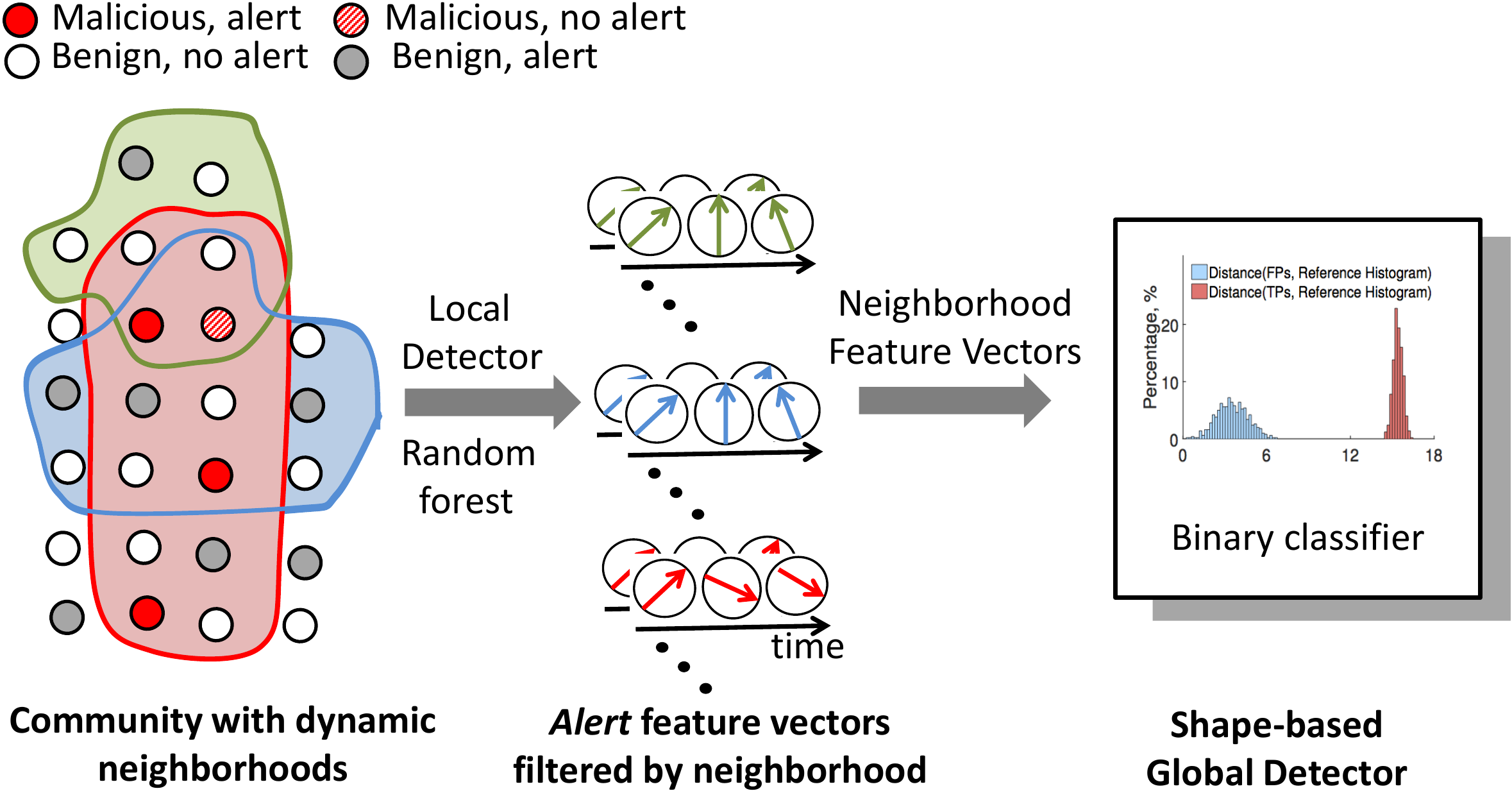}\par
%\caption{\emph{(L to R) Each circle is a node that runs a local malware
%detector (LD).  Our goal is to create a robust global detector (GD) from weak
%LDs.  We observe that nodes naturally form {\em neighborhoods} based on
%attributes relevant to attack vectors -- e.g., all client devices that visit a
%website W within the last hour belong to neighborhood $NB_w$, or all users who
%received an email from a mailing list M in the last hour belong to neighborhood
%$NB_m$.  We propose a new GD that groups together suspicious local feature
%vectors based on neighborhoods -- traditional GDs only analyze local alerts
%while we re-analyze feature vectors that led to the alerts.  Our GD then
%exploits a new insight -- the conditional distribution of true positive feature
%vectors differs from false positive feature vectors -- to robustly classify
%neighborhoods as malicious.}}
\caption{\emph{(L to R) Each circle is a node that runs a local malware
detector (LD).  Our goal is to create a robust global detector (GD) from weak
LDs.  We observe that nodes naturally form {\em neighborhoods} based on
attributes relevant to attack vectors -- e.g., all client devices that visit a
website W within the last hour belong to neighborhood $NB_w$. 
We propose a new GD that groups together suspicious local feature
vectors based on neighborhoods -- traditional GDs only analyze local alerts
while we re-analyze feature vectors that led to the alerts.  Our GD then
exploits a new insight -- the conditional distribution of true positive feature
vectors differs from false positive feature vectors -- to robustly classify
neighborhoods as malicious.}}
\vspace{-0.2in}
\label{fig:sys}
\end{figure}
%----------------

%\ignore{ Much prior research has focused on improving malware detectors
%deployed {\em locally} on each machine (e.g.,
%~\cite{Christodorescu:2008:MSM:1342211.1342215,kruegel-clustering,Canali2012}).
%This includes engineering better features (n-grams, histograms, markov models
%etc of system calls and network traffic) and composing them using ensemble
%methods~\cite{dmitry-ensemble} } Complementary research proposes a {\em
%global} detector (GD) that uses the outputs of a {\em community} of local
%detectors to boost the detection and false positive rate~\cite{cids-survey}.
%Such {\em collaborative} generate a global alert

Behavioral detectors are a crucial line of defense against malware. By
extracting features out of network packets~\cite{wenkee_lee_bothunter_2007,
paxson1999, Sommer2010, zhang2001}, system
calls~\cite{Canali2012,Christodorescu06,Hofmeyr1998, Mutz2006, Somayaji2000,
Warrender1999, Fredrikson2010}, instruction set~\cite{Juxtapp,
Christodorescu05}, and hardware~\cite{Demme2013, Tang2014, dmitry-ensemble}
level actions, behavioral detectors train machine learning algorithms to
classify program binaries and executions as either malicious or benign.  In
practice, enterprises extensively deploy behavioral detectors as
per-machine {\em local} detectors whose alerts are analyzed by an
enterprise-wide {\em global} detector
~\cite{cisco-amp,dell-ampd,osquery,GRR,hone,hone_code}.  Our goal is to design
a robust global detector that composes {\em weak} local detectors in a {\em noisy}
community. 

Behavioral detectors are weak -- i.e., have high false positives and negatives
-- because a large class of malware includes benign-looking behaviors, such as
encrypting users' data, use of obfuscated code, or making HTTP requests.
Further, machine learning-based detectors have been shown to be susceptible to
evasion attacks~\cite{evade-ml,evade-pdf,practical_black_box_attacks} that
either increase false negatives or force detectors to output more false
positives. In practice, global detectors in enterprises with $\sim$100K local
detectors have to process millions of alerts per day~\cite{attack_graphs} which
stresses heavy-weight program analyses and human analysts who investigate the
final alerts~\cite{graphistry}.  

Furthermore, local detector communities are {\em noisy}, where local machines
often fail to report alerts or report them (often, months)
late~\cite{wine,accenture-blackhat}. This noise is because machines often go out of network
access, users decline to send reports, etc.  Enterprise settings are also noisy
because attacks might target local machines in unpredictable ways -- 
%a community of machines in an enterprise who are potentially
in a `waterhole' attack~\cite{waterhole_attack} (where a compromised webpage
spreads malicious code to machines in the enterprise), a malicious
javascript-advertisement might be targeted by an ad-broker to only a fraction
of visitors to a set of webpages;  the specific exploit might only succeed on a
small fraction of recipient machines because of browser versions or patching
status or human-user actions, etc.

%Local Detection has high false positives []. A global detector like Tudor's
%has high false negatives [].  We observe that attack vectors that malware
%spreads along provide an opportunity to amplify the weak local detectors'
%outputs -- the challenge is that prior methods to do so are not robust to
%noise in the local detector community.

\noindent{\bf Challenges for prior global detectors.} Boosting weak detectors
using purely machine learning techniques is challenging.  The dominant
approaches are (a) clustering: combine feature vectors using some distance
metric to identify suspicious clusters of feature
vectors~\cite{Seurat2004,beehive2013,bot_graph,bot_grep}, and (b) counting:
train local detectors (LDs) such as Random Forest or gradient-boosted trees 
to generate local alerts,
and generate a global alert if there is a significant fraction of local alerts
in the
enterprise~\cite{Dash2006,BotSniffer,wenkee_lee_bothunter_2007,Shin_Infocom12_EFFORT}.
Both approaches have limitations that force enterprises to deploy brittle
rule-sets that explicitly correlate local detector alerts.%~\cite{rules}.

%Clustering high-dimensional data is well-known to be
%ineffective~\cite{todo}. Further, counting alert-intensity in a community of
%local detectors is ineffective in early stages of infection when the signal is
%weak. 

Clustering algorithms are well-known to be highly sensitive to noise,
especially in the high-dimensional
regime~\cite{donoho1983notion,huber2011robust,xu2013outlier}. Indeed, classical
approaches that attempt to detect or to score "outlyingness" of points (e.g.
Stahel-Donoho outlyingness, Mahalanobis distance, minimum volume ellipsoid,
minimum covariance determinant, etc) are fundamentally flawed in the
high-dimensional regime (i.e., theoretically cannot guarantee correct detection
with high probability).  \ignore{In practice, we see this in prior
work~\cite{beehive2013} where clustering is used primarily as a first-level
analysis to discover malicious incidents for a human analyst (i.e. requires
lower accuracy than a global detector).} In Appendix~\ref{sec:eval-clustering},
we demonstrate how a clustering global detector is ineffective in detecting a
waterhole infection -- i.e., clustering yields an Area Under
Curve (AUC) metric of only $\sim$ 48\% in the waterhole attack.
%\ignore{ interpret some clusters as outliers for human analysts to analyze
%further~\cite{Seurat2004,beehive2013}. Variants of these count- and
%clustering-based GDs %(e.g. count number of alerts, search for sizeable
%clusters in feature space) are deployed by security companies such as
%Cisco~\cite{cisco-amp}, Dell~\cite{dell-ampd}, Facebook~\cite{osquery}, etc.\\
%}

Count-based global detectors (Count-GD), on the other hand, suffer because they
need to know the size of local detector communities extremely accurately to
determine whether a significant fraction is raising alerts.  
%\mikhail{In the experiments with the Wine dataset, we found that neighborhood
%sizes vary by 1600\% around mean.}\mohit{Not sure measuring variance is useful
%here -- need to show these are hard to model?}
Fundamentally, even small errors in estimating the number of feature vectors in
the community {\em linearly} affects the global detector's decision thresholds.

\noindent{\bf Proposed Ideas -- Neighborhood filtering and Shape.} Our
intuition is that weak local detectors can be aggregated robustly by using
information about how the malware spreads.
%a weak signal indicative of malicious behavior still separates true- and
%false-positive feature vectors, even though local detectors classify both as
%malicious.  
Our proposed system (Figure~\ref{fig:sys}), \sysname{}, relies on two key insights to correctly identify
malicious feature vectors.

First, while attacks can take many forms, attack vectors are easier to
identify. For example, many attacks on Symantec's client machines rely on
`downloader trojans' to bring successive stages of payloads -- hence, {\em
downloader graphs}~\cite{downloader_graphs} on a machine are correlated with malware propagation. 
Similarly,
%attack vectors into a firewalled enterprise create short-lived and dynamic
%correlations across nodes -- e.g., 
in a firewalled enterprise, machines that visit a specific server (in watering
hole attacks) are more
likely to be compromised than a random machine in the enterprise.  
%Since an attacker cannot directly target a machine inside an enterprise,
Our key assumption is that {\em machines that have been exposed to a common
attack vector have correlated alerts} -- we call such a set of machines a {\bf
neighborhood}.  Grouping local detectors into 
neighborhoods (as they form dynamically) 
concentrates the signal of malware activity
that is otherwise not visible at the overall community level. 
%and can thus
%enable early detection of malware attacks.  
However, neighborhoods are extremely noisy due to 
exploit-types, machine status, and human usage 
and render cluster and count-based GDs ineffective --
hence we propose Shape-GD to aggregate local detectors' outputs.

The second insight behind Shape-GD is that the {\em distributional shape of a
set of suspicious feature vectors} can robustly separate true positive
neighborhoods from false positive neighborhoods.  Shape-GD analyzes only
those feature vectors that cause alerts by the local detectors ({\em
alert-FVs}) instead of analyzing all feature vectors. Alert-FVs thus represent
draws from one of two {\em conditional distributions} -- i.e., distribution of
malicious or benign feature vectors conditioned on being labeled as malicious
-- which are similar but not the same.  Next, while a single suspicious feature
vector is uninformative, a set of such feature vectors (i.e., alert-FVs from
a neighborhood) can indeed be tested to
come from one of two similar-but-distinct distributions.  

%To conduct this
%hypothesis test, Shape-GD introduces a quantitative scoring function that maps a
%set of feature vectors (the alert-FVs per neighborhood) into one
%scalar value -- the ShapeScore of the neighborhood.  

\noindent{\bf Case Studies.} We consider two distinct case studies where
\sysname is applied in noisy communities of weak behavioral detectors -- one
with long-term log entries from a commercial detector, and the other a
real-time attack simulated using enterprise traces.
  
Our first setting comprises of 5 million client machines monitored by malware
detectors (here, Symantec~\cite{wine}). A local detector
algorithm~\cite{vt_report_classification} that analyzes file attributes using
VirusTotal, when applied to this Symantec Wine dataset~\cite{wine}, achieves a
false positive rate of 5\% -- with 5 million local detectors in place, this
requires deeper human or program analysis of up to $\sim$1.1M files to detect
close to 137K malware files.
A recent local detector improves false positive rates down to 1\% 
by training on metadata,
such as features extracted from `downloader
graphs'~\cite{downloader_graphs,pup_ndss}, but this increases false negatives
since it only detects malicious downloaders (that install malware on devices)
which comprise only $\sim$32.7\% of the overall malware in the community. 

%at a low false positive rate of 1\%. 

Our second setting is an enterprise whose devices 
are infected through a compromised server
(waterhole attack), where each device also runs a local system-call based malware
detector~\cite{Canali2012} and  sends reports to a global detector.
We reimplemented system call based local detectors to achieve representative
detection rates~\cite{Canali2012} -- where a true positive rate of 92.4\%
yields a false positive rate of 6\%.

We show that in the Symantec Wine case study \sysname detects malicious
neighborhoods early -- with more than 5\% of malicious files -- 
at a false positive and true positive rate of 5.8\% and 84\% 
respectively. And it achieves 0.54\% false positive rate and 78\% true positive 
file-level detection results.
In the waterhole case study it detects malicious neighborhoods with less than 
1.1\% compromised nodes per neighborhood
at a false positive and true positive rate of 1\% and 100\% respectively.

\textit{Neighborhood filtering} and \textit{shape property} complement each
other -- neighborhoods concentrate the weak signal into a small but
unpredictable set of feature vectors while shape extracts this signal without
knowing the precise number of feature vectors.  
In contrast, our experiments show that when applied to noisy neighborhoods,
Count-GD's detection performance 
only
matches \sysname{}'s detection performance
if it can estimate neighborhood size to within
%if its underestimation error of the number of nodes where the exploit ran
%successfully (i.e., the community size) is lower than 
-30\% to +1\% for the Symantec case study and
-0.1\% to +13.8\% in the waterhole attack -- this makes
CountGD extremely fragile in real-world distributed
systems.
%Similarly, an overestimation error should be
%lower than 1\% (Symantec Wine) and 13.8\% (waterhole).  
To summarize, \sysname{} enables practitioners' 
insights about attack vectors to be captured algorithmically
and at scale.

\section{Overview of \sysname}
\label{sec:overview}

%Community deployment and threat model.
\noindent{\bf Threat model and Deployment.} We assume a standard threat model
where trusted local detectors (LDs) at each machine communicate with a trusted
global detector (GD) that receives alerts and other metadata from the local
detectors. The LDs are isolated from untrusted applications on local machines
using OS- (e.g., SELinux) and hardware mechanisms (e.g., ARM TrustZone), and
communicate with the enterprise's GD through an authenticated channel. 
%\st{The GD
%is trained as a standard anomaly detector -- using benign data generated from
%uninfected (e.g.  test/quality-assurance) machines that run LDs on benign
%software, or assuming the current state of the system as benign in order to
%detect future malware as anomalies.}

\sysname fits deployment models that are common today.  Currently, enterprises
use SIEM tools (like HP Arcsight and Splunk) to monitor network traffic and
system/application logs, malware analysis sandboxes that scan emails for
malicious links and attachments, in addition to host-based malware detectors
(LDs) from Symantec, McAfee, Lookout, etc. We use exactly these
side-information -- from network logs (client-IP, server IP, timestamp) and
email monitoring tools -- to instantiate neighborhoods and filter LDs' alert-FVs
based on neighborhoods (Algorithms~\ref{symantec_nbd_algo},~\ref{waterhole_nbd_algo}). 
Upon receiving
alert-FVs, \sysname runs its malware detection algorithm (Algorithm
\ref{ShapeGD_algorithm}) for all neighborhoods the alert-FVs belong to. If a
particular neighborhood is suspicious, then \sysname will notify a downstream
analysis (deeper static/dynamic analyses or human analysts) and forward
relevant information in the incident report. 

\ignore{
The key difference is that \sysname needs to know the alert feature vectors from the
LDs -- black-box LDs do not currently provide these. Hence (e.g., {\tt
osquery}-based) co-designed LD-GD detectors~\cite{osquery,GRR} are the most
appropriate deployment counterpart for \sysname{} -- this also motivates our
experimental setup combining host-level malware analysis and web-service/email
datasets.
}

%\st{
%Operationally, the LD at each machine transforms its input signal into an alert
%time series. This transformation consists of two steps: \textit{(a) Generate
%Feature Vectors:} convert the raw OS system calls trace into a feature vector
%(FV) time series, and \textit{(b) Generate Alerts:} Determine if each FV is
%malicious or not using a local detector (typically through random forests, SVM,
%etc.).  Crucially however, at an individual node level, it is impossible to
%distinguish a false alert (false positive) from a true alert (a true positive). 
%}

%Neighborhoods based on attack vectors.  \noindent{\bf Commmon attack vectors
%create neighborhoods.}
\noindent{\bf Inferring neighborhoods from common attack vectors.} \sysname
operates over dynamic neighborhoods, which are updated once per neighborhood
time window (NTW).  Neighborhoods within large communities are a set of
nodes that share a statically defined {\em action attribute} within the current
time window -- this allows an analyst to create neighborhoods of nodes based on
common attack vectors.
Below are some illustrative examples of communities and neighborhoods.
%\mikhail{that are responsible for a large fraction of malware in enterprises.}
% -- we
%focus our experiments on the first two examples that are responsible for a
%large fraction of malware in enterprises.

\noindent {\em 1.} Malware propagation across Symantec clients.
The community here consists of all Symantec clients.
Though attackers, when distributing malware through compromised websites, 
may not have an intention to target Symantec clients' machines, they get infected due to high
number of subscribers to Symantec malware detection service. In the Symantec dataset,
both benign and malicious files launch a chain of downloads.
Thus, a neighborhood can comprise a set of files transitively downloaded from
a suspicious domain (Section~\ref{subsec:neighborhood_instantiation}). 
As domains get periodically cleared out,
and their classification is not necessarily very robust, neighborhoods
only indicate a probability that files within them may be malicious.

\noindent {\em 2.} Waterhole attack. The community here consists of the
employees of an enterprise such as Anthem Health~\cite{anthem,
anthem-waterhole}.  In a waterhole attack, adversaries compromise a website
commonly visited by such employees as a way to infiltrate the enterprise
network and then spread within the network to a privileged machine or user.
Within this community, a neighborhood can be the set of nodes that visited the
same type of websites within the current neighborhood time window (for example,
some percentile of suspicious links rated by VirusTotal~\cite{virustotal} or
SecureRank~\cite{secure_rank}). Since these rankings themselves are fuzzy, and
the websites and their contents are dynamic, neighborhoods only indicate a
probability that the node was actually exposed to an exploit.
 
\ignore{
\noindent {\em 2.} \mikhail{drop?} Phishing attack over enterprise email networks.  The
community here consists of all employees within an enterprise.  A phishing
attack here would typically spread over email and use a malicious URL to lure
nodes (users) to drive-by-download attacks~\cite{phishing-1,phishing-2} or
spread through malicious attachments.  Here, a specific user's neighbors are
that subset of users with whom she/he exchanged emails with during the current
neighborhood time window.
}

%\mikhail{drop it? let's stay focused!}
%\st{
%Similar correlations occur in physical hardware attacks -- community here
%consists of all machines in a workplace that are physically proximal (e.g.
%machines in a specific hospital or bank determined using the configuration of
%LAN/WiFi infrastructure, GPS information etc).  The potential attack mode here
%is through physical hardware such as {\tt badUSB}. The neighborhood of a node
%is simply all other nodes in the neighborhood that were connected to similar
%external hardware (e.g. a USB drive) over the current neighborhood time window.
%}
%\st{
%Attacks that target specific app-stores (e.g., the Key-Raider attack in the
%Cydia app-store or the malicious Xcode attacks due to compromised mirror sites)
%also propagated across users with specific attributes (membership in a store or
%downloaded Xcode from specific sites) more likely than a random user in the
%network.
%}

%%%%%%%%%%%%%%%%
\noindent{\bf Intuition behind shape property.}
%\subsection{Intuition behind \sysname}
\label{sec:intuition}
\begin{figure*}[t]
\begin{minipage}[tbp]{0.24\linewidth}
   \includegraphics[width=\textwidth]{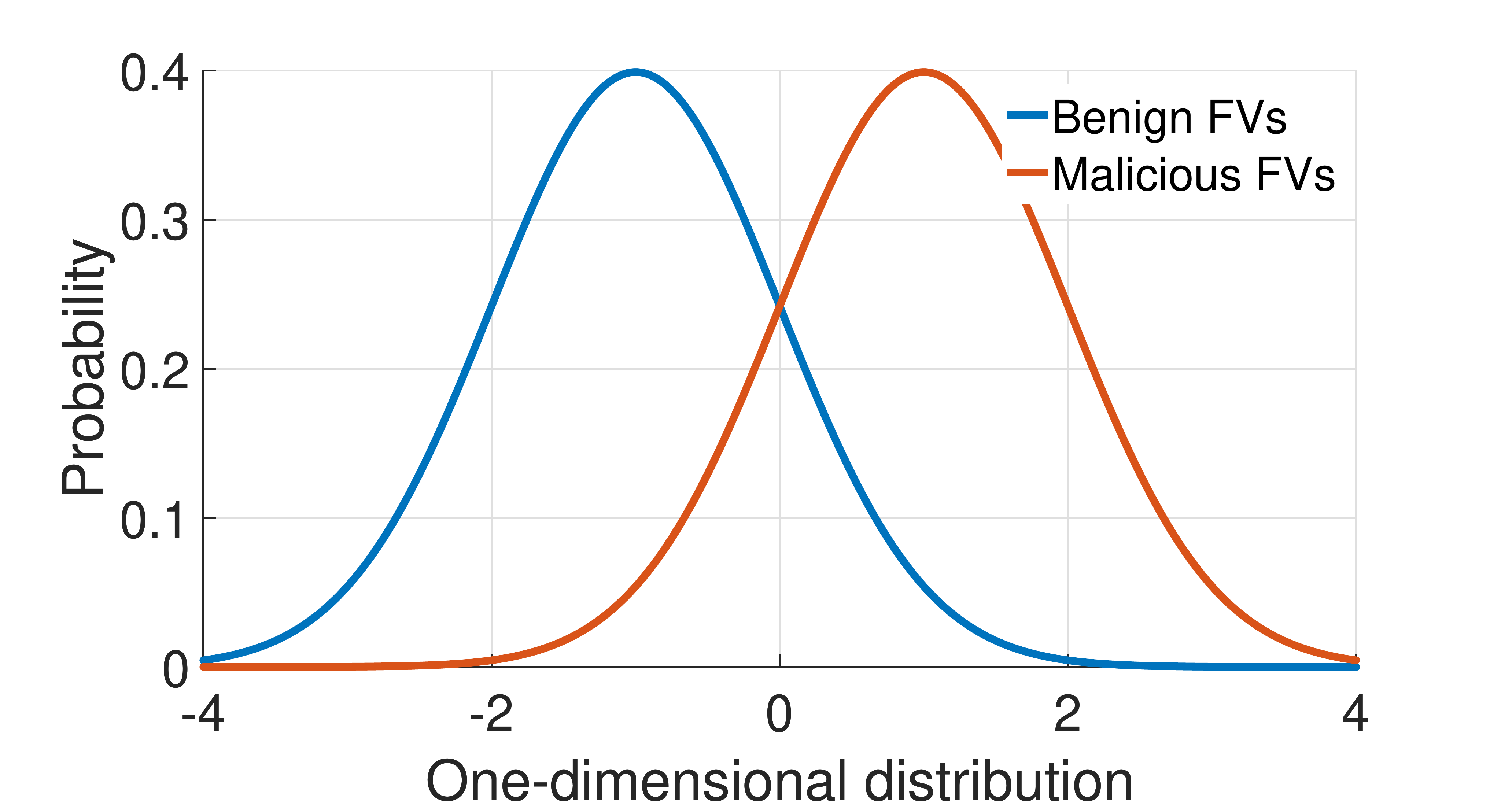}\par
   \label{fig:intro_plot_original_pdfs}
\end{minipage}
\begin{minipage}[tbp]{0.24\linewidth}
   \includegraphics[width=\textwidth]{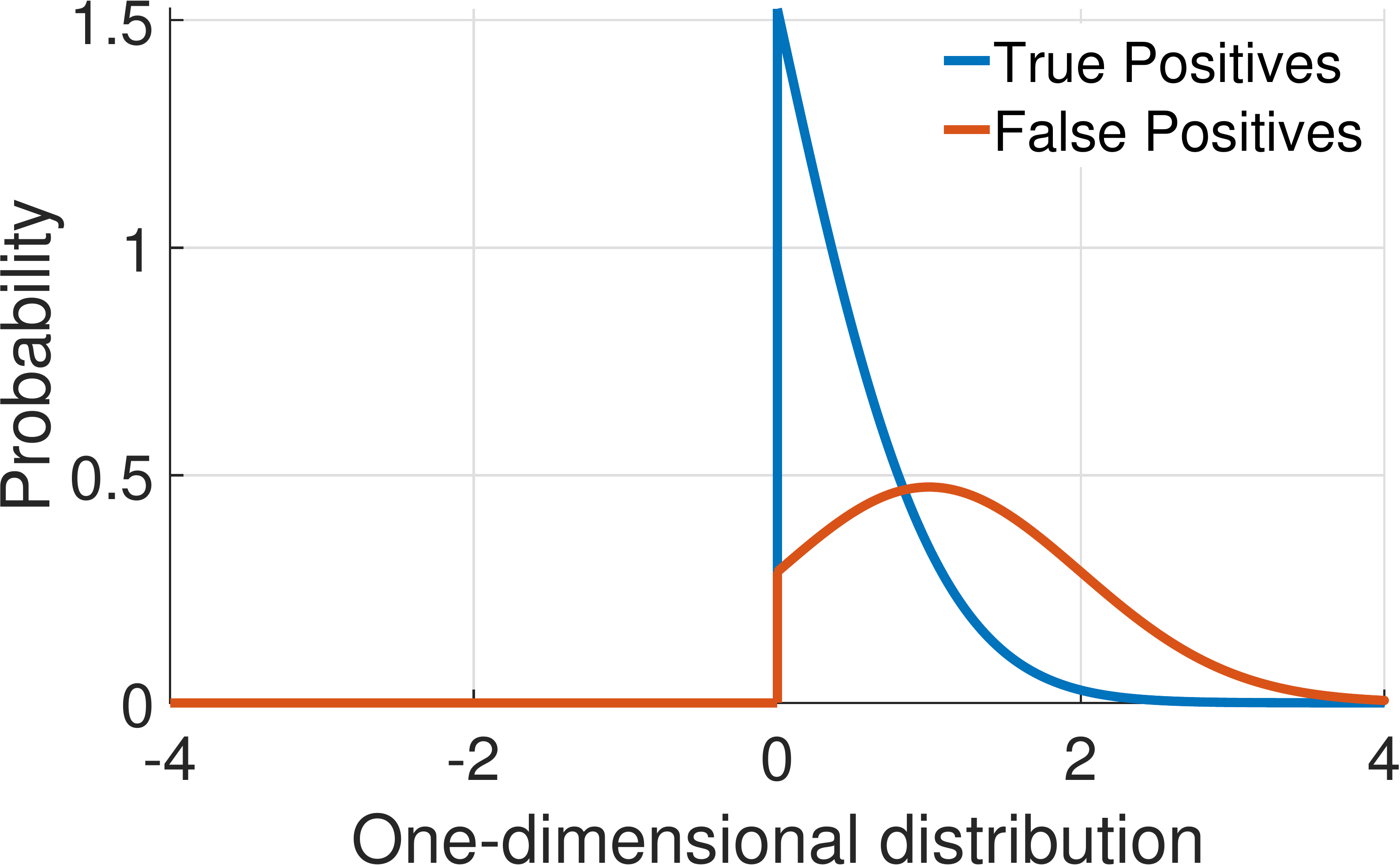}\par
   \label{fig:intro_plot_cond_pdfs}
\end{minipage}
\begin{minipage}[tbp]{0.24\linewidth}
   \includegraphics[width=\textwidth]{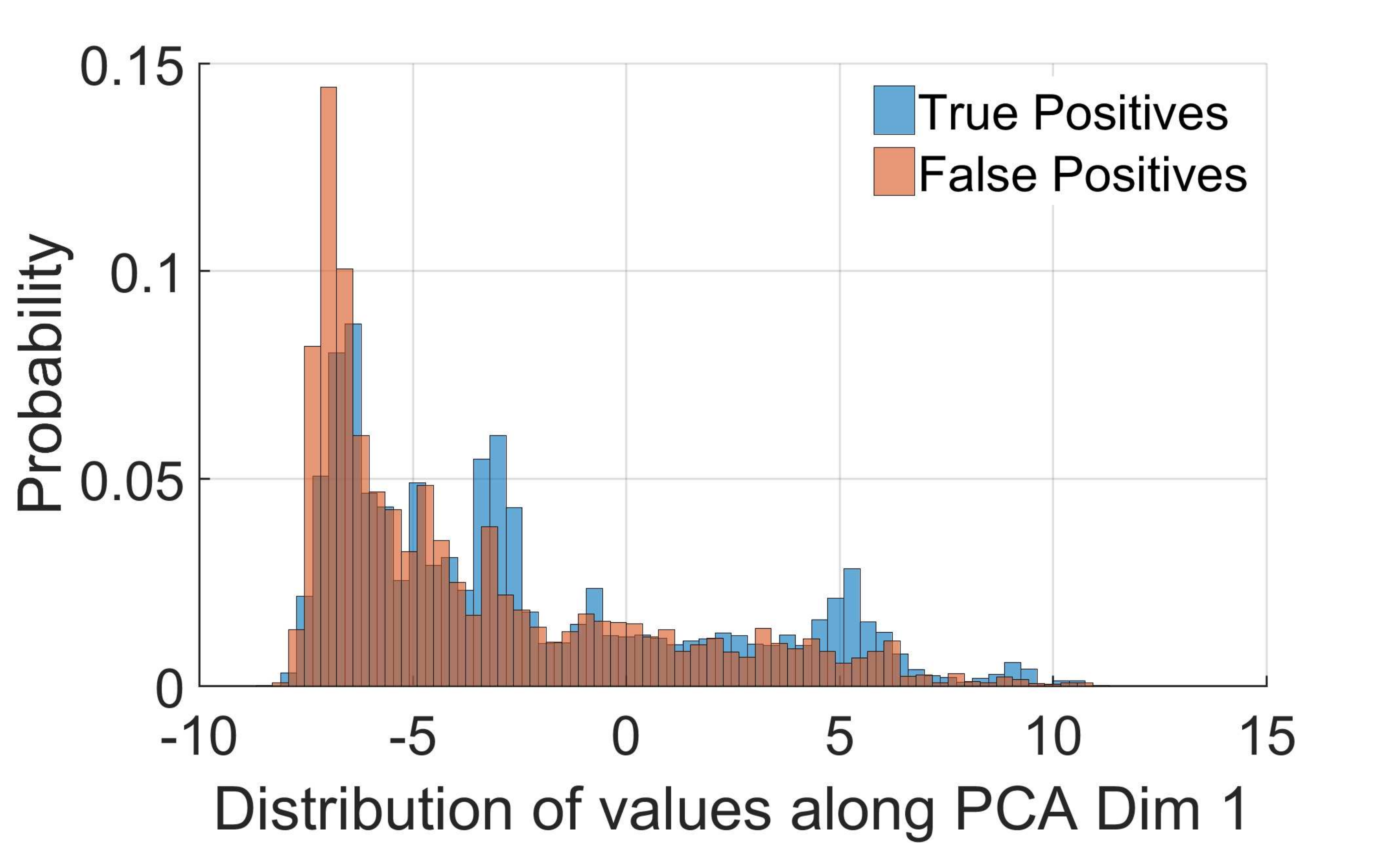}\par
   \label{fig:intro_plot_VT_cond_pdfs}
\end{minipage}
\begin{minipage}[tbp]{0.24\linewidth}
   \includegraphics[width=\textwidth]{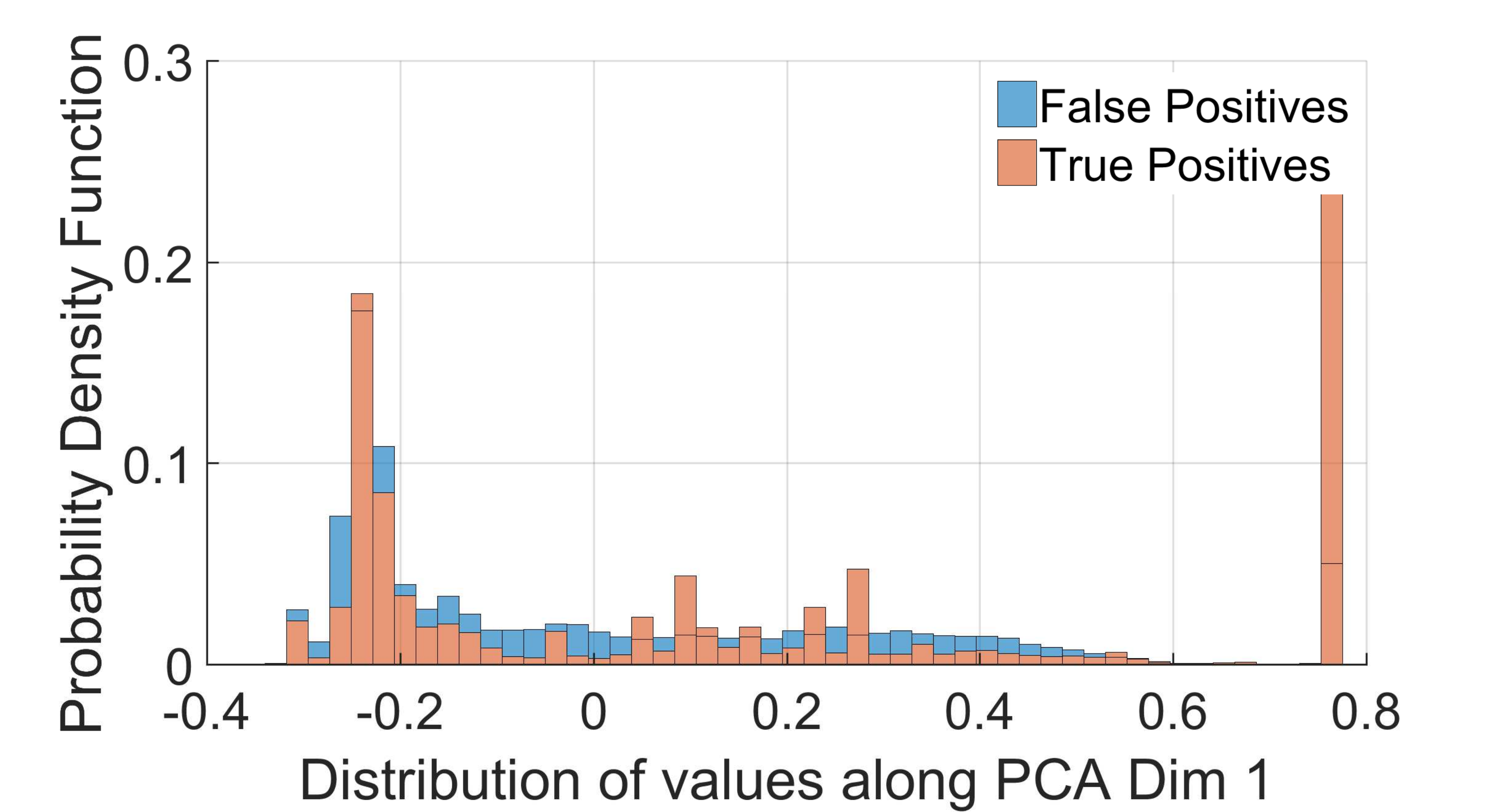}\par
   \label{fig:intro_plot_syscall_cond_pdfs}
\end{minipage}
%
%\caption{
%	{\em (Shape of conditional distributions) 
%	\mikhail{
%		The left figure is the probability density function (pdf) 
%		of benign/malicious feature vectors in a stylized example,
%		which are drawn from the Gaussian with mean `-1/+1'. 
%     The optimal
%     local detector at any machine would declare 'malware' if a sample's
%     value is positive, and declare 'benign' if a sample's value is
%     negative. The figure in the middle shows the pdfs of the same Gaussians
%     but now conditioned on the event that the sample is
%     positive -- the pdfs corresponding
%     to false positive and true positive feature vectors respectively 
%     have different shapes.	
%     And the figure on the right shows the difference 
%     between shapes of true positive and false positive feature vectors
%	 extracted from VirusTotal reports
%     and projected on the first principal component.
%	}
%     }}
\caption{
	{\em (Shape of conditional distributions) (L to R)
	Probability density function (pdf) of benign/malicious feature vectors (FVs) in a stylized example,
	which are drawn from the Gaussian with mean `-1/+1'.	
	PDFs of the same Gaussians, but now conditioned on a local detector raising an alert --
	PDFs of true/false positive FVs have \textbf{different shapes}.			
	Real-world PDFs of true/false positive multi-dimensional FVs projected on the first two principal 
	components in Symantec Wine and waterhole case studies respectively.	
%	\vspace{0.in}	
     }}
\label{fig:cond-dist}
\end{figure*}
The statistical shape of local detectors' false positives (FP
conditional distribution) differs from the corresponding shape for true
positives (TP conditional distribution) -- we use this property to aggregate
LDs' alert-FVs to find the shape of each neighborhood and then classify
neighborhoods based on their shapes.

The central question then is -- {\em why do true- and false-positive FVs'
shapes differ?}  To explain this and set the stage for \sysname, we consider a
stylized statistical inference example. Suppose that we have an unknown number
of nodes within a neighborhood. We want to distinguish between two extremes --
all nodes only run benign applications (benign hypothesis), or all nodes 
run malware (malware hypothesis). We look at a single snapshot of time
where each node generates exactly one feature vector. Under the benign
hypothesis, assume that the feature vector from each node is a (scalar valued)
sample from a standard Gaussian with mean of `-1'; alternatively it is standard
Gaussian with mean of `+1' under the malware hypothesis (Figure~\ref{fig:cond-dist}) (leftmost plot). 
The optimal local detector at any machine would declare 'malware' if a sample's
value is positive, and otherwise -- 'benignware'.
     
%\mohit{\st{At each node,
%malware behavior is only slightly different from benignware and leads to false
%positives -- aggregated over many nodes that each belong to a neighborhood, the
%slight difference gets concentrated and becomes visible as the infected
%neighborhood's shape being measurably different from a neighborhood that has
%only false positives.}}

%\mikhail{
Even though individual false and true positives are indistinguishable
at the local detector level, we can differentiate between them by approximating
distributions they come from. To do this we need to aggregate alerts at the neighborhood
level. These values represent independent draws
from a {\em conditional} distribution -- either the distribution of a normal
random variable of mean '$-1$' {\em conditioned on taking a nonnegative value},
or the distribution of a normal random variable of mean '$+1$' {\em conditioned
on taking a nonnegative value}. 
This conditioning occurs because of the local detector tags a sample as
an alert if and only if the sample drawn was non-negative.  
Thus, irrespective of the size of the neighborhood, the global
detector would ``look at the shape'' of the empirical distribution of the received FVs.
If it is ``closer'' to the distribution of false positives rather than
the distribution of true positives (Figure~\ref{fig:cond-dist}) (second from the left), it would declare
a neighborhood to be ``benign'', otherwise -- ``malicious''.

\noindent{\bf Shape in real datasets.}
Though in the stylized one-dimensional example it is straightforward to distinguish
between benign and malicious neighborhoods, real-world multidimensional distributions
in Symantec Wine and waterhole case studies
do not allow such simple interpretation (Figure~\ref{fig:cond-dist}) (two plots on the right).
Figure~\ref{fig:cond-dist} (two plots on the right) shows conditional distributions
of false and true positives projected on the first two principal components 
in Symantec Wine and waterhole case studies respectively.
We show that the intuition behind this simple example scales to
real malware detectors that use high-dimensional feature vectors. However, to
use this insight in practice, we need to address two issues: {\em (i)} while
corresponding conditional distributions are visually distinct, an algorithmic approach
requires a quantitative score function to separate between the (vector-valued)
conditional distributions generated from feature vector samples; and {\em (ii)}
the global detector receives only finitely many samples; thus, we can construct
(at best) only a noisy estimate of the conditional distribution.  We describe
\sysname{}'s details in Section~\ref{sec:model}.

%% file: related-work.tex
\subsection{Behavioral analysis}

Behavioral analysis refers to statistical methods that monitor signals from
program execution, extract features and build models from these signals, and
then use these models to classify processes as malicious.  Importantly, as we
discuss in this section, all known behavioral detectors have a high false
positive and negative rate (especially when zero-day and mimicry
attacks are factored in).

%such as system calls, network traffic, instruction-level
%behavior etc, 

% System calls and parameters have been studied extensively
%Signals for behavioral analysis include
System-calls and middleware API calls have been studied extensively 
as a signal 
%are very commonly used signals 
for behavioral detectors~\cite{Forrest1996,
Wagner2002,
kruegel2007,
Christodorescu:2008:MSM:1342211.1342215,
Fredrikson2010,
Canali2012, robertson_anomaly}.
%Burguera2011,Kolbitsch:2009:EEM:1855768.1855790,Xie2010,
%rieck2008,bose2008,Mutz2006}
%,Forrest1996,Hofmeyr1998,Warrender1999,Somayaji2000}.
%Specifically for Android, Crowdroid\cite{Burguera2011} 
%uses system call traces sourced from
%several devices as the sole signal in its anomaly detection.  
Network intrusion 
detection systems~\cite{paxson1999}
%HIDE~\cite{zhang2001}, EMERALD~\cite{porras1997}, BotSniffer~\cite{Gu2008}, and 
%NetBiotic \cite{vlachos2004} 
analyze network traffic to detect known malicious
or anomalous behaviors.
More recently, behavioral detectors use signals such as power
consumption\cite{clark2013}, CPU
utilization, memory footprint, and hardware
performance counters\cite{Demme2013,Tang2014}. 

%\ignore{Specifically on 
%Android, Andromaly\cite{Shabtai2012}
%uses power consumption and CPU utilization, in addition to syscalls,
%network flow, and control flow, as signals that are
%tracked at runtime.}

%Features go Here Rather than just saying whats popular, lets talk about how
%these features are useful in the statistical model building process Up the
%citation count
Detectors then extract {\em features} from these raw signals.  For example, an
n-gram is a contiguous sequence of n items that 
%is useful for their simplicity and their ability to 
captures total order relations~\cite{BotSniffer,Canali2012}, n-tuples are ordered
events that do not require contiguity, and bags are simply histograms.
%Note, the dimensionality of the feature space is directly proportional to the
%value of n, however this comes at the expense of increased complexity and
%memory requirements. Dimensionality these features to compress the feature
%space into a smaller form that remains representative.  N-tuples are a strict
%relational ordering of n-grams, without requiring contiguity. n-tuples are
%useful in non-trivial combinations of
These can be combined to create bags of tuples, tuples of bags, and tuples of
n-grams~\cite{Canali2012,Forrest1996}
%,Hofmeyr1998,Warrender1999}
often using principal component analysis to reduce dimensions.
%Bags are simply an unordered count of their contents. For example, bags may
%contain n-grams, tuples, or simply the original signal. Bags are useful
%because they describe the data in terms of its frequency
%distribution~\cite{Canali2012}. %Notable examples To reduce dimensions,
%analysts uses techniques such as include~\cite{Burguera2011,Canali2012,
%rieck2008,Warrender1999}. Rieck et al.~\cite{rieck2008} proposed a framework
%for automatically classifying malware samples into groups with similar
%behaviors.  
Further, system calls with their arguments form a dependency graph structure 
%spatio-temporal relations among features.  As~\cite{Canali2012} pointed out,
%complex combinations of N-grams, Tuples, and Bags can approximate these
%graphs. However, there are algorithms from graph
that can be compared to sub-graphs that represent malicious
behaviors~\cite{Christodorescu:2008:MSM:1342211.1342215,Kolbitsch:2009:EEM:1855768.1855790,kruegel2007}.
%theory that are useful for comparing graphs and subgraphs

% Anomaly start by listing overlapping methods Talk about hardware performance
% counters in the signals section
%Anomaly detection research use many different algorithms to extract alerts
%from
Finally, detectors train models to classify executions into malware/benignware
using supervised (signature-based) or unsupervised (anomaly-based) learning.
These models range from distance metrics, histogram
comparison, hidden markov
models (HMM), and neural networks (artificial neural
networks, fuzzy neural networks, etc.), to 
more common classifiers such as
%classification 
kNN, one-class SVMs,
decision trees, and ensembles thereof. 
%Andromaly evaluates histogram comparison, k-Means, logistic regression,
%decision trees, and baysian networks algorithms for its anomaly detection. 
%\ignore{
%pBMDS\cite{Xie2010} use HMMs to correlate
%%probabilistic approach to anomaly detection by 
%user inputs with system calls, whereas Xu et al.~\cite{Xu2004} apply sequence
%alignment techniques to known and unknown malicious binaries to detect
%polymorphic strains.
%}

%Results
%Mobile platforms face resource limitations not evident in their PC
%counterparts. Although smartphones have become more powerful in recent years,
%minimizing analysis and detection overheads is still a top priority. Host-based

Such machine learning models, however, result in high false positives and
negatives.  
Anomaly detectors can be circumvented by mimicry attacks where
malware mimics system-calls of benign applications~\cite{Wagner2002} or
hides within the diversity of benign network traffic\cite{Sommer2010}.
Sommer et al.~\cite{Sommer2010} additionally highlights 
several problems that can arise due
to overfitting a model to a non-representative training set, suggesting signature-based
detectors as the primary choice for real deployments. Unfortunately, 
signature-based detectors cannot detect new (zero-day) attacks.
%show how mimicry attacks can circumvent system-call based detectors.
%Similarly, 
On Android, both system
calls~\cite{Burguera2011} and hardware-counter based
detectors~\cite{Demme2013} 
%results for hardware performance counters to detect
yield $\sim$20\% false positives and $\sim$80\% true positives.

Finally, with their
ability to extract highly effective features, deep nets {\em may} provide a
new way forward for creating novel behavioral detectors. At the global
level, however, what is needed is a data-light approach for global
detection by composing local detectors, tailored to be agile enough to
do global detection in a fast-changing (non-stationary) environment.

\subsection{Collaborative Intrusion Detection Systems (CIDS)}
\label{sec:cids}

% Can You infect me now?
% Malware Propagation in Large Scale Networks
% Smartphone Malware and Its Propagation Models: A Survey
% Modeling Malware Propagation in Smartphone Social Networks
% Why mobile-to-mobile Malware wont cause a storm
% BIG: Modeling and Restraining Mobile Virus Propagation

%Intrusion detection systems (IDS) monitor local or network operations for
%suspicious activities. 
Collaborative intrusion detection systems (CIDS) provide an architecture where
LDs' alerts are aggregated by a {\em global detector} (GD).
%correlate information across many per-device local detectors of these
%intrusion \emph{monitors}, improving the detection of intelligent local
%attacks. In particular, local detectors emit a stream of alerts and network
%overlays allow the system to determine neighborhoods.  The global detector
%takes the local detector streams and the neighborhoods from the network
%overlays and correlates the local detectors' alerts, considering only a
%neighborhood at a time.  Local Detector
GDs can use either signature-based or 
%kannadiga2005 
%snapp1991, cuppens2002,
%yegneswaran2004, 
%duma2006},
anomaly-based\cite{zhang2001,vlachos2004}, or even a combination
of the two~\cite{krugel2001} to generate global alerts.
%Signature-based detectors are similar to malware detectors in that ther are
%unable to address unknown attacks. Anomaly detectors, however, are able to
%handle unknown attacks but usually suffer from a high false positive rate.
%Simple local detectors such as those in DIDS\cite{snapp1991} and
%DIDMA\cite{kannadiga2005} throw labeled flags for all suspicious activity
%(e.g. failed login attempts, modification of system files, telnet usage,
%rlogin usage, etc.). On the other hand, systems like HIDE\cite{zhang2001} emit
%statistical alerts in the form of random variables.
Additionally, the CIDS architecture can be centralized, hierarchical,
or distributed (using a peer-to-peer overlay network)~\cite{zhang2001}.

In all cases, existing GDs use some variant of either clustering or count-based
algorithms to aggregate LDs' alerts. Count-based GD raises an alert once the
number of alerts exceeds a threshold within a space-time window, while
clustering-based GD may apply some heuristics to control the number of alerts
~\cite{Dash2006,Seurat2004,BotSniffer,wenkee_lee_bothunter_2007,Shin_Infocom12_EFFORT}. 
%Beyond the GDs mentioned in the introduction, here we describe other seminal
%work \ignore{In all cases, existing GDs use some variant of count-based
%algorithms to aggregate LDs' alerts} \ignore{once the number of \st{alerts
%exceeds a threshold within a space-time window, the GD raises an intrusion
%alert.}
In HIDE~\cite{zhang2001}, 
%uses a multi-tiered system where 
the global detector at each hierarchical-tier is a neural network trained on
network traffic information.  Worminator\cite{Locasto2005} additionally uses
bloom filters to compact LDs' outputs and schedules LDs to form groups in order
to spread alert information quickly through a distributed system.  All count-
and clustering-based algorithms are fragile when the noise is high (in the
early stages of an infection) and when the network size is uncertain.
%Our GD uses shape information and dynamically censors the output
In contrast, our neighborhood filtering and shape-based GD is robust against
such uncertainty.

%We focus on early stage detection of zero-day infections and on
%aggregating alerts in 
%a network where the membership information is unreliable. 

Note that distributed CIDSs are vulnerable to probe-response attacks, where the
attacker probes the network to find the location and defensive capabilities of
an LD~\cite{Shmatikov2007,Bethencourt2005,Shinoda2005}. These attacks are
orthogonal to our setting since we do not have fixed LDs (i.e. all nodes are
LDs).

%% file: algorithm_3.tex
\begin{figure}[tbp]
   \vspace{-0.0in}
   \centering
   \includegraphics[width=0.45\textwidth]{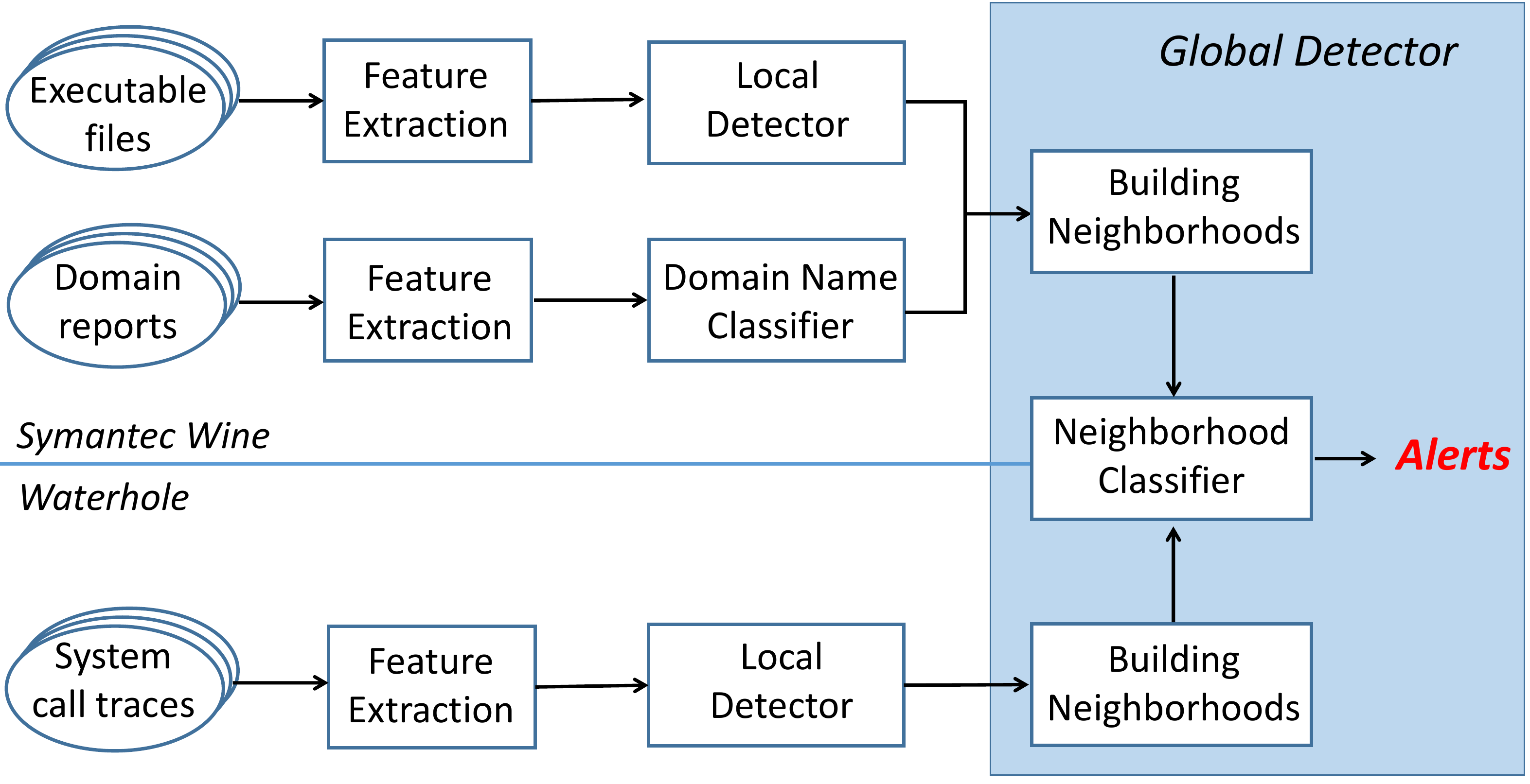}\par
   \caption{Application of \sysname{} to malware detection in Symantec Wine and waterhole case studies.}
   \vspace{-0.2in}
\label{fig:algo_schema}
\end{figure}

The algorithm consists of feature extraction, local detectors (LDs), and the
global detector (GD).
Figure~\ref{fig:algo_schema} shows how to apply \sysname to malware detection in Symantec Wine and waterhole case studies. 
Our key innovations are in the GD.
LDs' design is inspired by prior work, therefore
we discuss it in details in the Appendix~\ref{app:local_detectors} and briefly summarize LDs detection 
performance in Sections~\ref{sec:results-symantec},~\ref{sec:results-waterhole}.

\subsection{\sysname Classifiers}
\label{sec:local_detectors}
\sysname utilizes two types of local detectors that analyze executable files 
and a domain name classifier that analyzes domain metadata.
%three types of local detectors: two of them analyze executable files 
%and the other one analyzes domain metadata to predict whether a domain is likely to distribute malware samples.
To perform file analysis, we adapt local detectors from prior work. 
The Symantec Wine data set lacks executable files (it includes only their hashes), therefore 
we use VirusTotal~\cite{virustotal} file analysis reports. 
Our detector combines feature extraction described in a prior work~\cite{vt_report_classification} and
a standard machine learning classifier -- XGBoost~\cite{xgboost}.
In the waterhole case study, we develop a detector that extracts feature vectors from dynamic sequences of
executed system call traces and uses Random Forest algorithm for classification. 
Overall, it achieves performance comparable 
to the best classifier from a prior survey~\cite{Canali2012}.

To use \sysname a human analyst needs to supply a description of neighborhood attributes.
They can be as simple as a list of high valued servers (waterhole case study) or they can be derived
using a machine learning algorithm. In the Symantec Wine case study we use a domain name classifier, which
consumes VirusTotal domain reports as input, to detect suspicious domains that are used to form neighborhoods.

\subsection{Neighborhood Instances from Attack-Templates}
\label{subsec:neighborhood_instantiation}
Within each neighborhood time window (NTW), \sysname generates neighborhood instances
based on statically defined attack vectors --
each attack vector is a ``Template'' to generate neighborhoods with.
%The $NTW$ parameter provides the trade-off between the time to detection and required computational resources. Thus, the lower $NTW$ parameter, the faster \sysname catches 
%malware entering the network, i.e. the lower detection time.
The goal of partitioning data into neighborhoods is to create predominantly benign or malicious neighborhoods.
The algorithm runs once per neighborhood time window (NTW).
Hence the partitioning algorithm is radically different across the case studies.
%Conceptually the algorithm partitions data into neighborhoods that are likely 
%to contain higher concentration of either benign or malicious files/machines than on average.
%However, the partitioning step is radically different across our case studies.

%\begin{figure}[tbp]
%   \vspace{-0.0in}
%   \centering
%   \includegraphics[width=0.45\textwidth]{figs/symantec/downloader_graph.pdf}\par
%%   \caption{Example of a downloader graph.}
%   \vspace{-0.2in}
%\label{fig:downloader_graph}
%\end{figure}
\noindent{\bf Symantec Wine.} The Algorithm~\ref{symantec_nbd_algo} partitions downloaded files into multiple neighborhoods.
It uses the following intuition: 
if a domain is malicious, then the files transitively downloaded from such a domain are likely to be malicious.

%The key idea of the partitioning algorithm (Algorithm~\ref{symantec_nbd_algo}) is to form neighborhoods around suspicious domains following the intuition 
%that if a domain is malicious, then the files transitively downloaded from such a domain are likely to be malicious.

%The algorithm uses a predefined 150-day long neighborhood time window (NTW) because we found in our experiments 
%that malicious domains distributing more than one malware sample have an average lifespan of 156 days (Section~\ref{}).
%Therefore, we consider downloads within a 150-day window to be correlated.

For ease of explanation, we treat the previously introduced domain name classifier as a predicate (line 1).
At each iteration the algorithm starts with identifying a set of suspicious domains within the current NTW (lines 4--5), which is done using the domain name classifier. Then the algorithm uses each suspicious domain as 
a seed to initiate the neighborhood formation process (lines  6 -- 12).
Next, for each suspicious domain it searches for the files within the current NTW that access 
that particular domain (either download other files from it or being downloaded from it) -- the set $F$ (line 7). 
By following downloader graph edges
the algorithm selects files transitively downloaded by the files in the set $F$ (line 10) and filters out
those that do not access any of the suspicious domains (line 11). The files that have not been excluded are
added to the current neighborhood (line 12).

\begin{figure}
\removelatexerror
\begin{algorithm}[H] 
 \small
 \DontPrintSemicolon
 \newcommand\mycommfont[1]{\footnotesize\ttfamily\textcolor{blue}{#1}}
 \SetCommentSty{mycommfont}
 \SetKwInOut{Input}{Input}
 \SetKwInOut{Output}{Output}
 \Input{Downloader graphs}
 \Output{Neighborhoods}
 \BlankLine
 \codecomm{Domain name classifier}\;
 \textbf{Let} DNC ($domain$): $domain$ is malicious\;
 \BlankLine
  
\codecomm{execute once per NTW}
 
\While{True}{
 \codecomm{create an empty list of neighborhoods}\;
 $nbds \leftarrow \emptyset$\;
 \BlankLine
 
 \codecomm{identify active domains within the current NTW}\;
 $D$ $\leftarrow$ domains accessed within the current NTW\;
 \codecomm{identify suspicious domains}\;
 $D'$ $\leftarrow$ \{d $\in$ $D$ $\vert$ DNC($d$)\}
 \BlankLine
  
 \ForEach{suspicious domain $d_i \in D'$}{
    \codecomm{identify files accessing the domain $d_i$}\;
	$F$ $\leftarrow$ files accessing the domain $d_i$

    \codecomm{initialize an empty neighborhood}\;
	$nbd$ $\leftarrow$ $\emptyset$
    \BlankLine
 	
	\ForEach{file $f_i \in F$}{	
	    \codecomm{search for transitively downloaded files}\;
		$F_i$ $\leftarrow$ files transitively downloaded by  $f_i$

	    \codecomm{retain suspicious files}\;
	    $F'_i$ $\leftarrow$ \{$file \in F_i$ $\vert$  $\exists d \in file.domains \quad DNC(d)\}$\;
    	$nbd \leftarrow N$ $\bigcup$ $F'_i$
	}
	$nbds \leftarrow nbds$ $\bigcup$ $nbd$\;
 }
}
 
 \caption{Symantec Wine: Neighborhoods from Attack-Vectors}
 \label{symantec_nbd_algo}
\end{algorithm}
 \vspace{-0.2in}
\end{figure}

%After mapping each suspicious domain to a set of suspicious files within a neighborhood (lines 6 -- 12), 
%the algorithm aggregates files that are transitively reachable in terms of the downloader graph abstraction 
%from suspicious domains.

Note that the algorithm formation process may generate many small neighborhoods. 
An estimate of the conditional distribution using such feature vectors (Section~\ref{subsec:neighborhood_feature_extraction}) 
is usually susceptible to high variance, thus neighborhoods containing 
an insufficient number of files may have negative 
impact on the accuracy of the neighborhood classifier (Section~\ref{subsec:neighborhood_feature_extraction}).
To reduce variance and achieve robust classification of neighborhoods, the algorithm merges them such 
that final neighborhoods are greater than some predefined minimum size.
Empirical analysis of the accuracy of the neighborhood
classifier shows that it achieves robust classification of neighborhoods containing more than 1000 files.

In order to maintain neighborhood effect after merging, i.e. to have mostly homogeneous neighborhoods -- either benign or malicious, the merging algorithm ranks neighborhoods in terms of maliciousness, where malicious score is defined
as the relative number of LDs' alerts within a neighborhood. 
After that the algorithm sorts neighborhoods based on their malicious score and
proceeds merging them in a decreasing order of their malicious scores. 
Note that malicious score estimation may be incorrect if we incorrectly estimate the neighborhood size, 
but \sysname tolerates such errors.

\noindent{\bf Waterhole.}
The algorithm (Algorithm~\ref{waterhole_nbd_algo}) to form a neighborhood to detect a waterhole attack significantly differs
from the one used in the Symantec Wine experiment.
It creates a neighborhood from client machines that access a server or a group of servers within a neighborhood
time window.

\begin{figure}
\removelatexerror
\begin{algorithm}[H] 
 \small
 \DontPrintSemicolon
 \newcommand\mycommfont[1]{\footnotesize\ttfamily\textcolor{blue}{#1}}
 \SetCommentSty{mycommfont}
 \SetKwInOut{Input}{Input}
 \SetKwInOut{Output}{Output}
 \Input{Network flow data}
 \Output{Neighborhoods}
 \BlankLine
 \textbf{Let} predicate(A:Client, B:Servers) := $A$ accesses $B$\;
 \BlankLine
  
 \codecomm{execute once per NTW}\;
 
\While{True}{
 \codecomm{create an empty list of neighborhoods}\;
 $nbds \leftarrow \emptyset$\;
 \BlankLine
 
 $V$ := client machines\codecomm{*}\;
 $S$ := accessed servers\codecomm{*}\;
 \BlankLine

 \codecomm{partitioning a set into non-disjoint sets to incorporate}\;
 \vspace{-.03in}
 \codecomm{structural filtering}\;
  \vspace{-.03in}
 ${P_{1}, P_{2}, ..., P_{N}}$ $\leftarrow$ partition-set($S$), where $S$ = $\bigcup\limits_{i=1}^{N}    P_{i}$\;
 
 \ForEach{partition $P_i$}{
 	\codecomm{form neighborhoods $nbd_i$ using partitions $P_i$}\;
	 $nbd$ $\leftarrow$ $\{V |$ predicate($V$, $P_i$)$\}$\;
	 $nbds \leftarrow nbds$ $\bigcup$ $nbd$\;
 }
 }
 \BlankLine
 \codecomm{*active within the time window NTW}\;
 \caption{Neighborhoods from Attack-Vectors}
 \label{waterhole_nbd_algo}
\end{algorithm}
 \vspace{-0.2in}
%\label{NI_algorithm}
\end{figure}

%The partitioning algorithm (Algorithm~\ref{waterhole_nbd_algo}) executes once per NTW. 
%The best results achieved with \todo{6-second} long
%NTW (see Section~\ref{}).

To abstract away from technical details, we define the predicate (line 1) which is
true if a client $A$ accesses a server $B$.
Each iteration starts with defining the set $V$ of client machines that are active within the current NTW
and the set $S$ of servers that those clients access within the NTW (line 4 -- 5).
Then the algorithm proceeds with partitioning the set $S$
into one or more disjoint subsets $P_i$ (line 6). This is to incorporate
`structural filtering' into the algorithm, allowing an analyst to create
neighborhoods based on subsets of servers (instead of all servers in case of
waterhole). %, which are likely to be attacked. 
Structural filtering boosts detection under certain conditions (Appendix~\ref{sec:time-struct}).
The neighborhood instantiation algorithm builds a neighborhood for each partition
$P_i$ (line 8) and, finally, it adds the just formed neighborhoods to $nbds$ list (line 9).

\subsection{Shape Property for Malware Detection}
\label{subsec:neighborhood_feature_extraction}
After identifying neighborhoods, the next step is to detect neighborhoods with high malware concentration. 
In order to accomplish this, we introduce \textit{a novel approach to extracting neighborhood features} 
that formalizes \textit{shape property}.

The \textit{key algorithmic idea} is to map all alert-FVs within a neighborhood to a {\em single}
\textbf{vector-histogram} which robustly captures the neighborhood's statistical
properties. 
\textit{Such transformation allows us to analyze the joint properties of all alert-FVs generated
within a neighborhood without requiring FVs to be clustered or alerts to be
counted.}
After that, \sysname feeds neighborhood-level feature vectors into a binary classifier 
to identify malicious neighborhoods. 
We use two types of binary classifiers: boosted decision trees in the Symantec Wine case study 
and a Wasserstein distance-based threshold test in the waterhole experiment.

\noindent {\bf Generating a vector-histogram from alert-FVs.} 
The algorithm aggregates
$L$-dimensional projections of alert-FVs on per neighborhood basis into a set
$B$ (Algorithm \ref{ShapeGD_algorithm}, line 3).
%The algorithm processes each neighborhood separately: it aggregates alert-FVs
%on per neighborhood basis and projects them on an $L$-dimensional PCA basis to
%reduce dimensionality (Algorithm \ref{ShapeGD_algorithm}, lines 3--4).  Thus,
%the set $\mathcal{H}_{PCA}$ contains $L$-dimensional alert-FV vectors.
After that, \sysname converts low dimensional representation of alert-FVs, the
set $B$, into a single $(L, b)$-dimensional vector-histogram denoted by $H_B$
(line 4).  The conversion is performed by binning and normalizing $L$-dimensional vectors
within the set $B$ along each dimension.
%In each of the L-dimensions, the L-dimensions
%scalar-histogram of the corresponding component of the vectors is binned and
%normalized.  
Effectively, a vector-histogram is a matrix $L$x$b$, where $L$ is
the dimensionality of alert-FVs and $b$ is the number of bins per dimension. 
Further implementation details can be found in the Appendix~\ref{vector_histogram_mplementation}.

We use standard methods to determine the size and number of bins. % and note that
%the choice of Wasserstein distance in the next step makes \sysname robust
%against variations due to binning.  
In particular, we tried square-root choice,
Rice rule, and Doane's formula~\cite{binning} to estimate the number of bins,
and we found that 20--100 bins yielded best results. %separable histograms (as in Figure

\noindent{\bf Neighborhood classifier.}
\sysname may use any binary classifier (Algorithm~\ref{ShapeGD_algorithm}, line 4) as a neighborhood classifier. 
We use the following two classifiers -- boosted decision trees (XGBoost~\cite{xgboost}) 
and a specially designed Wasserstein distance-based distance ('ShapeScore')
(Appendix~\ref{sec:shape-score-function}) 
in the Symantec Wine and waterhole case studies respectively.
The main advantage of using XGBoost is its ability to learn complex decision boundary 
and it can be trained in a non-parametric mode (we completely automated parameter search process).
However, in comparison to ShapeScore, XGBoost algorithm requires 
both benign and malicious data for training purposes.
Thus, the threshold test can be trained using only benign data and it acts as an anomaly detector.
In our experiments, we found that XGBoost outperforms the ShapeScore function
in the Symantec Wine case study,
while the ShapeScore yields good detection accuracy in the waterhole case study.

Note like any other machine learning classifier, the binary classifier employed by \sysname 
needs to be retrained periodically to account for constantly evolving statistical software properties.

\begin{figure}
\removelatexerror
\begin{algorithm}[H]
\small
 \DontPrintSemicolon
 \SetKwInOut{Input}{Input}
 \SetKwInOut{Output}{Output}
 \Input{Suspicious neighborhoods $nbds$}
 \Output{Malicious neighborhoods}

 \For{each nbd in nbds}{
   %// collect alert-FVs from the nodes in the neighborhood\;
   \codecomm{aggregate L-dim projections of alert-FVs on per neighborhood basis}\;
%   $\mathcal{B}$ 
  $B$ $\leftarrow$ $\{alert-FVs$  $|$ $alert-FV$ $\subset$ $nbd\}$\;
   %\BlankLine

   %// reduce dimensionality\;
%   $\mathcal{B}_{PCA}$ $\leftarrow$ project $\mathcal{B}_{FVs}$ on $L$-dim PCA basis\;
   %\BlankLine

   %// perform binning with \textit{b} bins along each dimension\;
   \codecomm{build an $(L, b)$-dim. vector-histogram}\;
%   $\mathcal{H}_{\mathcal{B}}$ 
   $H_B \leftarrow$ bin \& normalize $B$ along each dimension\;
	
   \codecomm{classify the neighborhood}\;
   \If{Neighborhood Classifier($nbd$)} {
     label $nbd$ as \textit{malicious}\;
   }
 }
 \caption{Neighborhood Classification}
 \label{ShapeGD_algorithm}
\end{algorithm}
 \vspace{-0.2in}
%\label{ShapeGD_algorithm}
\end{figure}

%% file: experimental-setup.tex
%%%%%%%%%%%%%%%%%%%%%%%%%%%%%%%%%%%%%%%%%%%%
%\subsection{Case for a New Methodology}
%\label{sec:exp_justification}
%\input{exp_justification}

%\mohit{The two settings should be mentioned in the 
%algo section already. Also, this para below might be
%better fit for results.}

We evaluate \sysname using two publicly available datasets.
First, in the Symantec Wine dataset~\cite{wine}, \sysname
uses malware reports from Symantec client devices and reduces
the LDs' false positives from $\sim$1M down to $\sim$110K, while 
retaining 107K out of 137K malware files. 
Second, we simulate a waterhole attack using Yahoo's web-service network logs~\cite{yahoo-G4}
overlayed with host-level malware and benighware traces~\cite{kruegel-bare-metal}.
%where servers are marked compromised at arbitrary times and place
%where a server
%is marked compromised at an arbitrary time and places 
%\mikhail{sentence needs to be refactored}
%malware~\cite{} or benign~\cite{}
%execution traces on client machines in the Yahoo log -- 
In this testbed, \sysname detects an attack within a few seconds and with only about
100 compromised machines out of over 550,000 potential compromises).
In both settings, \sysname successfully amplifies 
the weak signal inherent to malware propagation.

\subsection{Wine dataset}
\label{sec:wine_dataset}

\begin{figure}[tbp]
   \vspace{-0.0in}
   \centering
   \includegraphics[width=0.4\textwidth]{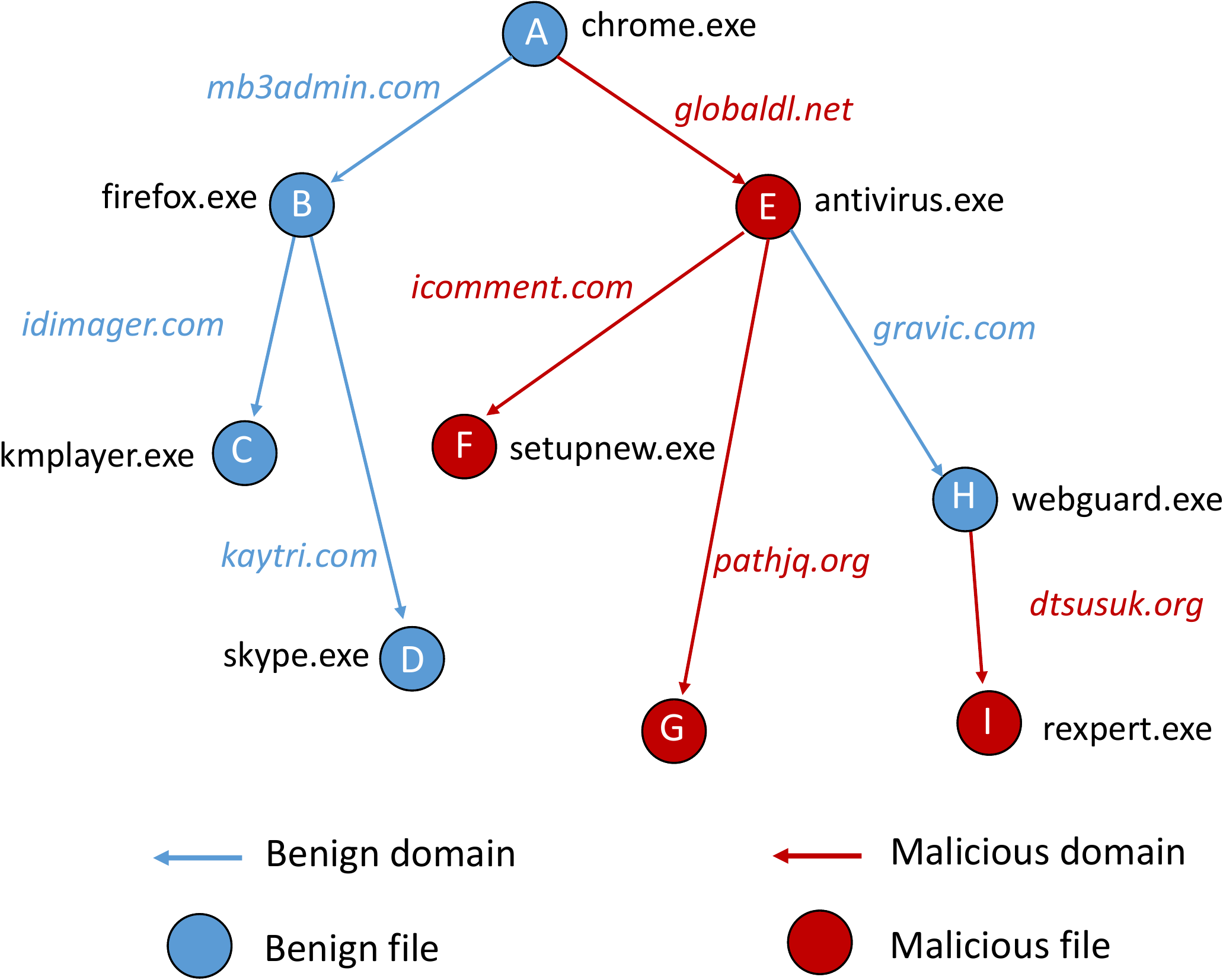}\par
   \caption{Example of a downloader graph.}
   \vspace{-0.2in}
\label{fig:downloader_graph}
\end{figure}

Wine dataset~\cite{wine_dataset,downloader_graphs,pup_ndss} contains telemetry
information collected by Symantec's intrusion prevention system and Symantec
antivirus product over 5 year period -- from 2008 until 2013.  The dataset
summarizes file downloader activities across 5M Windows hosts around the
world.  File downloads are represented in the form of downloader graphs (the
abstraction introduced by Kwon et al.~\cite{downloader_graphs}) -- one per end
host.  A graph node represents a downloaded file (SHA256 file hash) and a
directed edge between two nodes $n_a$ and $n_b$ indicates that the file $n_a$
has downloaded the file $n_b$ from a domain $D$ on the corresponding host
machine, where $D$ is the edge's label.  
%Wine dataset shared with us by Kwon
%et al.~\cite{downloader_graphs} contains information about $\sim$20.7M
%unique files which were downloaded 67 million times from $\sim$353K
%domains.

Figure~\ref{fig:downloader_graph} depicts an example of a downloader graph.
Each node is labeled with a corresponding file name, and each edge bears a
domain name from where a file has been downloaded. We also overlay ground truth
on the nodes and edges: red color means that a file or a domain is malicious,
while the blue color means that a file/domain is benign.

%\textbf{Ground truth.} As Wine dataset lacks any ground truth information, i.e.
%which files/domains are benign and which are malicious,

We used the VirusTotal (VT) service to obtain ground-truth information about
the 20.3M file-hashes downloaded 67M times and all 353K domain names in Wine
(Table~\ref{fig:file_agg_analysis}).
Though file-level VirusTotal reports contain results of signature-based malware detection, 
we do not use them for within \sysname (except for computing the ground truth).
Hence, information within VirusTotal domain reports might be affected by post-analysis 
performed by commercial antivirus vendors.
However, there are alternative approaches to establish domain reputation~\cite{predator_domain_reputation} that
outperform our domain name classifier by using a different set of domain features,
which are unavailable in the Symantec Wine dataset.

For files (corresponding to a file-hash)
or domain names that VT has information for, it used 62 different anti-viruses
and other heuristics to generate a report -- this report is used to train the
file-behavior and domain-name classifiers.  
%To assign a single label (either benignware or malware) to a file based on a
%VT report, we follow the methodology established in Downloader
%Graphs~\cite{downloader_graphs}: a file is 
We consider a file to be malicious if more than 30\% of antivirus products
label it as malware~\cite{downloader_graphs}.  This yields 2.6M reports for
file-hashes, with 137K confirmed to be malicious, and 301K reports for domain
names.  We label all remaining files and domain names (i.e., that are not
confirmed to be either malware or benign by VT) as benign -- this is a
conservative step that weakens the malware propagation signal in the dataset
and is also representative of real deployments where information about
suspicious files/domain-names is often delayed or unavailable.

\subsection{Modeling Waterhole Attacks}
%%%%%%%%%%%%%%%%%%%%%%%%%%%%%%%%%%%%%%%%%%%%

\noindent \textbf{Waterhole attack.} To model a waterhole attack, we use
Yahoo's ``G4: Network Flows Data"~\cite{yahoo-G4} dataset, which contains
communication data between end-users and Yahoo servers.  The 41.4 GB (in
compressed form) of data were collected on April 29-30, 2008.  Each netflow
record includes a timestamp, source/destination IP address, source/destination
port, protocol, number of packets and the number of bytes transferred from the
%source to the destination.  %(note that all IP addresses in the dataset are
%anonymized using a random %permutation algorithm, thus it is impossible to
%trace them back to the real %servers).  We model the setting where heuristics
%such as SecureRank~\cite{secure_rank} are applied to identify suspicious
%servers and %\sysname is provided with list of suspicious servers to create
%neighborhoods %with.  In practice, numerous heuristics can be applied to
%reduce the number of %monitored servers, and one of the popular ones is
%%SecureRank~\cite{secure_rank}.  we assume that \sysname monitors the top
%(here, 50) suspicious servers based on SecureRank's scores.  Specifically, we
%use 5 hours of network traffic (208 million records) captured on April 29,
%2008 between 8 am and 1 pm at the border routers connecting Dallas Yahoo data
%center (DAX) to the Internet. 
source to the destination~\footnote{All IP addresses in the dataset are
anonymized using a random permutation algorithm, thus it is impossible to trace
them back to the real servers}.

Specifically, we use 5 hours of network traffic (208 million records)
captured on April 29, 2008 between 8 am and 1 pm at the border routers
connecting Dallas Yahoo data center (DAX) to the large Internet. 
%The data within a selected time interval include 208 millions of records --
%connections of end-users to DAX center. 
The selected 50 DAX servers communicate with 3,181,127 client machines over
14,249,931 requests.

We assume that an attacker compromises one of the most frequently accessed DAX
server -- 118.242.107.76, which processes $\sim 752,000$ requests within 5-hour
time window ($\sim 43.7$ requests per second).  In our simulation it gets
compromised at random instant between 8am and 10.30am.  Hence, Shape GD can use
the remaining 2.5 hours to detect the attack (our results show that less than a
hundred seconds suffice).  Following infection, we simulate this `waterhole'
server compromising client machines over time with an infection probability
parameter -- this helps us determine the time to detection at different rates
of infection.  The benign and compromised machines then select corresponding
type of execution trace (i.e., a sequence of FVs generated below) 
and input these to their LDs.

\noindent{\bf Benign and malware applications.}
\label{sec:apps}
%\begin{table*}[t]
%  \centering
%\begin{tabular}{ l l l l l}
%  % \label{table:app-table}
%  %\multicolumn{2}{c}{User Traces} \\
%  \hline
%  Program type & \# of executed programs & Avg. execution time, sec & Avg. \# of FVs per run & Avg. \# of syscalls per run \\ \hline
%  Benignware\cite{kruegel-bare-metal} & 1,889 & $7.5\pm 3.0$ & $8.2\pm 3.4$ & $13.5K\pm 32K$ \\            %\hline
%  Malware\cite{kruegel-bare-metal} & 1,311 & $6.1\pm 3.4$ & $6.1\pm 3.6$ & $15.2K\pm 58K$ \\            %\hline
%  Malware\cite{kaspersky-windows-2015} & 2,364 & $8.6\pm 2.9$ & $8.6\pm 2.7$ & $10.1K\pm 35.2$ \\            %\hline
%  \hline
%\end{tabular}
%\label{table:syscall-stat}
%\end{table*}
We collect data from thousands of benign applications and malware samples.  To
avoid tracing program executions where malware may not have executed any stage
of its exploit or payload correctly, we set a threshold of 100 system calls per
execution to be considered a success.  Our experiments successfully run 1,311
malware samples from 193 malware families collected in July
2013~\cite{kruegel-bare-metal}, and 2,364 more recent samples from 13 popular
malware families collected in 2015 \cite{kaspersky-windows-2015}, to compare
against traces from 1,889 benign applications.

%Thus, we are able to collect data from 1,889 out of 2,000 benign applications
%and 1,311 out of 2,000 malware samples from 193 malware families collected in 
%July 2013~\cite{kruegel-bare-metal} and from 2,364 out of 3,225 more recent samples from 13
%popular malware families collected in 2015 \cite{kaspersky-windows-2015}.}

%\mikhail{\st{We collect data from 1,889 benign applications
%and 1,311 malware samples from 193 malware families collected in 
%July 2013~\cite{kruegel-bare-metal}, and add 2,364 samples from 13
%popular malware families from 2015 \cite{kaspersky-windows-2015}. 
%We set a threshold of 100 system
%calls per execution to avoid tracing executions
%where malware may not have executed 
%any stage of its exploit or payload correctly.}}

%Also we added to the malware set 6,500 recent samples from 13 families
%dominated in 2015 \cite{kaspersky-windows-2015}.

We record time stamped sequences of executed system calls using Intel's Pin
dynamic binary instrumentation tool.  Each Amazon AWS virtual machine instance
runs Windows Server 2008 R2 Base on the default T2 micro instances with 1GB
RAM, 1 vCPU, and 50GB local storage.  The VMs are populated with user data
commonly found on a real host including PDFs, Word documents, photos, Firefox
browser history, Thunderbird calendar entries and contacts, and social network
credentials.  To avoid interference between malware samples, we execute each
sample in a fresh install of the reference VM.
%clean environment that we achieved by cloning a virtual machine image.  
As malware may try to propagate over the local network, we set up a sub-net of
VMs accessible from the VM that runs the malware sample.
%simulated it with several VMs without internet access, but accessible to other
%VMs running malware.  
In this sub-net, we left open common ports (HTTP, HTTPS, SMTP, DNS, Telnet, and
IRC) used by malware to execute its payload. We run each benign and malware
program 10 times for 5 minutes per run 
%for 5 minutes 
for a total of almost 53,000 hours total compute time on Amazon AWS.

% including 10-fold cross-validations and repetitions.  

Overall, benignware and malware were active for 141,670 sec and 283,270 seconds
respectively, executing an average of 11,900 and 13,500 system calls per second
respectively.  Using 1 second time window (Section \ref{sec:model})
and sliding the time windows by 1ms, we extract histograms of system calls
within each time window as the ML feature, and finally pick 1.5M benign and 1M
malicious FVs from this dataset for the experiments that follow. Importantly,
we do not constrain the samples on neighboring machines to belong to the same
families -- as described above, 
malware today predominantly spreads through malware distribution networks
where a downloader trojan (`dropper') can distribute arbitrary and unrelated 
payloads on hosts. We want to test \sysname in the extreme case
where malicious FVs can be assigned from any malware execution to any machine.

\ignore{
\noindent \textbf{Phishing attack.} We simulate a phishing attack in a medium size
corporate network of 1086 nodes that exchange emails with others in the
network.  To model email communication, we pick 50 email threads with 100
recipients each from the publicly available Enron email dataset
\cite{enron-dataset} (the union of all email threads' recipients is 1086). 

We start the simulation with these 50 emails being sent into the 1086-node
neighborhood, and seed only {\em one} email out of 50 as malicious.
%into this 1086-node neighborhood, 
We then model the infection speading at different rates as this malicious email
is opened by its (up to 100) recipients at some time into the simulation and is
compromised with some likelihood when the user `clicks' on the URL in the
email.  Our goal is to measure the number of compromised nodes before Shape GD
declares an infection in this neighborhood. All nodes that open and `click' the
link in the malicious email will select malware FVs from Section~\ref{sec:apps}
as input to their corresponding LDs, while the remaining nodes select benign
FVs.  

%\sanjay{For each node, the corresponding time-series of FVs are simulated
%by using the pre-recorded FVs, as described in Section~\ref{sec:apps}.}

%Malware spreads as an attachment/link to only one of the emails within a network
%and can reach at most 100 recipients.  The total number of benign emails in a
%simulation is irrelevant because they affect neither generation of LD's FP or
%TPs.  Each email's recipient performs two actions: open an email and click on
%an attachment.  Only if they are executed sequentially, does a host get
%compromised.
%Such realistic modeling is required to faithfully estimate GD's detection
%capabilities.

To simulate the infection spreading over the email network, we need to (a)
model when a recipient `opens' the email:
%Despite short response times in the corporate world, recipients do not open
%emails immediately upon receiving them.  Delays associated with opening emails
%cause malware infection to be distributed over time and, as a consequence, add
%extra noise to GD's input data.  
we do so using a long tail distribution of reply times where the median open
time is 47 minutes, 90-percentile is one day, and the most likely open time is
2 minutes~\cite{email-resp-dist}; and (b) model the `click' rate (probability
that a recipient clicks on a URL): we vary it from 0\% up to 100\% to control
the rate of infection.
%And the minimum click rate or alternatively the minimum number of malicious
%nodes in a neighborhood such that Shape GD reliably spots malware directly
%characterizes its sensitivity to malware spread in a network.
For example, within 1-, 2-, 3-hour long time interval only 55\%, 65\%, and 70\%
of recipients of a malicious email open it, which corresponds to 55, 65, and 75
infected machines respectively at 100\% click rate.
%Therefore, malware infects only a small portion of a network.

Overall, these two scenarios differ in their time-scales (seconds v. hours)
and in the relative rate at which benign and malicious neighborhoods grow.
As we will see, these parameters have a significant impact on the composition
of neighborhoods and the Shape GD's detection rate.
}

%\subsection{Methodology}
%\mohit{Re-do this -- very brief and clean. No innovation here.}
%\mikhail{TODO: separate Symantec from waterhole}

\ignore{
To estimate \sysname{}'s detection performance, we apply the standard methodology -- 10-fold cross validation --
to all local detectors and to the neighborhood classifier in the Symantec Wine case study.
In the waterhole experiment, we partition the data into a training and testing data sets.
%To train the reference histogram, $H_{{\rm ref}}$, we select
%15K FVs from the training data set.
We repeat each experiment 100 times using a random subset of training data set for training and a random subset 
of the testing data set for testing.
}

%We report averaged results from repeating each
%experiment multiple times with random initialization parameters.  In
%particular, we use 10-fold cross validation %for machine learning experiments
%\mikhail{to train machine learning classifiers}
%%(Figure \ref{fig:windows-LD-roc})
%, 500 randomly sampled benign/malicious
%neighborhoods with 10 repetitions to compute average (Figures
%\ref{fig:windows-hist}%, \ref{fig:windows-hist-stability}
%), 100 repetitions of
%each malware infection experiment (Figures
%\ref{fig:windows-shape_vs_count},\ref{fig:windows-shape_vs_count-waterhole},%\ref{fig:windows-noisy_count},
%\ref{fig:windows-noisy_count-waterhole}),
%and 100 repetitions of infection with 10 repetitions per data-point (Figures
%\ref{fig:windows-time-NF-phishing},\ref{fig:waterhole_time_NF},\ref{fig:windows-detection_vs_censoring_threshold},\ref{fig:waterhole-structural-NF}).
%To train the reference histogram, $H_{{\rm ref}}$, we select
%15K FVs and 100K FVs from the training data set in the waterhole experiment.
%%phishing and waterhole experiments respectively.
%All Shape GD's parameters are chosen based on a training data set (used for
%Figures~\ref{fig:windows-hist}% and~\ref{fig:windows-hist-stability})
% -- we then
%evaluate Shape GD (in the remaining figures) using a completely separate
%testing data set.  
 
%Moreover, the data used for the experiments in Section \ref{sec:power-shape},
%which guided the choice of Shape GD's parameters, is not reused for Shape GD's
%performance results.

%Throughout the paper global false positive rate is fixed at 1\% level.

%%%%%%%%%%%%%%%

\ignore
{
%%%%%%%%%%%%%%%
\subsection{Virtual Machine Experimental Setup}
\label{sec:impl}
%\ignore{
For our large-scale experiments, we deploy our data collection system on Amazon Web Services (AWS).
Experiment nodes run Windows Server 2008 R2 Base on default T2 micro instances with 1 GiB RAM, 1 vCPU, and 50 GB local storage. 
We populate the experiment environment with user data commonly found on a real host including: 
PDFs, word documents, photos, Firefox browser history, Thunderbird calendar entries, and Thunderbird contacts.
We then create a snapshot image of the virtual machine environment and clone this image for each experiment.
For simplicity, all malware experiments are conducted with a fresh clone of the base image.

Client code runs on-host and runs each experiment and communicates results with the central database.
Each experiment is run via a custom Intel Pin tool, which launches the target applications and intercepts system calls. 
This tool records timing information and system calls with arguments, and is configured to follow spawned child processes.
In comparison to full system emulation approach, Pin incurs moderate overhead while transparently instrumenting binaries.
%it instruments binaries without modification in a format comparable to \tttext{strace} in Linux. 

%Simulated local network
The worker instances run in a virtual private cloud (VPC).
We simulate a local network via two subnets in the VPC. 
\textit{Neigboring machines} exist in a private subnet with no internet access, but are accesible by the experiment machines.
If malware happens to propogate to one of these vulnerable machines then it is, by default, quarantined.
Experiment machines run on the second subnet with limited public internet access. 
The second subnet allows malware to communicate over common C\&C ports and the client code to post results.
Specifically, we allow access for HTTP, HTTPS, SMTP, DNS, Telnet, and IRC.

%%%%%%%% oakland text below

\input{experimental-setup/intro}

\subsection{Benign Applications}
\input{experimental-setup/about-the-apps}

\label{sec:benign}

\subsection{Diverse Malicious Behaviors}
\input{experimental-setup/synthetic-malware-gen-2}

\label{sec:synthetic_malware}

%\subsection{Malware Payloads}
%\input{experimental-setup/synthetic-malware-gen}
%\label{synthetic_malware}

%\subsection{Benign Applications}
%\input{experimental-setup/about-the-apps}

\subsection{Local Detectors and Feature Extraction algorithms}
\label{sec:LDs}
\input{experimental-setup/LDs}

\subsection{Synthetic and Real Networks}
\input{experimental-setup/synthetic-network-gen}

\subsection{Implementation Details}
\input{experimental-setup/data-collection}
}
% \subsection{Malware Propogation Models}
% \input{experimental-setup/malware-propogation-model}

%% file: experimental-setup/about-the-apps.tex
% Choice of apps to repackage

\input{experimental-setup/app-table2.tex}

Our platform currently includes 10 benign apps chosen from the top 100 Android
apps from the Google Play store.  
%We selected
Table~\ref{table:benign} shows the apps and their aggregate system calls -- it
is clear that the chosen apps are popular as well as diverse in each system
call category.
%functionally rich apps with diverse dynamic behaviors to stress-test \sysname. 
In particular, we included apps such as online shopping clients (Amazon,
Walmart, eBay), anti-virus software (AVG) and a wallpaper app (Zedge) that are
popular carriers of malware, an entertainment news app (Buzzfeed), an
encyclopedia (Wikiepdia), online advertisement (Yelp), HTML5 reference app, and
a Bible. 

%All of them represent
%popular usage, therefore they could be potential targets for malware
%developers. 

For each benign application, we supplied realistic and diverse user input to
observe variety of its dynamic behaviors. For this purpose we relied on two
different tools: TestDroid~\cite{testdroid} and Android Monkey
tool~\cite{adb-monkey}. The former can be used to record {\em real user input}
as a sequence of UI events and then to replay it multiple times preserving time
intervals between subsequent events. Monkey randomly generates user input
events such as touch, scroll, push events. We intentionally did not use
automatic input generation tools \cite{dynodroid,concolic-android} to explore
app's state space because they are still much worse than real user input and do
not work with custom app's layout.

We arranged 20 human participants to record dozens of 3-min long
interactions with four apps using TestDroid. 
Humans can produce a semantically meaningful sequence of inputs
(e.g. to log-in, satisfy ordering constraints when filling out a sequence
of fields in forms, etc).
Later, we replayed
these recordings multiple times to collect system call data. 

However, TestDroid has several limitations -- it often fails to replay UI sequences, especially when an app
extensively uses Android WebViews, 
or frequently causes an app under analysis to
crash -- and
we were unable to expand this approach to the other apps. 
Hence, we chose six other popular apps whose realistic usage can be closely
approximated by inputs generated by Monkey.

\noindent {\bf Realistic user data.} 
%In addition to careful modeling of malicious behaviors, we have invested a lot
%of efforts into 
We populate our emulators with synthetic data that is statistically close to
the real user data, following data set sizes estimated by Kazdagli et
al~\cite{morpheus-hasp}. Without user data such as Contacts to steal, 
even a functional information stealer malware may not execute 
malicious code.

%% file: experimental-setup/app-table2.tex
\begin{table*}[t]
\centering
\begin{tabular}{ p{1.5cm}  p{3.3cm} p{1.2cm} p{1.5cm} p{1.5cm} p{1.5cm} p{1.5cm} p{1.25cm}}
  % \label{table:app-table}
  \multicolumn{8}{c}{User Traces} \\
  \hline
  App & Description of User Activity & Exec time (min) & File Ops  & Network Ops & Process Ctrl & System Mgmt & Misc \\ \hline
  Wikipedia & Browse and search multiple articles & 480 & 1723054 &124003 &36649 &1508412 &5 \\            %\hline
  Allrecipes & Search for recipes, scroll through recipe photos & 510 & 24598084 & 11515472 &60076 &4757467 & 0 \\               %\hline
  Yelp & Browse local catering, read reviews & 500 &4270481 &949696 &178299 &4685311 & 28\\                       %\hline
  HTML5 Reference & Repeatedly open and read documentation & 545 &1745087 &264893 &110896 &4208092 &654 \\  %\hline
  \hline
  & \\                                          %\hline
  \multicolumn{8}{c}{Monkey} \\                \hline
App & Description of App & Exec time (min) & File Ops  & Network Ops & Process Ctrl & System Mgmt & Misc \\ \hline
  Walmart & Online shopping & 540 &4361902 &324263 &50866 &3937191 & 0\\                  %\hline
  AVG & Free antivirus & 800 & 16339990&826870 &245933 &8317314 &634 \\                       %\hline
  Buzzfeed & Social news & 520 &15733289 &1100059 &154094 &9380923 &0 \\                     %\hline
  eBay & Online Auction & 510 & 2178359&307148 &58990 &4716632 &0 \\                      %\hline
  Zedge & Wallpaper and Ringtones & 675 & 3779037&137739 &79588 &6052683 &3 \\            %\hline
  Holy Bible & Holy Text & 840 & 1229925&761161 &97201 &10703697 &10 \\                 %\hline
  \hline \\
\end{tabular}
\caption{Characterization of Applications. All applications chosen from Google Play Store top 100 free apps. Here, system calls are grouped into logical categories. For example, the \texttt{mount} system call is categorized as system-wide, and the \texttt{timer\_create} system call is classified under misc. }
\label{table:benign}
\end{table*}

%% file: experimental-setup/synthetic-malware-gen-2.tex
% Synthetic Malware Generation 
%After many unsuccessful attempts to run real malware in our experimental setup, we reimplemented malicious functionality in the form of synthetic malware and use it throughout the paper. Carefully crafted synthetic malware benchmark is designed to closely resemble real-world malware and to provide precise control over its parameters. Also, synthetic malware includes a futuristic feature not seen in the wild -- it can lower its intensity to remain unnoticed by local detectors.

%After in-depth analysis of hundreds of mobile malware samples in the wild, we
%have come up with a new methodology that is a few steps ahead of the current
%approach to evaluation of malware detectors -- this is the most crucial
%component of our experimental setup.

We evaluate our GD algorithms not simply against off-the-shelf malware samples,
but against malware samples that are guaranteed to execute correctly and have a
computationally diverse payload.
%tries to escape detection by diversifying its payloads' executions. 
We now describe the range of malware behaviors and our process to construct a
parameterized, state-of-the-art malware sample.
%guaranteed to execute correctly.

%hiding behind the variations of benign executions.
%Instead of testing the \sysname against old malware, we have developed a
%We begin with an off-the-shelf malware sample -- Geinimi.a
%benchmark that extends existing malicious behaviors by including the new ones
%-- not seen in the wild -- in order to be more powerful in breaking the
%detection schema.

\noindent {\bf Limitations of off-the-shelf malware.} Most malware samples
available
online~\cite{dissect_malware_2012,contagiodump,malware.lu,virusshare.com} do
not execute correctly. This is because malware may require older, vulnerable
versions of Android OS; they are designed to run only in a specific
geographical location; include anti-emulation
defenses~\cite{anti_emulation_android_vidas}; passively wait for commands from
command and control (C\&C) servers that are temporarily or permanently down;
react to specific user actions.  Another challenge of using off-the-shelf
malware is that
%understand malicious behavior that stands behind 
malware code-names assigned by anti-virus (AV) companies -- like
\textit{CruseWin} and \textit{AngryBirds-LeNa.C} -- do not inform an analyst
what payload the malware actually executes. Hence we evaluate our detector
against malware that we have reverse-engineered and ensured to work correctly
(on both the device and C\&C server sides); whose payloads we understand in
behavioral terms (such as stealing files or Contacts); and whose payloads we
actively diversify to stress test our detector.

%Moreover, sticking to a handful of malware samples that are able to
%run in a lab environment is likely to introduce bias into the experimental
%results because such malware samples usually do not exhibit complex behavioral
%patterns due to being trivially simple.

%The main issue with malware samples available online~\cite{dissect_malware_2012,contagiodump,malware.lu,virusshare.com} is that they often do not execute 'properly'. Malware may require older, vulnerable versions of Android OS, run only in a specific geographical location, include anti-emulation defenses~\cite{anti_emulation_android_vidas}, passively wait for commands from command and control (C\&C) servers that are temporarily or permanently have been brought down, react to specific user actions. Moreover, sticking to a handful of malware samples that are able to run in a lab environment is likely to skew the resulting data set because such malware samples usually do not perform complex actions, thus we would not be able to observe complex malicious behaviors.

%they should expect from those samples. 
%Finally, inconsistency between naming conventions used by different AV
%companies aggravates the problem.

\noindent {\bf Executing diverse malicious behaviors.} 
We analyze 229 malware samples from 126 families drawn from public
repositories~\cite{contagiodump,malware.lu,virusshare.com} (dating from
2012--2015).  We identify common malware payload behaviors and design patterns 
and extend a Geinimi.a sample~\cite{geinimi} to implement these payloads in a
parameterizable manner.

%\noindent {\bf Malware design pattern.}
%When designing our synthetic malware, we have applied client-server paradigm that is common across all real-world malware samples. Malware running on a device performs malicious actions in response to the commands that it receives from its C\&C server and communicates back to the server to confirm successful completion. Synthetic malware allows numerous network-related parameters to be fine-tuned: frequency of communication with C\&C, network bandwidth, the size of data packets sent by client side, interpacket delays between them, device-level intensity of malicious actions.
%
%The client-side code is organized using delegation software design pattern together with Android message passing interface. When a client receives a notification from C\&C in the form of an xml file, it parses server's request, configures itself and launches one or more Android services responsible for carrying out a malicious task. Depending on C\&C's request, malware may run those services sequentially or concurrently. Some services act as an intermediate dispatcher -- they perform initialization and then spawn parallel threads to conduct actual malicious activity asynchronously. For example, the service performing click fraud activity creates a thread pool and supplies a list of URL links to the workers in the pool that they must access.
\input{experimental-setup/malware-table}

%Malicious code can perform several independent tasks that fall into one of the
We find that malware payloads fall into {\em three} orthogonal behavioral
categories:
% that we have found when analyzing real malware
%samples: 
{\em information stealers, networked nodes, and compute nodes}. C\&C server may
instruct client-side code to execute any combination of atomic tasks drawn from
these categories. 

%Their implementation is described later in this section.

\noindent {\bf 1. Information stealers.} Malware stealing personal information
usually focuses on accessing the following data: contacts for spamming
purposes, text messages for breaking two-factor authentication (e.g. bank
trojans), location and files on device for spying on users, phone IDs for
legitimizing stolen devices.  The Geinimi.a sample we start from already
implemented commands to execute these behaviors -- we add the tasks' intensity
and volume as parameters as shown in Table~\ref{table:malware}.

Contact stealing service requires two parameters: the number of
contacts being exfiltrated and delay between sending queries to the contact
provider. SMS stealers in real malware samples act either as a batch
stealer or as an intelligent stealer (e.g. bank trojans). 
Batch stealers transmit all the text
messages to their C\&C server, while the latter register an Android listener to
intercept incoming SMS messages and scan them for the presence of
authentication codes. Alternatively, they can perform intelligent text search
within the database of already received messages and upload 
only sensitive ones to the C\&C server.

To steal location information in Android, our malware can be configured to
register an Android listener that receives location updates or request location
directly from the Android middleware.  In the latter case, Android returns a
previously cached location that may have been determined using cellular network
(approximate mode) or GPS (precise mode) if it is available on a device.

Our extended Geinimi.a malware supports all file operations: it can steal
directory contents, upload a particular file or several files to the C\&C
server, or download a file received from a server to the device.  The
download-file functionality is widely used by malware developers to install new
(likely malicious) apps.

Finally, malware steals device IDs such as IMEI and IMSI codes, OS info
available via the static class android.os.Build, and other miscellaneous device
specific data such as browsing history and bookmarks, description of installed
apps, call logs. This information, especially IMEI and IMSI IDs, is used to
impersonate other devices --- specifically, to legitimize stolen devices before
selling them on a black market and for targeted advertising.

In summary, our Geinimi.a sample allows an analyst to specify the
following parameters: amount of data being read (e.g. 1 or 10
contacts/SMSs/files), intensity of data accessing operations (e.g. read a file
by 1K chunks), delay between successive reading operations (e.g. sleep for 100
ms between retrieving subsequent contacts/SMSs).

\noindent {\bf 2. Networked nodes.} 
%The second behavioral malware category --
This category exploits the phone as a device on a network
of nodes, e.g., to run click fraud or distributed denial-of-service (DDoS) 
attacks~\cite{malware_blackhat_15, prolexic_q4_2013,
android_ddos_2012}. 
%It is getting more and more widespread as network bandwidth keeps
%rising up. 
%Malware developers are lured by frequently changing 
Mobile devices are particularly attractice since their 
IP changes frequently, making it hard to blacklist one.
%the ease of distributing malicious activity across many network nodes (devices)
%that are unlikely to be blacklisted. 

%We choose click fraud and DDoS attacks
Click fraud is a popular, revenue-generating payload. The malicious service
receives a list of URLs from a C\&C, periodically fetches webpages specified in
the URL list and traverses their DOM structure. To speed up the process, our
Geinimi.a extension can be configured to launch several parallel threads.
%The other type -- DDoS attacks from compromised mobile devices -- are getting
%more popular. 
A majority of DDoS attacks abuse the HTTP protocol (e.g. 80\%
according to \cite{ddos_threat_spectrum_2012}). Our malware is able to mount
two attacks: GET flood and SlowLoris~\cite{slowloris_ddos}. GET flood comprises
a series of GET requests sent by compromised network nodes. As opposed to GET
flood, SlowLoris attack is far less computationally expensive because it tries
to exhaust server pool of available connections by opening numerous connections
and sending data very slowly to keep connections alive over a long time period.
In our experiments, synthetic malware establishes 500 connections. If some
connections fail, they are automatically reopened to keep the total number of
active connections constant.

\noindent {\bf 3. Compute nodes.} The last category of malicious behaviors
includes computationally intensive malware. Unlike the previous two, it may or
may not leave distinguishable system-call fingerprint. A typical
example of such malware is a bitcoin miner \cite{mobile-bitcoin-miner}. We
approximate this category by the code cracking SHA1 passwords.

\ignore{
\noindent {\bf Realistic user data.} 
%In addition to careful modeling of malicious behaviors, we have invested a lot
%of efforts into 
We populate our emulators with synthetic data that is statistically close to
the real user data, following data set sizes estimated by Kazdagli et
al~\cite{morpheus-hasp}. Without user data such as Contacts to steal, 
even a functional information stealer malware may not execute 
malicious code. 
%if, for example, it is designed to steal data (e.g. photos,
%contacts, text messages and etc).
}

\noindent {\bf Repackaging Android apps with malware payloads.} 
We embed malware payloads
into benign underlying apps using the same methods as malware developers:
disassembling an apk using \textit{apktool}, copying malicious code into the apk,
and modifying \textit{Manifest.xml} to extend the list of required permissions and
to statically register malicious components. Finally, we reassemble the
decompiled app with \textit{apktool} and sign it using \textit{jarsigner}.

\ignore{
\noindent {\bf Ensuring correct execution.} One of our key design decisions 
is to augment malware to notify the platform 
when events happen (e.g. C\&C request has been successfully
parsed, malware starts/finishes execution). This is very important because
android emulators often crash, thus we can identify and rerun failed
experiments. We also use malware start and stop time stamps to start and stop
collecting malicious feature vectors.  We also make sure that the notification
mechanism does not pollute the collected data.
}

%When designing malware, we have applied client-server paradigm that is common
%across all real-world malware samples. 
Malware payload on a device perform malicious actions in response to the
commands that it receives from its C\&C server and communicates back to the
server to confirm successful completion.  For example, the service performing
click fraud activity creates a thread pool and supplies a list of URL links to
the workers in the pool that they must access.  Most Android apps (both benign
and malicious) are obfuscated using standard Android tool --
Proguard~\cite{proguard} -- which renames classes, fields, and methods with obscure
names. We have applied Progaurd to our modified Geinimi.a malware as well.

\noindent{\bf Diversity of malware payloads.} Table~\ref{table:malware}
shows the parameters used for each payload category 
and the effect of payloads on system call traces (in aggregate). It is clear
that the payloads differ in system call intensities and (as we find in practice)
yield a diverse set of machine learning features.

%Our malware allows
%numerous network-related parameters to be fine-tuned: frequency of
%communication with C\&C, network bandwidth, the size of data packets sent by
%client side, interpacket delays between them, device-level intensity of
%malicious actions.

%The client-side code is organized using delegation software design pattern
%together with Android message passing interface. 
\ignore{When a client receives a
notification from C\&C in the form of an xml file, it parses server's request,
configures itself and launches one or more Android services responsible for
carrying out a malicious task. Depending on C\&C's request, malware may run
those services sequentially or concurrently. Some services act as an
intermediate dispatcher -- they perform initialization and then spawn parallel
threads to conduct actual malicious activity asynchronously. For example, the
service performing click fraud activity creates a thread pool and supplies a
list of URL links to the workers in the pool that they must access.
}

%It is worth noting that m

%\mikhail{we need a table with malware parameters + one paragraph to describe them}

%% file: experimental-setup/malware-table.tex
\begin{table*}[tb]
\centering

\begin{tabular}{p{3cm} >{\raggedright\arraybackslash}p{2.25cm} p{1.75cm} >{\raggedright\arraybackslash}p{1.5cm} >{\raggedright\arraybackslash}p{2cm}}
Syntetic Malware & Parameters (per payload action) & Malware Spec. Delay (ms) & Exec. Time (min) & Num. Syscalls (Million)\\ 
\hline

File Stealer (4.2MB each) &1, 10, 50 &0, 5k & 3215 & 493.8\\
Contact Stealer & 25, 250 &0, 50 & 1520 & 238.7\\
SMS Stealer & 50, 1000&0, 50 & 1525 & 236.4\\
ID, GPS Stealer &data size fixed &0, 200 & 540 & 89.5\\
Click Fraud (webpages) & 20, 300&0, 3k & 1490 & 223.7\\
DDos (slow loris) & 500 Connections & 1, 200 & 495 & 73.9\\
SHA1 passwd. Cracker & 10k, 1.5M&0, 50 & 1540 & 242.3\\
\hline \\
\end{tabular}
\caption{Summary of synthetic malware payload configurations and resulting system call traces. Note, the total execution times and system call counts were counted only for the one-second intervals where malware payload was active.}
\label{table:malware}
\end{table*}

%% file: experimental-setup/LDs.tex
\begin{figure*}[t]
  \begin{minipage}[tbp]{0.33\linewidth}
    \centering
    \includegraphics[width=\textwidth]{figs/ROC/ROC_wiki_pca.pdf}\par
    \subcaption{Histogram FE algorithm.}
    \label{fig:roc_pca_wiki}
  \end{minipage}
  \begin{minipage}[tbp]{0.33\linewidth}
    \centering
    \includegraphics[width=\textwidth]{figs/ROC/ROC_wiki_ngrams.pdf}\par
    \subcaption{N-gram FE algorithm.}
    \label{fig:ROC_ngram_wiki}
  \end{minipage}
  % \hfill
  \begin{minipage}[tbp]{0.33\linewidth}
    \centering
    \includegraphics[width=\textwidth]{figs/ROC/ROC_wiki_intuitive_basis.pdf}\par
    \subcaption{Histogram of manually classified system calls.}
    \label{fig:ROC_intuitive_basis_wiki}
  \end{minipage}
  \caption{
(Wikipedia app) True positive v. False positive (ROC) curves shows
    detection accuracy of six  local detectors. In terms of detection accuracy
    and computational efficiency, Random Forest together with histogram feature
    extraction (FE) algorithm outperforms the rest classifiers and FE
    algorithms.
    However, any of these local malware detectors has unacceptably high false
    positive rate if one wants to achieve at least 80\% TP rate.}
  \label{fig:ROC}
\end{figure*}

%Despite many years of research into generic machine learning-based malware
%detectors, they still remain ineffective due to high false positive rate. 

Our first step is to establish a good local detector (LD) for Android devices.
In particular, we choose system call based LDs since the system call interface
has visibility into an app's file, device, and network activity and can thus
capture signals relevant to malware executions.  Interestingly, 
%while system call based detectors have been evaluated for desktop systems,
%detection rates 
unlike system call detectors for Linux and Windows, detectors on mobile
platforms like Android have not been evaluated by prior work.  
Doing so is important since
Android adds several middleware services that consolidate 
system call traces traces from third-party applications -- for example, services that 
manage
Contacts, SMSs, sensors, etc. Further, Android also 
provides a different event-driven
programming model and software stack compared to Windows and Linux. 

We experiment with an extensive set of system-call LDs -- our takeaway is that
even the best LD we could construct operates at a true- and false-positive
ratio of 80\%:15\% and is not deployable by itself.

\ignore{
The high false positive rate is likely due to the nature of mobile malware
as well as our evaluation method that emphasizes realistic usage of benign 
applications.\mohit{point to experimental setup here}
%the dataset reflects correct (and diverse) benignware and
%malware executions (see Section~\ref{todo}), and this diversity is hard to
%capture precisely in machine learning models. 
Note that alternative LDs based on
hardware~\cite{Demme2013,ponomarev-hpca-15} or Android middleware~\cite{Enck2009,Peng2012} also report
similarly high false positives -- all such LDs' results can be improved using
our proposed Global Detector (GD).
%feature extraction and machine learning models to construct as good an LD as
%possible.  not able to achieve high detection rate on our data set while
%maintaining low false positive rate. 
%This high false positive rate
%motivates our global detector (GD) algorithms, and we quantify its performance
%in subsequent sections. 
Further, improvements to LDs are orthogonal to \sysname and only serve to
further improve the time to detect a malware outbreak.
}

Each LD 
%can be considered as a tuple that includes 
comprises of a machine learning (ML) classifier and a feature extraction (FE)
algorithm. 
%Design of an FE algorithms and machine learning classifiers is not a part of
%our contribution, however, we have analyzed several such algorithms in order
%to choose a representative LD. 
We choose LD candidates from general machine learning (ML) models that are
computationally efficient to train -- such as SVMs, random forest, k-Nearest
Neighbors, etc and not including the more complex artificial neural networks or
deep learning algorithms.  We deliberately avoid handcrafted ML algorithms and
hardcoded detection rules because such custom models usually are overfitted to
a specific dataset, and because they require humans to continuously adjust the
detector to new types of malware.

%Our choice of FE algorithms is guided by previous work and computational
%constraints of mobile devices. In particular, 
To extract features, we experiment with
histograms, n-grams, and also using a human-understandable system call basis.
When using histograms and n-grams, we compute the (top ten) principal
components using PCA analysis to reduce dimensionality of the data. 
For the human-understandable \intbasis, we group system calls into
33 dimensions each related to file access, networking, process control, etc.
We use \intbasis as a candidate because it enables malware analysts to 
map malware payloads back to intuitive tasks such as stealing SMSs or snooping
on location data.

Using these features, we experiment with six state-of-the-art ML algorithms: random
forest, 2-class SVM, kNN, decision trees, and their ensemble versions --
boosted decision trees with AdaBoost algorithm and Random SubSpace ensemble of
kNN classifiers~\ref{fig:ROC}. We also evaluated 1-class SVM as an anomaly
detector and Naive Bayes classifier -- however, both yielded an extremely high
FP rate and we exclude them from further discussion.

In total, we consider these 18 potential LDs (three FE algorithms x six ML
classifiers) to choose the best one.  We compare all of them using ROC
curves~\ref{fig:ROC} that plot true positive v. false positive rates, so that
the area under the ROC curve (AUC) is a quantitative measure of LD's
performance: the larger the AUC, the more accurate the detector.
%that show how true positives vary with false positives.  where projecting a
%point on ROC curve on the horizontal axis shows corresponding false positive
%rate, while its projection on the vertical axis is true positive rate.
To construct ROC curves we apply 10-fold cross validation to avoid any bias
from the dataset, i.e. our classifier predicts labels of in-fold observations
using a model trained on out-of-fold observations. 
%The results are grouped based on FE algorithm. 

%% file: experimental-setup/synthetic-network-gen.tex
%We designate 2 primary categories of malware propogation networks: first are
%small world and scale free community models, and the second are latice-based
%and tree-based structures. The former captures social network \textit{nodule}
%behaviours, where a single vertex may have a high degree where general vertices
%do not. For example, in so called \textit{waterhole attacks}, many users may
%become infected if they go to the same infected shared resource. Contacts
%graphs are another example of community social networks and can be useful when
%modeling spam networks. However, lattice and tree based graphs best represent
%some notion of \textit{locality} of members. For this reason, we can categorize
%approximate peer to peer networks, compromised peripheral devices, and
%geographic/proximity attacks via grid or tree-based structures.

\ignore{
We now present our choice in representative graphs for describing malicious
activity on networks. We propose two naive synthetic networks,
Erd\H{o}s-R\'{e}nyi and grid graphs, and generate independent communities
connected with a known probability. For simplicity, each community contains
approximately an identical number of nodes.
This conservatively models the situation where we operate on a sub-graph of a
larger network with coarse grained community detection, which is the worst case
scenario for our detection system. In addition, we test on several real world
graphs taken from the SNAP database \cite{snapnets}.      
}

The first type of community graph is an Erd\H{o}s-R\'{e}nyi model with $|V|$
nodes and $|E|$ edges.  Given the size of the community as $A_c = \{ v \in V :
v \in c\}$ then we estimate the probability of intra-community connection
probability to be $p = 2*|E|/|A_c|^2$.  We choose $p$ such that the degree of
the network, $M$, is between three and four\footnote{Nodes with zero degree are
removed.} so that the network is not too dense, mimicking many realistic
graphs. We generate a random graph with 2 communities of (approximately) 1k
nodes such that the communities are equal in size. The resulting graph contains
958 nodes, 3108 edges, and approximately 600 nodes per community.

The second community graph we construct is a grid graph. For each community, we
generate a lattice-mesh with a known number of vertices and edges. 
%This guarantees every node has equal degree. 
We generate a grid with 2 communities,
1250 nodes, 4950 edges, and 625 nodes per community.  
  
Lastly, we use a subset of the Youtube graph from the Stanford SNAP database
\cite{snapnets}.  
%The SNAP data set provides various types of physical networks such as contacts
%graphs, social networks, physical network structures, and ground-truth
%community graphs. In particular, these datasets include the most
%representative 5000 ground-truth communities within a variety of networks. For
%example, both YouTube and Orkut graphs capture the concept of a community. In
%YouTube, for instance, a community corresponds to a channel which users may
%subscribe.  For efficient processing of the SNAP datasets, 
We choose a sub-graph as follows: First, we pick a random node and find its
k-hop neighbors, building a sub graph $G'$ from these nodes to preserve
community structure; next we choose a random sub-graph from $G'$, removing all
the zero degree nodes until the desired number of nodes is reached.  Sub-graphs
picked in this way preserve the community structure of the network
% and provide
%us with a realistic network to test on. 
Using this algorithm on the YouTube social network, we extract a subgraph of
1005 nodes, 2906 edges, and 19 communities.

\noindent{\bf Malware Spread Dynamics.}  The malware spreads according
to a standard SI model \cite{Durrett07}. In brief, the SI process is a
random infection process spreading on the graph. Each edge in the
graph has a time-interval associated with it (exponential, independent
across edges). Initially, all nodes are tagged `0' (uninfected),
except a source node which is tagged `1' (infected). The edge weights
(time intervals) now propagate this `1' throughout the network. The
time at which a specific node becomes `1' is simply the sum of the
weights along the shortest weighted path on the graph connecting the
source node to this specific node.

% In our case, we run two SI processes, one originating from `infected
% source' and the other originating from the `benign source'. Every node
% in the (connected) graph thus gets a pair of timestamps: the time at
% which the SI process from `infected source' reached it, and the time
% at which the SI process from `benign source' reached it. We tag a node
% as `infected' if the `infected source' reached it first, and `benign'
% other-wise. 

% We model malware propagation as an exponential process. First we pick subsets
% of nodes to represent malicious sources and benign sources accordingly. Next we
% assign edge weights in the graph by sampling an exponential distribution. We
% chose an exponential distribution because it is \emph{memoryless}. Each
% non-source node in the network is assigned the state of its closest source
% node; nodes closest to benign sources become benign, and nodes closest to
% malicious sources become malicious. We determine the closest source by
% computing the distances to all nodes from each source node via the Bellman-Ford
% shortest paths algorithm and comparing the results.

%% file: experimental-setup/data-collection.tex
% Data Collection phase goes here
For our large-scale experiments, we deploy \sysname on Amazon Web Services (AWS).
Experiment nodes run Windows Server 2008 R2 Base on default T2 micro instances with 1 GiB RAM, 1 vCPU, and 50 GB local storage. 
We Populate the experiment environment with user data commonly found on a real host including: 
PDFs, word documents, photos, Firefox browser history, Thunderbird calendar entries, and Thunderbird contacts.
We then create a snapshot image of the virtual machine environment and clone this image for each experiment.
For simplicity, all malware experiments are conducted with a fresh clone of the base image.

Client code runs on-host and runs each experiment and communicates results with the central database.
Each experiment is run via a custom Intel Pin tool, which launches the target applications and intercepts system calls. 
This tool registers timing information and system calls with arguments, and is configured to follow spawned child processes.
We chose Pin because it instruments binaries without modification in a format comparable to \tttext{strace} in Linux. 

Simulated local network
The worker instances run in a virtual private cloud (VPC).
We simulate a local network via two subnets in the VPC. 
\textit{Neigboring machines} exist in a private subnet with no internet access, but are accesible by the experiment machines.
If malware happens to propogate to one of these vulnerable machines then it is, by default, quarantined.
Experiment machines run on the second subnet with limited public internet access. 
The second subnet allows malware to communicate over common C&C ports and the client code to post results.
Specifically, we allow access for HTTP, HTTPS, SMTP, DNS, Telnet, and IRC.   

%% file: results_symantec_2.tex
We now quantify how \sysname concentrates malware in Symantec's Wine dataset
into neighborhoods. By using downloader graphs as a weakly correlated 
attribute, \sysname identifies malicious files and infected machines with
significantly lower false positives than using LDs~\cite{vt_report_classification} alone 
%(while losing
%only a few true positive malware that didn't get captured into neighborhoods).
%On the other hand, the state-of-the-art 
and far higher true-positives than a downloader-graph based
detector~\cite{downloader_graphs,pup_ndss} alone. 
%s limited to detecting downloaders (to reduce
%false positives) -- in comparison, \sysname has significantly higher
%true-positives at a comparable false positive rate. 

In addition, neighborhoods and shape together are good predictors of malware
behavior -- hence \sysname does not have to wait until the entire sequence of
malware payloads have been downloaded to declare a downloader or a machine as
malicious. We find that on average, \sysname can identify a file as malicious
only $\sim$20 days after it enters the Wine dataset and $\sim$345 days
before VirusTotal confirms it as malware. Table~\ref{fig:file_agg_analysis}
%and~\ref{fig:machine_agg_analysis} 
summarizes these results.

%%%%%%%%%%%%%%%%%%%%%%%
\subsection{\sysname Classifiers}
\label{sec:symantec_local_detectors_results}

\begin{figure}[tbp]
\begin{minipage}[tbp]{0.49\linewidth}
  \includegraphics[width=\textwidth]{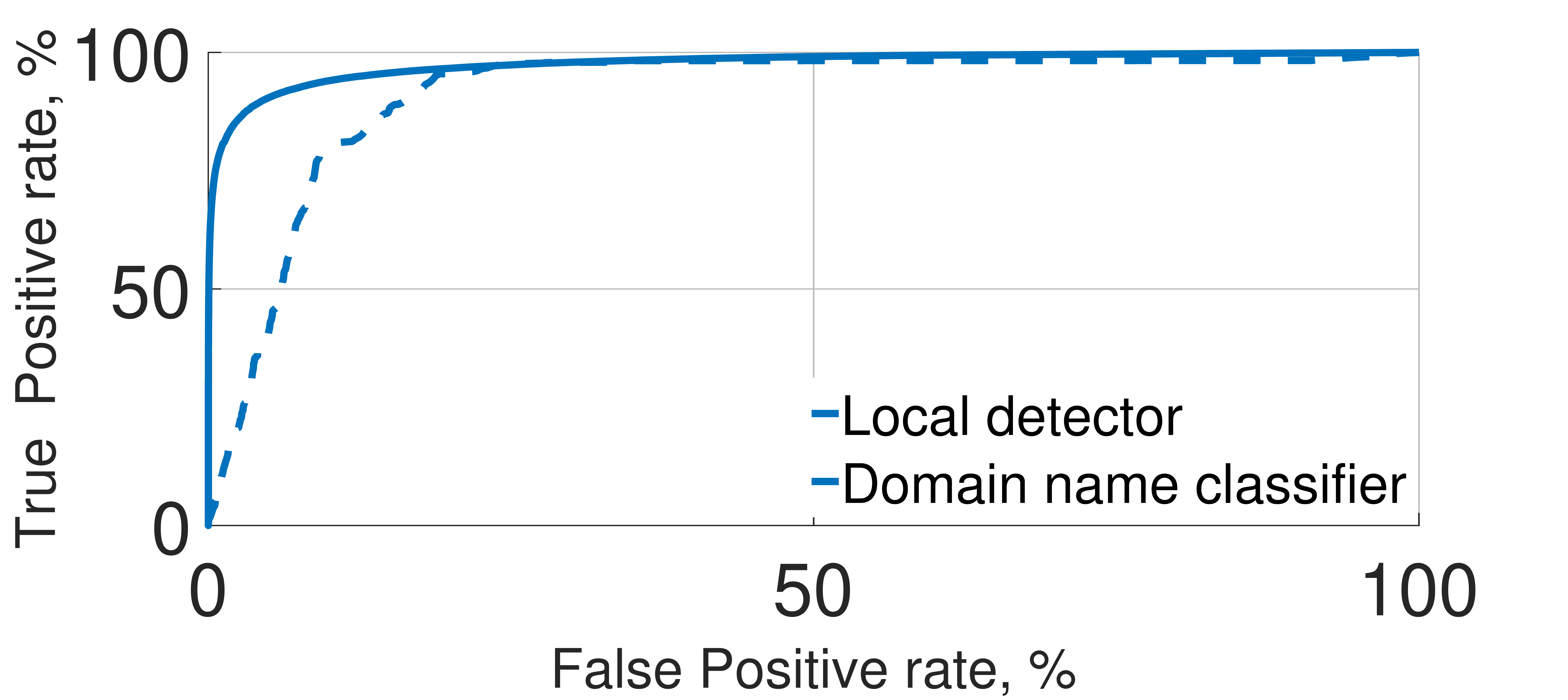}\par
   \label{fig:ld_roc}
\end{minipage}
\begin{minipage}[tbp]{0.49\linewidth}
   \includegraphics[width=\textwidth]{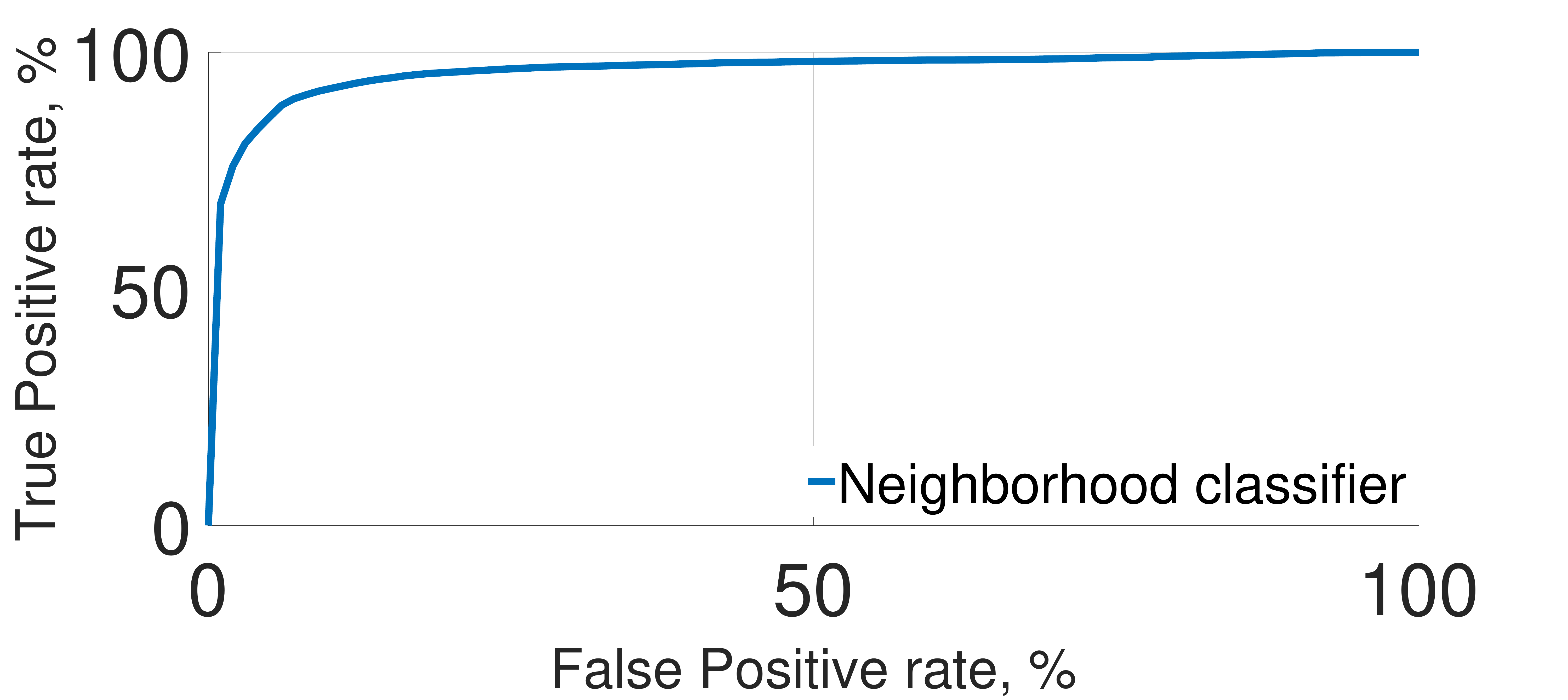}\par
   \label{fig:nbd_roc}
\end{minipage}
\caption{{\em (Left) Receiver operating curve (ROC) of the local detector and the domain name classifier.
(Right) ROC of the neighborhood classifier.}}
\label{fig:roc-curves-symantec}
\vspace{-0.2in}
\end{figure}

\noindent \textbf{Local detectors.} We start with the evaluation of local
detectors (Section~\ref{sec:local_detectors}). Each local detector algorithm
comprises two parts -- feature extraction and a binary classifier 
(XGBoost in our prototype).
We train a local detector on the set of 2.6 million VirusTotal reports using
10-fold cross validation.  The detector achieves 97.61\% area-under-the-curve metric
(Figure~\ref{fig:ld_roc}), and we chose the operating point of 5.0\% false
positive rate and 90.47\% true positive rate. Note that due to the high number of
benign files in the dataset, a 5.0\% false positive rate corresponds to more than
1M misclassified files, which is likely to prevent practical deployment 
of such a local detector.  In subsequent experiments, we use out-of-fold
predictions made by the detector.

\noindent \textbf{Domain name classifier.} We train and evaluate the classifier
(Section~\ref{sec:local_detectors}) on 251K VirusTotal domain
reports using 10-fold cross validation to achieve an 91.58\% AUC
(Figure~\ref{fig:roc-curves-symantec}). We specifically choose an operating point of
19.03\% false positives and 95.41\% true positives.

The domain name classifier is `weak' because it is conservative while labeling
domains --  an entire domain is considered malicious if it serves at least one
malware sample. However, even malicious domains serve several benign files, and
the local detector (above) that analyzes file-level features using VirusTotal
contradicts the domain name classifier. Adding more information about the URL
can improve the classifier -- however, even the weak signal in domain names is
sufficient for \sysname to significantly improve the local detectors.
Interestingly, since the domain name classifier is only used to create
neighborhoods (and not alerts), it can operate at a conservative setting
and rely on the shape-based neighborhood classifier to weed out false
positives.

%Even though the domain name classifier is not deployable by itself due its high
%false positive rate, it is an integral part of \sysname and 
The domain name classifier lets \sysname
efficiently filter out domains that are unlikely to distribute malicious files.
Specifically, it removes from further consideration 68.62\% (214,884 out of
313,133) completely benign domains that are responsible for  delivering 80.70\%
(16,222,941 out of 20,103,211) benign files.  At the same time the
classifier retains 75.86\% (30,448 out of 40,134) malicious domains
responsible for delivering 88.31\% (94,457 out of 106,959) malicious files.

\noindent{\bf Neighborhood classifier.} 
The neighborhood classifier (Algorithm~\ref{ShapeGD_algorithm}) performs 
neighborhood-level feature extraction and feeds resulting feature vectors 
into an XGBoost classifier.
We estimate its detection capabilities using 10-fold cross validation.
The ROC plot (Figure~\ref{fig:nbd_roc}) shows that the classifier
achieves 96.13\% AUC score, and we choose the following operating point: 5\%
false positives and 91.83\% true positive rate. 

A neighborhood-level alert is different from an the above file- and domain-name
based local detectors' alerts -- it signifies that a set of files that have
suspicious behavior have been downloaded from suspicious links, and hence
identifies the large majority of files that were false positives at the local
level. First, we measure the degree to which our neighborhood classifier
removes benign files, and then show that by re-examining files in suspicious
neighborhoods (using the file-based LD), we can capture 78.03\% of true
positives.

%\mikhail{the 1st sentence contradicts the subsecton's title; need to add smth at the beginning}
%After removing domains serving only a single
%malicious file, the average lifespan of a malicious domain is 157 days; hence we set the
%neighborhood time window to 150 days. We slide this 150-day NTW by 30 days over
%the 5-year period under evaluation (having observed no significant change in
%results by changing the sliding window duration).
%
%The ROC plot (Figure~\ref{fig:nbd_roc}) shows that the neighborhood classifier
%achieves 96.13\% AUC score, and we choose the following operating point: 5\%
%false positives and 91.83\% true positive rate.  
%Note that since this classifier labels files downloaded from a domain as either benign 
%or malicious, it
%each true- or
%false-positive applies to a set of nodes -- 

%Note that since this
%classifier labels a neighborhood of files as malicious, each true- or
%false-positive applies to a set of nodes -- 

%At this point we can not make any conclusions about the performance of the
%entire malware detection framework because we do not know how many malicious
%and benign files are covered by the neighborhoods labeled as malicious.
%interestingly, we noticed that due to correlation of 
%because benign domains that are classified as a false positive also create the
%5\% of misclassified neighborhoods, the files in these neighborhoods largely
%overlap rather being completely distinct. 
%\todo{Mikhail, numbers here on
%distinct files v. total files in FP nbds.}

%%%%%%%%%%%%%%%%%%%%%%%%
\subsection{Neighborhoods Concentrate Malware}
\label{sec:nbd-conc}

\begin{figure}[tbp]
   \vspace{-0.0in}
   \centering
   \includegraphics[width=0.45\textwidth]{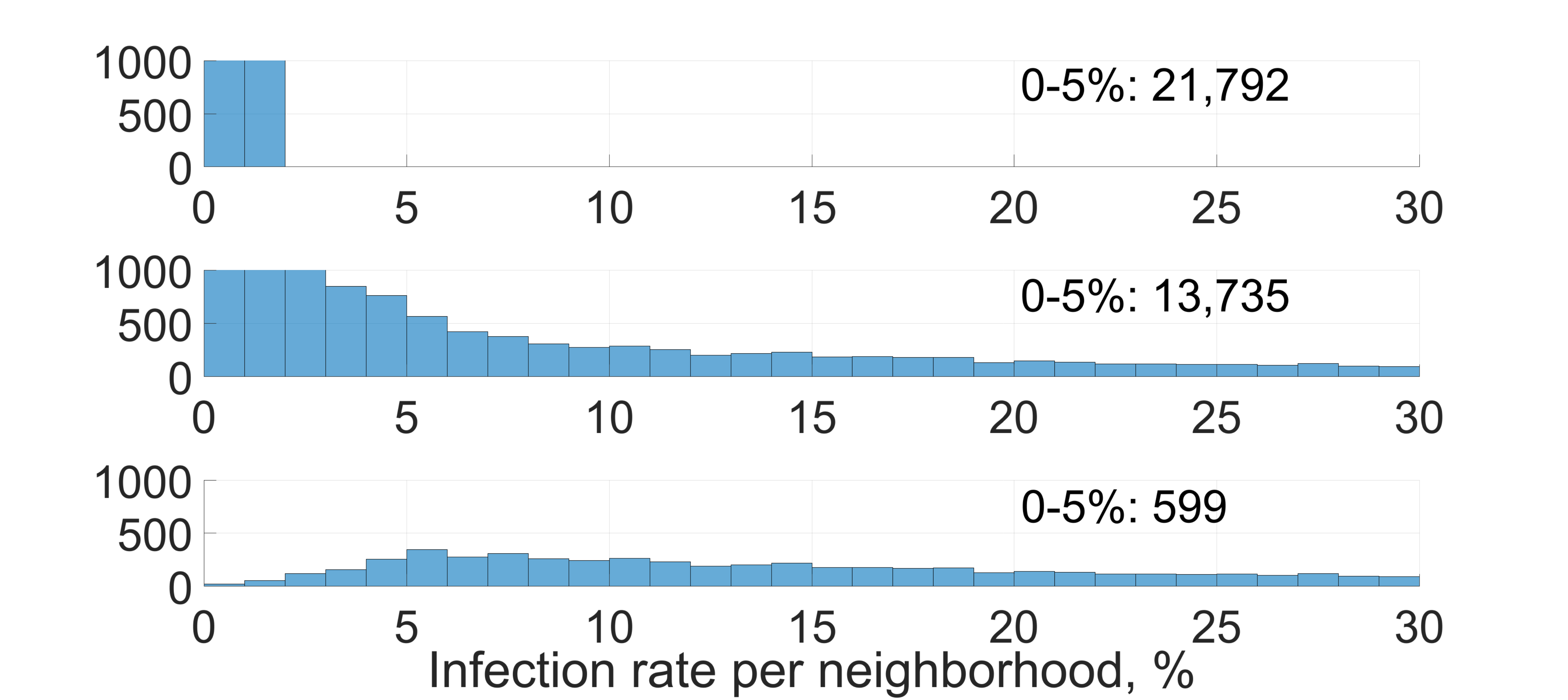}\par
   \caption{{\em Neighborhood classifier acts as a malware concentrator. 
   (Top) Distribution of infection rates of randomly grouped files.
   (Middle) Distribution of neighborhoods' infection rates.
   (Bottom) Distribution of neighborhoods' infection rates after filtering out low-infected neighborhoods. 
   The neighborhood classifier retains only highly infected neighborhoods.
   (Distributions are capped at 1,000 level.)
   }}
   \vspace{-0.2in}
\label{fig:nbd_concentrator}
\end{figure}

First, we measure the effect of using domain names from downloader graphs 
as an attribute to create neighborhoods.

The original malware concentration in the Wine dataset is only 0.663\%, as
shown in the top-most plot of Figure~\ref{fig:nbd_concentrator}.  If a random
subset of files are grouped into a neighborhood, each neighborhood will have
considerably less malware than the false positive rate of the malware detectors (5\%)
-- i.e., creating neighborhoods randomly does not concentrate malicious
activity. This is the baseline against which downloader graph based
neighborhood creation and shape-based neighborhood classifier have to be
compared -- the neighborhoods labeled as malicious have to contain more than
5\% malicious files while achieving high malware coverage overall.

\sysname first uses the domain-name classifier to prune out files downloaded
from benign domains -- this increases the 0.663\% infection rate to
9.49\% (middle plot, Figure~\ref{fig:nbd_concentrator}). %However, the spike
%on the left-hand side indicates that a large majority of neighborhoods have
%less than 1\% of malware files in them.
However, high bars on the left-hand side (they are cut off at 1,000 file level) 
indicate a large majority of neighborhoods have relatively low concentration of malicious files in them.

\sysname then uses the shape-based neighborhood classifier to identify infected
neighborhoods. This dramatically changes the distribution of neighborhood
infection rates, i.e. the peak shifts to the right -- from 1\% to 5\% (lowest
plot in Figure~\ref{fig:nbd_concentrator}).  The neighborhood classifier brings the
average malware concentration in a neighborhood from 9.49\% to 24.6\%, an
increase of 37.1$\times$ compared to randomly grouping files into neighborhoods.

Specifically, the number of neighborhoods with the infection rate less than 1\%
drops by 437.6 times (from 8752 on the upper plot to 20 on the lower plot).
%Overall, the neighborhood classifier reduces the number of low infected
%neighborhoods (neighborhoods with less than 5\% of malicious files) by 22.9
%times (from 13,735 to 599).
Overall, the neighborhood classifier together with the domain-name classifier
reduce the number of low infected neighborhoods (neighborhoods with less than 5\% of malicious files) 
by 36.4 times (from 21,792 to 599).

%%%%%%%%%%%%%%%%%%%%%%%%%
\subsection{Aggregate Detection Results}
\label{sec:symantec-detection}
 
We now quantify the detection performance of the complete pipeline -- i.e., by
applying the malware classifier to files inside infected neighborhoods.
By identifying malicious neighborhoods, \sysname effectively weeds out
many files that trigger false alerts -- hence, the alerts within infected
neighborhoods are $\sim$37 times more likely to be malware (true postive).

\begin{figure}[tbp]
   \vspace{-0.0in}
   \centering
   \includegraphics[width=0.45\textwidth]{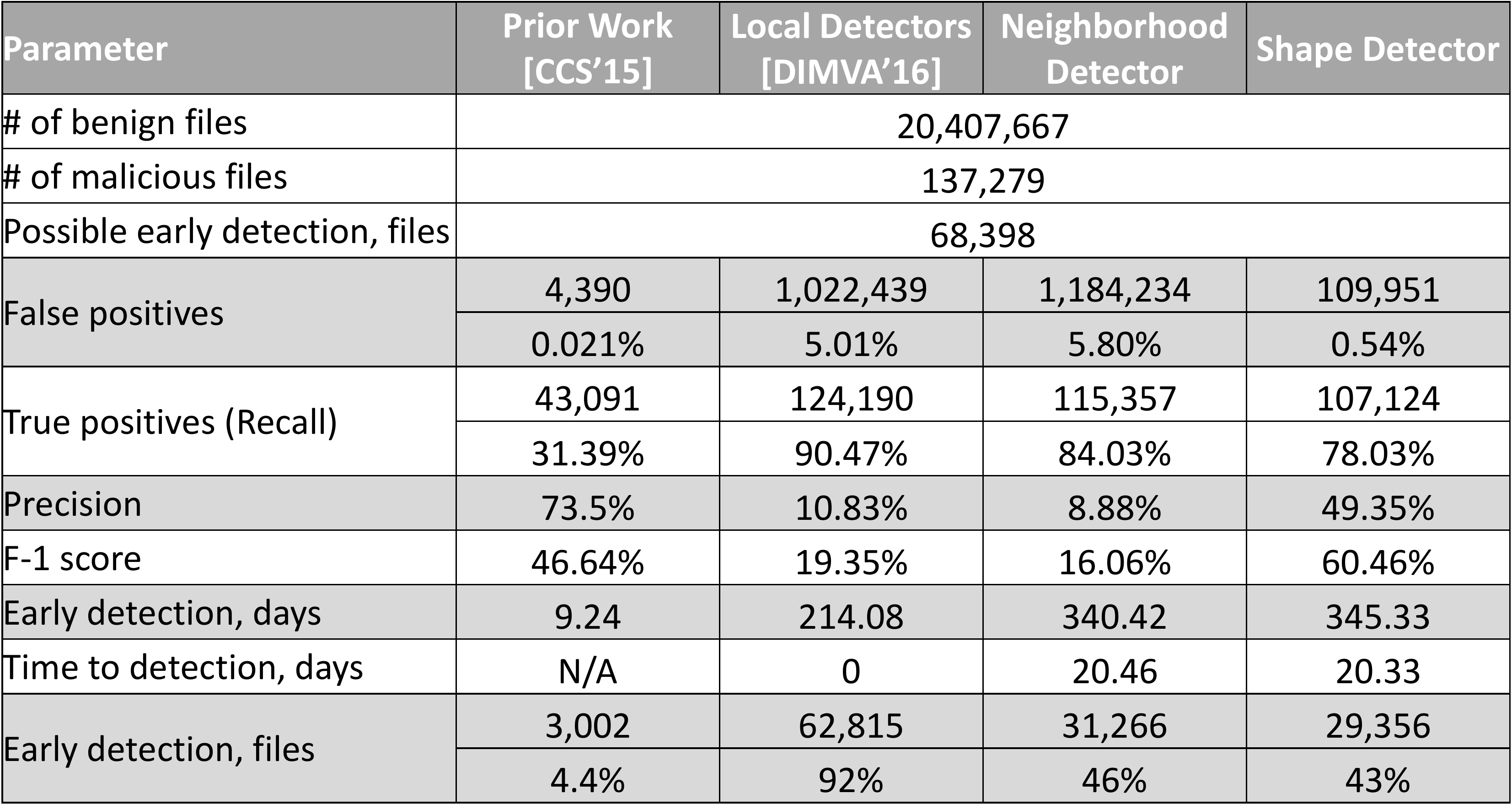}\par
   \caption{File-level aggregate results.}
   \vspace{-0.2in}
\label{fig:file_agg_analysis}
\end{figure}

%Specifically, we focus our analysis on how well \sysname detects malicious files.
%the following two characteristics -- how well
%\sysname detects malicious files and how well it detects infected machines, i.e. 
%machines that have downloaded at least one malicious file.
%
%\mikhail{After removing domains serving only a single
%malicious file, the average lifespan of a malicious domain is 157 days; hence we set the
%neighborhood time window to 150 days. We slide this 150-day NTW by 30 days over
%the 5-year period under evaluation (having observed no significant change in
%results by changing the sliding window duration).}
%
To perform real-time analysis, we replay the 5-year long history of download events in the Wine dataset 
(each event has a timestamp associated with it)
and execute \sysname every 30 days. 
We set the neighborhood time window (NTW) parameter to 150 days because we found that the average
lifespan of malicious domains is 157 days.
In our experiments we observed that
shorter period between consecutive runs of \sysname does not significantly affect
results, it only improves time to detection and early detection parameters (Table~\ref{fig:file_agg_analysis}). 
We intentionally stick to a 30-day period between consecutive runs of \sysname{} 
to keep execution time ($\sim$12 hours) and resource consumption manageable.

We compare \sysname that comprises of the neighborhood classifier and local detectors 
with prior work -- local detectors~\cite{vt_report_classification} and
the state-of-the-art malware detector in the Wine dataset~\cite{downloader_graphs} -- as well as
a neighborhood detector. For comparison we use standard machine learning metrics:
$precision=\frac{TP}{TP + FP}$, $recall=\frac{TP}{TP + FN}$, 
and $F-1=2\cdot\frac{precision\cdot recall}{precision+ recall}$.

%We compare real-time behavior of three detectors -- local detectors~\cite{vt_report_classification}, 
%a neighborhood detector, which conservatively labels all the files within a malicious neighborhood as malicious, 
%and \sysname that comprises of the neighborhood classifier and local detectors 
%-- in terms of precision, recall, and F-1 score, which is a commonly used way to combine precision and recall.
%$
%\displaystyle
%\begin{aligned}
%Precision = \frac{TP}{TP + FP}
%\end{aligned}
%$
%
%$
%\displaystyle
%\begin{aligned}
%Recall = \frac{TP}{TP + FN}
%\end{aligned}
%$
%
%$
%\displaystyle
%\begin{aligned}
%F1 = 2 \cdot \frac{precision\cdot recall}{precision+ recall}
%\end{aligned}
%$

%\subsection{Aggregate detection}

Though \sysname is designed to act as a real-time malware detector, i.e. output detection results
every time it is executed, in this section we only focus on the file-level aggregate results (Table~\ref{fig:file_agg_analysis}) 
in order to compare with a prior work. For completeness we describe machine-level aggregate results 
in the Appendix~\ref{agg_machine_detection} and the real-time detection results in the Appendix~\ref{app:real_detection_symantec}.
The aggregate results  are computed by merging malware detection results across independent executions
of a malware detector. Note that we count each file exactly once, for example, if a malware detector 
detects the same malicious file over multiple NTWs, we count it only once.

%To make comparison with prior work easier, we discuss only the aggergate detection results that do not
%include temporal aspect of real-time detection. %Such analysis allows us to compare \sysname with
%the state-of-the-art malware detector for the Wine dataset~\cite{downloader_graphs}.
%We compute aggregate statistics by merging malware detection results across independent executions
%of a malware detector. %In comparison to the real-time detection, which may produce biased results if
%a detector keeps detecting the same malware samples over multiple runs, 
%Aggregate results
%demonstrate what percentage of malware is actually detected across all runs.
%When presenting aggregate results, we count each file exactly once, i.e. whether it is
%detected or not, irrespective of that it may be present within multiple neighborhood 
%time windows.

\noindent{\bf False positive rate.}
The downloader detector~\cite{downloader_graphs} achieves the lowest FP rate.
It raises 1.0\% false positives on the set of downloaders, however, 
downloaders constitute only a small portion of the entire dataset 
(439K out of 20.55M files).
Thus, its effective FP rate comes down to 0.021\%, which is reached at the cost of excluding
more than 20 million (or more than 97.3\%) files from the analysis.
The other prior work -- a local detector~\cite{vt_report_classification} -- has a fixed false positive
rate of 5\% that we set up in our experiments to make it achieve above 90\% TP rate.

Surprisingly, the neighborhood detector's FP rate is only marginally worse than the local detector's
FP rate -- 5.8\% in comparison to 5\%. However, it filters out significant portion of benign files, which
helps \sysname to reduce FP rate by 10.7 (5.8\% vs. 0.54\%) times by using the neighborhood detector as a file filter.
In comparison to the local detector, \sysname has 9.3 lower FP rate (5\% vs. 0.54\%), thus 
it brings the absolute number of false positives from $\sim$1.2 million down to $\sim$109.9 thousand.
Therefore, the deeper (even human-level) analysis becomes feasible, i.e.
$\sim$109.9 thousand false alerts over a 5-year period correspond to 60 false alerts per day on average.

\noindent{\bf True positive rate.} 
The downloader detector~\cite{downloader_graphs} has the lowest TP rate due to
its inherent inability to analyze non-downloaders.
Therefore, it discovers
96\% malicious downloaders, but only 31.39\% all malware samples -- 
it misses 94K out of 137K malware samples.
Note the Wine dataset may be skewed in the favor of malicious downloaders, 
i.e. approximately one third of malware samples in the dataset are malicious downloaders.
Thus, the downloader detector may have even lower TP rate in a real deployment setting.

The neighborhood detector achieves a slightly lower TP rate ($\sim$84\%) 
because it erroneously filters out some malicious files while
the local detectors analyze all of them. Specifically, if the neighborhood detector
fails to correlate malicious downloads appropriately, it may distribute malware samples 
across multiple predominantly benign neighborhoods. Due to low malware concentration
they may be excluded from the further analysis by the neighborhood classifier.
The other reason why the neighborhood classifier misses some malware 
may come from labeling some malicious domains as benign. Thus, malware samples downloaded
from such domains are excluded from the further analysis.

In terms of true positives, \sysname inherits limitations of the neighborhood detector.
it loses a few more percent due to running imperfect local detectors within the neighborhoods
that capture only 84\% malware, which results in 78\% TP rate.
On the contrary, local detectors demonstrate the highest TP rate (90.5\%) because
they are tuned to achieve higher than 90\% TP rate.

\noindent{\bf F-1 score.}
All four detectors explore different operating points in the FP/TP design space.
To compare them, we use a standard machine learning metric -- F-1 score. 
The F-1 score is bounded by 100\%, which is achieved only if a detector 
has 100\% TP rate and 0\% FP rate.

\sysname achieves the highest F-1 score (60.46\%) because it detects a large portion of
malware samples in the dataset ($\sim$78\%) and it maintains the low FP rate (0.54\%).
The next closest competitor -- the downloader detector~\cite{downloader_graphs} --
achieves only 46.64\% F-1 score due to its low TP rate. 
Interestingly, the local detector~\cite{vt_report_classification} demonstrates $\sim$2.3 times worse
results than the downloader detector because of much high FP rate.

%\subsection{Symantec Count-GD fragility}
%
%\begin{figure*}[t]
%\begin{minipage}[tbp]{0.24\linewidth}
%  \includegraphics[width=\textwidth]{figs/symantec/count-fragility/nbd-count-gd-fragility.eps}\par
%   \label{fig:ld_roc}
%\end{minipage}
%\begin{minipage}[tbp]{0.24\linewidth}
%   \includegraphics[width=\textwidth]{figs/symantec/count-fragility//file-count-gd-fragility.eps}\par
%   \label{fig:domain_roc}
%\end{minipage}
%
%\caption{Symantec Count-GD fragility .Left - Right: NBD vs File results.}
%\label{fig:symantec-count-gd-fragility}
%\end{figure*}

%% file: results_v3.tex
\sysname identifies malicious neighborhoods with {\em less than
1\% false positive and 100\% true positive rate when the neighborhoods produce
more than 15,000 FVs within a neighborhood time window (i.e., $|B| > 15,000$ in
Algorithm~\ref{ShapeGD_algorithm}}).  Recall that at 60 FVs/node/minute, it
takes 1000 nodes only 15 seconds to create 15,000 FVs. For LDs like ours with
$\sim$6\% false positive rate, this corresponds to 900 alert-FVs.  We then
simulate realistic attack scenarios and find that \sysname can detect malware
when only 108 of 550K possible nodes are infected through a
waterhole attack using a popular web-service. %Finally, \sysname is

\subsection{Time to detection using temporal neighborhoods}
\label{sec:time-nf}
%Neighborhood filtering aggregates suspicious FVs across the nodes within a
%neighborhood window.

\begin{figure}[tbp]
   \vspace{-0.0in}
%\vspace{-0.2in}
   \centering
   \includegraphics[width=0.45\textwidth]{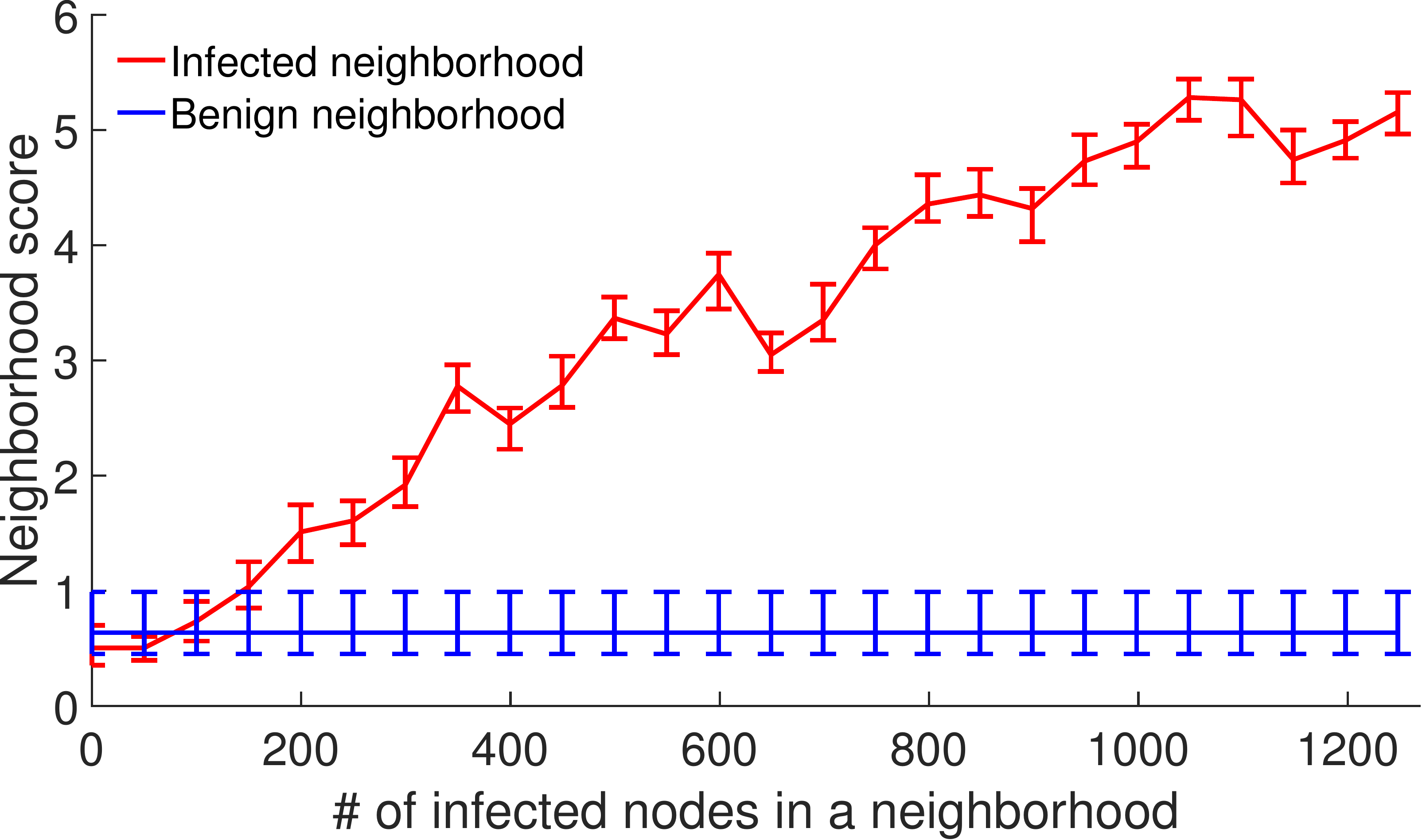}\par
   \caption{{\em (Waterhole attack: Time-based neighborhood filtering) 
     Dynamics of an attack:
     While the portion of infected nodes in a neighborhood increases
     over time reaching 1248 nodes on average, ShapeScore goes up
     showing that \sysname becomes more confident in labeling
     neighborhoods as `malicious'.  It starts detecting malware with
     at most 1\% false positive rate when roughly 200 nodes get compromised. 
     The neighborhood includes 17,178 nodes on average and
     spans over 30 sec time interval.}}
   \vspace{-0.2in}
\label{fig:windows-shape_vs_count-waterhole}
\end{figure}

Temporal filtering creates a neighborhood using only the nodes that are {\em
active} within a neighborhood time window (NTW). For example, a temporal
neighborhood for the waterhole attack scenario
would include all client devices that accessed {\em any} server within the last 
NTW into one neighborhood ($\sim17,000$ nodes on average in 30 seconds). 
This neighborhood filtering models a CIDS
designed to detect malware 
%Such NF encodes the fact that the 
whose infection exhibits temporal locality (and obviously does not detect
attacks that target a few high-value nodes through temporally uncorrelated
vectors). 

%Temporally correlated nodes are likely to exhibit coherent behaviors.

%In the phishing experiment, a neighborhood includes all the 1086 nodes in a
%simulated network that exchange emails within a 1-hour neighborhood time
%window (NTW).  Each neighborhood in the waterhole scenario 
%comprises of the client machines that access
%top 50 servers from Yahoo DAX dataset within 30-sec neighborhood window.
%\todo{clean this para up based on expt. setup.}

Interestingly, waterhole attacks exhibit 'bursty' nature: in our experiments,
a popular waterhole server quickly infects a large number of clients within a
short period of time -- thus, we vary the waterhole NTW from 4 seconds up to 100 seconds.

\noindent{\bf Shape GD's time to detection for one NTW.} 
We fix NTWs (30 seconds) and vary
a parameter that represents a node's likelihood of infection from 0\% up to 100\% -- 
modeling whether a drive-by exploit succeeeds in a waterhole attack.   

Figure~\ref{fig:windows-shape_vs_count-waterhole}  
plots the neighborhood score v. the average number of infected
nodes within benign (blue curve) and malicious (red curve) neighborhoods 
-- the two extreme points on the X-axis corresponds
to either none of the machines being infected (the left side of a figure) or
the maximum possible number of machines being infected (the right side of the figure).  
In this experiment, the waterhole server can infect at most 1250 nodes 
in the 30 seconds NTW.
Every point on a line is the median neighborhood score from 100 experiments with
whiskers set at 1\%- and 99\%- percentile scores. In each experiment we
use a random subset of training data for training purposes 
and a random subset of testing data for testing.

%Intermediate values correspond to partially infected neighborhoods.
%For each parameter's value we compute an average number of infected nodes in a
%neighborhood, which differs by an order of magnitude in phishing and waterhole
%simulations (Figure \ref{fig:windows-shape_vs_count},
%\ref{fig:windows-shape_vs_count-waterhole}).

%The intermediate values of the parameter `click rate' correspond to the
%scenario when malware infects only some machines.
%Figures~\ref{fig:windows-shape_vs_count}
%and~\ref{fig:windows-shape_vs_count-waterhole} shows the neighborhood scores at
%varying degrees of infection (red curve) compared to the score for benign
%neighborhoods.
%compromised nodes in a neighborhood degree of anomaly.  The blue curve shows
%distribution of neighborhood scores for completely benign neighborhoods, while
%the red curve corresponds to the scores assigned to partially infected
%neighborhoods.  

When increasing the number of infected nodes in a neighborhood, as expected,
the red curve larger deviates from the blue one.  Therefore, \sysname becomes
more confident with labeling incoming partially infected neighborhoods as
malicious.  \sysname starts reliably detecting malware very quickly -- 
when only 200 nodes have been infected.  
We also experimented with
other sizes of neighborhood window -- the plots we obtained showed similar
trends.
\ignore{min-max values for reliable detectio?}

\begin{figure}[tbp]
   \vspace{-0.0in}
   \centering
   \includegraphics[width=0.45\textwidth]{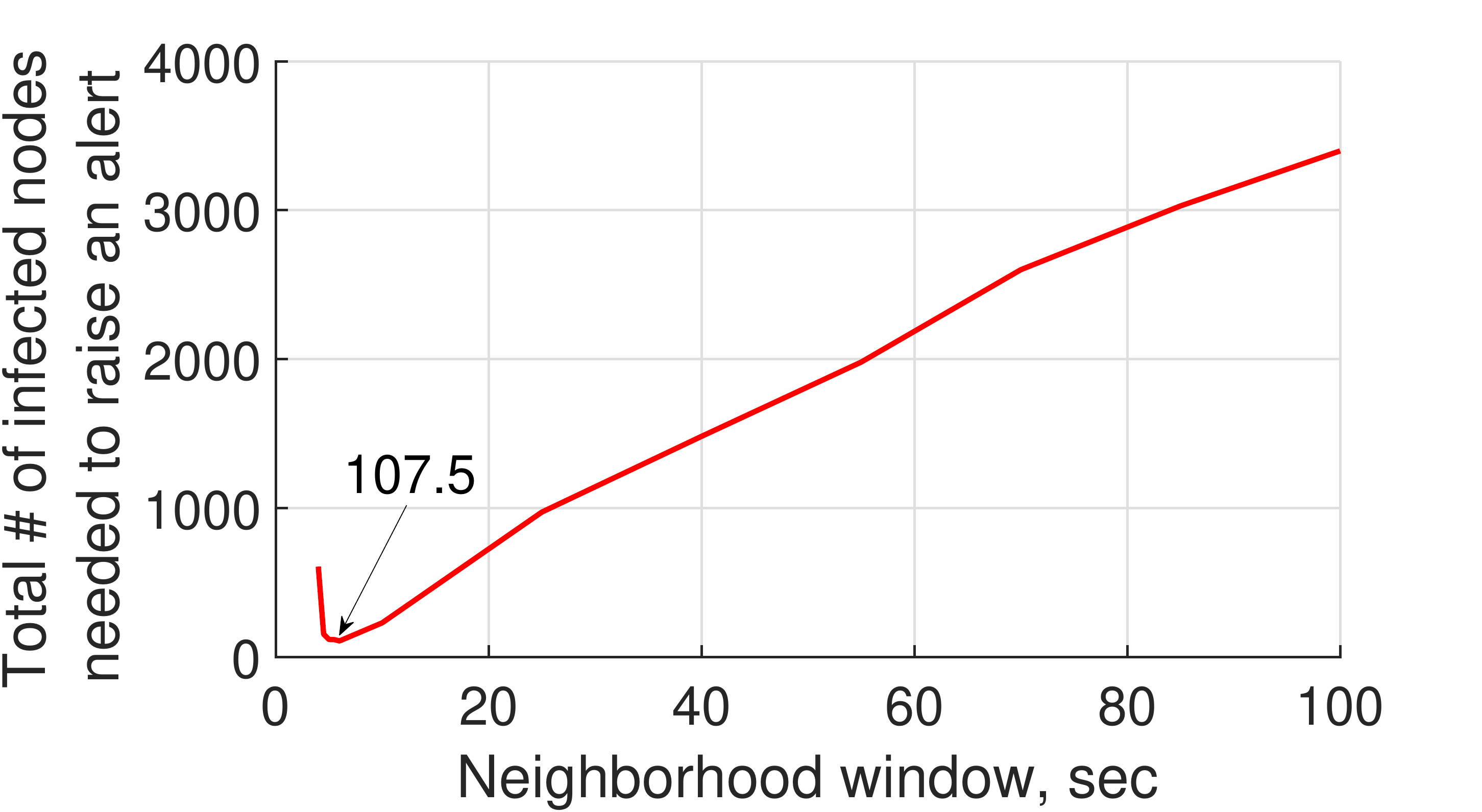}\par
\caption{{\em (Waterhole attack: Time-based neighborhood filtering) 
Shape GD's performance deteriorates linearly when increasing the size of a neighborhood window from 6 sec to 100 sec.
}}
   \vspace{-0.2in}
\label{fig:waterhole_time_NF}
\end{figure}

\noindent{\bf Shape GD's sensitivity to NTW.} We show that the size of a
neighborhood is important for early detection -- the minimum number of
nodes that are infected before \sysname raises an alert -- in
Figure~\ref{fig:waterhole_time_NF}.
Varying the NTW essentially competes the rates at which both malicious and
benign FVs accumulate.

We vary the NTW from from 4 sec to
100 sec and record the number of infected nodes when \sysname can
make robust predictions (i.e. less than 1\% FP for almost 100\% TP).
The results are averaged across 100 experiments.

In a waterhole scenario, the number of client devices active within a time
window (and hence the false positive alert-FVs from the
neighborhood) grows much faster than the malware can spread 
(even if we assume 
that {\em every} client that visits the waterhole server gets infected.
%, i.e. an infection rate of 100\%).  
Here, a large NTW aggregates many more
benign (false positive) FVs from clients accessing non-compromised servers.
Hence, increasing the NTW degrades time to detection. 
\sysname works best with an
NTW of 6 seconds -- only 107.5 nodes on average become infected out of a
possible $\sim$550,000 nodes.
%that would be infected if malware propagation was not detected at an early
%stage.  
Note that a very small NTW (below 6 seconds) 
either does not accumulate enough FVs for analysis --
if so, \sysname outputs no results --  or 
creates large variance in the shape of benign
neighborhoods and abruptly degrades detection performance.
%a portion of 
%malicious FVs is not enough for the detector to
%label the NTW as malicious.
%Very small NTWs (below 6 seconds) 

%In comparison to phishing where neighborhood window can be fixed in advance,
Note that \sysname requires a minimum number of FVs per neighborhood to make
robust predictions -- at least 15,000 FVs %based on
%Appendix~\ref{sec:nbd-size} 
-- hence, the \sysname has to set 
%designed to detect waterhole attacks can
NTWs based on the rate of incoming requests and access
frequency of a particular server. For example, if a server is not very popular
and is likely to be compromised, the \sysname could increase this server's NTW 
to collect more FVs for its neighborhood.

%\begin{figure}[tbp]
%   \vspace{-0.0in}
%%\vspace{-0.2in}
%   \centering
%   \includegraphics[width=0.45\textwidth]{figs/windows/enron/time-based-NF.eps}\par
%\caption{{\em(Phishing attack: Time-based NF algorithm)  Shape GD's performance improves by 18.5\% (20.24 and 17.08 infected nodes) when increasing the size of a neighborhood window from 1 hour to 3 hours.
%}}
%   \vspace{-0.2in}
%\label{fig:windows-time-NF-phishing}
%\end{figure}

\subsection{Fragility of Count GD}
\label{sec:eval-count}

\begin{figure}[tbp]
\vspace{-0.0in}
   \centering
   \includegraphics[width=0.45\textwidth]{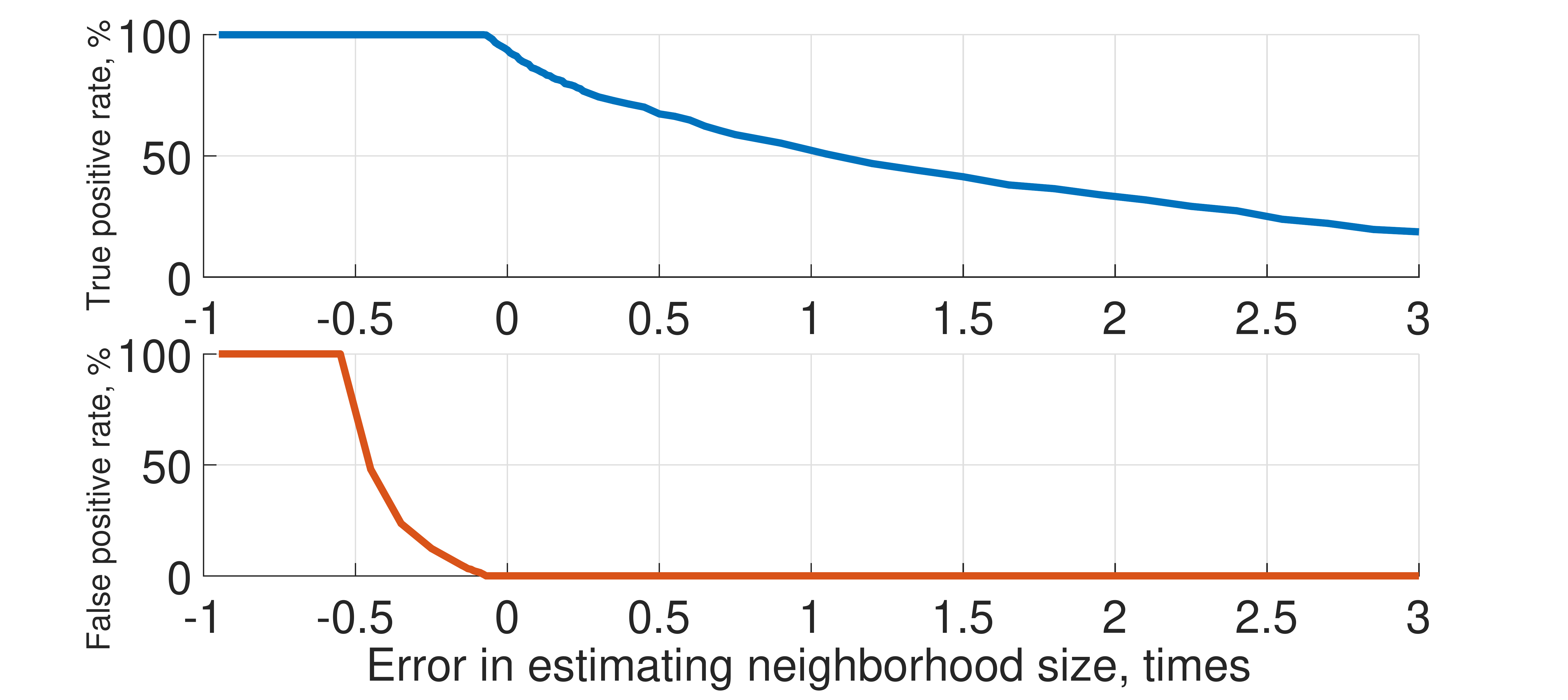}\par
\caption{{\em(Symnatec Wine dataset) An error in estimating neighborhood size dramatically affects Count GD's performance. 
It can tolerate at most 30\% underestimation errors and 1\% overestimation errors to achieve comparable with Shape GD performance.
}}
   \vspace{-0.2in}
\label{fig:symantec-noisy_count}
\end{figure}

\begin{figure}[tbp]
\vspace{-0.0in}
   \centering
   \includegraphics[width=0.45\textwidth]{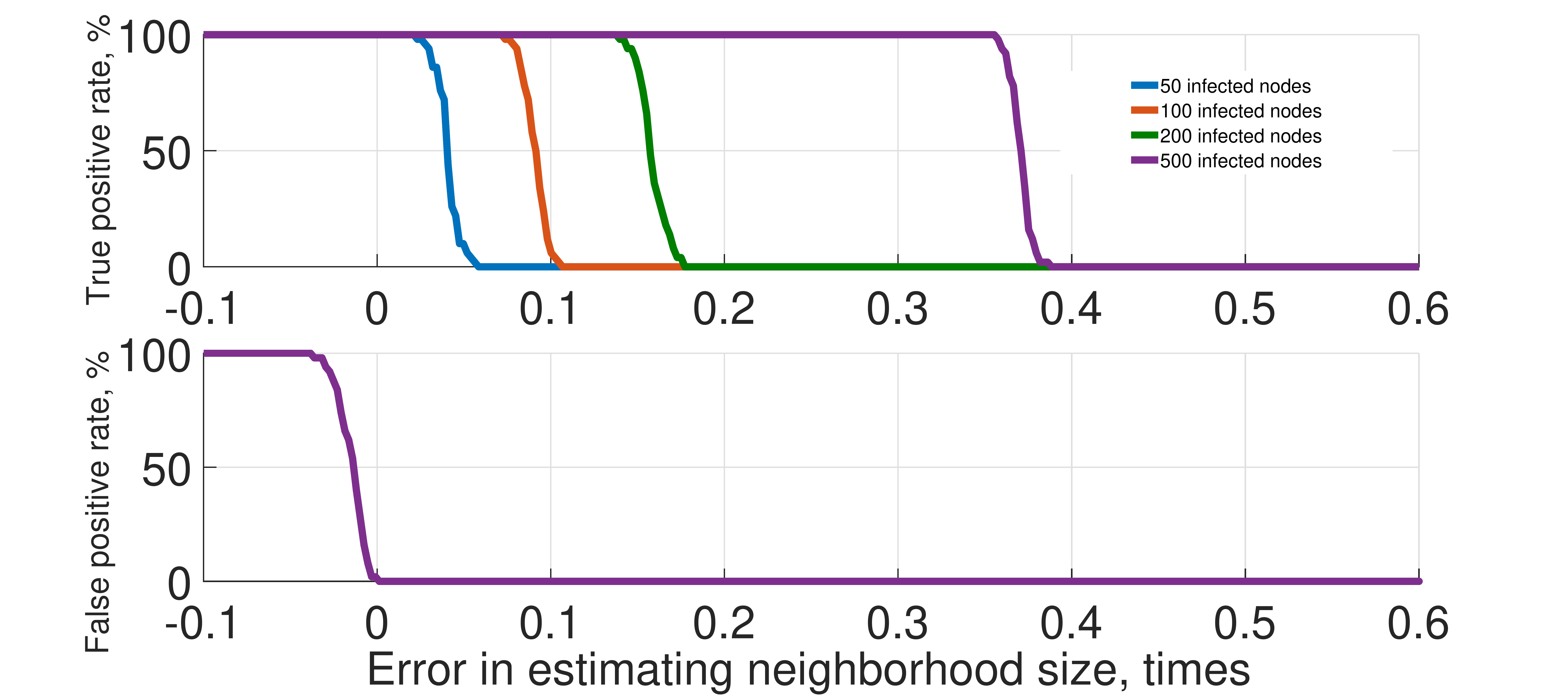}\par
\caption{{\em(Waterhole attack) An error in estimating neighborhood size dramatically affects Count GD's performance. 
It can tolerate at most 0.1\% underestimation errors and 13.8\% overestimation errors to achieve comparable with Shape GD performance.
}}
   \vspace{-0.2in}
\label{fig:windows-noisy_count-waterhole}
\end{figure}

A Count-GD algorithm counts the number of alerts over a neighborhood
and compares it to a threshold to detect malware. 
This threshold scales linearly in the size of the neighborhood -- 
we now experimentally quantify the error Count-GD can tolerate in
Symantec Wine (Figure~\ref{fig:symantec-noisy_count}) and
waterhole (Figure~\ref{fig:windows-noisy_count-waterhole})  settings.  
Note that the
error in estimating neighborhood size can be double sided --
underestimates (negative error) can make neighborhoods look like alert
hotspots and lead to false positives, while overestimates (positive
error) can lead to missed detections (i.e., lower true positives).

We run Count-GD in the same setting as \sysname.
In the Symantec Wine case study we adjust Count-GD's threshold to match 
the performance of \sysname{}'s Neighborhood classifier (true positive rate of 95.41\%, Section~\ref{sec:symantec_local_detectors_results})
with zero neighborhood estimation errors (Figure~\ref{fig:symantec-noisy_count}).
%In the Symantec Wine case study we adjust Count-GD's threshold to 
%yield 0\% false positive rate and higher than 90\% true positive rate
%%match 
%%the performance of \sysname{}'s Neighborhood detector (true positive rate of 84.03\%, Section~\ref{})
%with zero neighborhood estimation errors (Figure~\ref{fig:symantec-noisy_count}).
In the waterhole case study we evaluate Count-GD under the same conditions as 
\sysname when presenting the results of time-based neighborhood filtering
(Section \ref{sec:time-nf}) -- 30-sec long neighborhood including 17,178 nodes
(Figure \ref{fig:windows-noisy_count-waterhole}).
In comparison to the Symantec Wine experiment, whose parameters are fixed,
in the waterhole experiment we vary infection probability such that the
number of infected nodes in a neighborhood changes
from 0 to 500 (waterhole) in four increments -- note that
only a small fraction (2.9\%) of nodes per neighborhood get infected
in the worst case.

In this setting, recall that the neighborhood detector has a maximum global false
positive rate of 19.03\% and 1\% and a true positive rate of 95.41\% and 100\%
respectively in the Symantec Wine and waterhole case studies.
To maintain a similar detection performance,
our experiments show that the Count-GD can only
tolerate neighborhood size estimation errors within a very narrow
range -- [-30\%, 1\%] (Symantec Wine) and [-0.1\%, 13.8\%]
(waterhole).
A key takeaway here is that underestimating a neighborhood's size
makes Count-GD extremely fragile (-30\% in Symantec Wine and -0.1\% for
waterhole). On the other hand, overestimating neighborhood sizes
decreases true positives, and this effect is catastrophic.

We comment that this effect is important in practice. 
In the example of a Fortune-500 company, we observed 
that commercial SIEM tools often
do not report alerts in a timely manner and may delay delivering alerts 
by up to 2 months due to unpredictable infrastructure failures and 
due to a local IT service intervening into the analysis of alerts.
Also given the practical deployments where nodes get infected out of band (e.g.,
outside the corporate network), go out of range (with mobile devices), the
tight margins on errors can render Count-GD extremely unreliable. 
Even with sophisticated size estimation algorithms, recall that the
underlying distributions that create these neighborhoods 
(e.g. number of clients per server) have sub-exponential heavy
tails -- such distributions typically result in poor parameter
estimates due to lack of higher moments, and thus, poorer statistical
concentrations of estimates about the true value~\cite{fkz11}.
Circling back, we see that by eliminating this size dependence compared to
Count-GD, our \sysname provides a robust inference algorithm.

%% file: discussion_main_text.tex
\noindent \textbf{Evasion attacks.}
\sysname requires a human analyst to correctly specify attack vectors. 
If a new attack vector emerges (ex. {\tt badUSB}), then the corresponding 
attack may go undetected. 
However, attack vectors such as URLs or emails or physical devices 
along which malware propagates 
are far fewer than vulnerabilities, exploits, or malware samples.
%Thus, the majority of attackers keeps exploiting well-known 
%attack vectors such as external URLs, email attachments and etc. 
Further, individual local detectors may be susceptible to evasion attacks, which may
negatively affect \sysname's detection. 
However, designing evasion resistant local detectors~\cite{taming_evasions_in_ml} 
is outside the scope of this paper.

%% file: conclusion.tex
%\vspace{-0.1in}
Building robust behavioral detectors is a long-standing problem,
especially in large distributed systems where false positives can 
be overwhelming.  We observe
that attacks on enterprise networks induce low-dimensional neighborhoods
on otherwise high-dimensional feature vectors, but such neighborhoods
are unpredictable and thus hard to exploit.
%Reanalyzing the , \sysname
\sysname amplifies malware signal through neighborhoods 
and exploits their shape to identify infected ones early.
Automating the 
search for new neighborhoods, i.e. new attack vectors, 
that correlate with confirmed infections, 
would be a natural next step towards deployable behavioral
detectors.

%% file: scratchpad.tex
\section{Clustering Results}
\label{sec:eval-clustering}
\input{clustering_results}

\vspace{0.03in}
\section{\sysname Classifiers}
\label{app:local_detectors}
In both case studies local detectors (LDs) analyze executable files, 
however, they use different file abstractions -- static file analysis (Symantec Wine) 
because the original files are unavailable and
dynamic traces of executed system calls (waterhole case study).
%
%In the Symantec Wine case study we use an additional local detector -- domain name classifier, 
%which detects suspicious domains by analyzing their VirusTotal domain reports.
%
%In the Symantec Wine experiment LDs focus on static features of executable files because 
%the files are not available -- we have only an access to VirusTotal reports only that summarize them. Also we use an additional 
%local detector -- domain name classifier, which detects suspicious domains by analyzing
%their VirusTotal domain reports.
%And in the waterhole case study LDs extract features from 
%dynamic system call traces. 
%
%Even though particular feature extraction steps are dataset specific, both LDs use similar algorithms
%to spot suspicious feature vectors -- XGBoost~\cite{xgboost} and Random Forest~\cite{}.
Both LD algorithms leverage the state-of-the-art techniques in
automated malware detection. % to generate an sequence of {\em alerts} from the feature vector (FV) sequence. 
Specifically, the LD algorithm  uses both its internal state and the {\em current} feature vector (FV)
to generate an alert if it thinks that this FV corresponds to
malware. %, and produces no alert if it thinks that the current FV is benign. 
%Henceforth, we define an {\bf alert-FV} to be an FV that
%generates an LD alert (either true or false positive). 

Despite performing case study specific feature extraction, LDs employ similar algorithms 
as a binary classifier for malware detection: 
Boosted Decision Trees~\cite{xgboost} (Symantec Wine case study) and
Random Forest (waterhole case study).
These algorithms achieve the best performance on the training data set 
among the classifiers from a prior survey~\cite{Canali2012} 
and scale up well to process millions of FVs.
%they have been shown to be robust to adversarial inputs~\cite{},

\noindent{\bf Symantec Wine.}
%In the Symantec Wine case study, we use two types of local detectors: 
%a file-level LD and a domain name classifier. Both of them work with VirusTotal reports that
%summarize the results of file analysis and domain analysis respectively.
%
%\noindent{\bf File-level LD.}
We adapt an LD from the prior work~\cite{vt_report_classification}.
It primarily relies on a lightweight file analysis, which scales well when processing millions of downloads per day.
Specifically, the LD extracts syntactic features from a file and applies a binary classification algorithm 
that labels a file as either malicious or benign.

The LD is designed to run existing commercial tools such as TRID, ClamAV, Symantec on a binary file, analyze statically imported libraries and functions, detect common packers, check whether a file is digitally signed or not, and collect its binary metadata. 
VirusTotal provides outputs of these tools as a single file-level report, which
we directly use as LD's input.
%
%\mikhail{REMOVE
%The core LD's component is a feature extraction algorithm, i.e. an algorithm converting textual VirusTotal reports into
%fixed-length feature vectors used as input by a classifier. Even though the original algorithm ~\cite{vt_report_classification}
%allows the classifier to achieve high accuracy, it is inappropriate in our case because
%it produces very high-dimensional feature
%vectors that have limited usability due to the "curse of dimensionality" 
%and due to high memory consumption.
%We use a common method for dimensionality reduction -- feature hashing~\cite{feature_hashing}. 
%Empirically we found that the output dimensionality of 1024 allows us to 
%achieve the best trade-off between LD's accuracy and resource consumption. 
%}
%Finally, the LD uses an xgboost classifier~\cite{xgboost}, which is trained using 10-fold cross-validation 
%and area-under-the-curve (AUC) as the classifier's objective to deal with class imbalance.
%It achieves 97.61\% AUC, and we chose the operating point of 5.0\% false
%positive rate and 90.47\% true positive rate.

\noindent{\bf Waterhole.} 
In the waterhole case study, the LD analyzes system calls executed by a program.
It transforms continuously evolving 390-dimensional time-series of Windows system calls into
a discrete-time sequence of feature-vectors (FVs). 
This is accomplished by chunking the continuous time series into $r-$second
intervals, and representing the system call trace over each interval as a
single $L-$dimensional vector. 
$L$ is typically a low dimension, reduced down from 390 using PCA analysis, to (in
our experiments) $L=10$ and $r = 1$ second.

\noindent{\bf Domain name classifier.}
In addition to file-level LDs, we employ a domain name classifier in the Symantec Wine case study 
to extract attributes to form neighborhoods.
The domain name classifier analyzes VirusTotal domain reports and identifies domains that are likely to distribute malicious files.
It uses as input VirusTotal domain-level reports that aggregate domain classification produced by other commercial tools such as
Dr. Web, Websense ThreatSeeker, and VirusTotal. % (Table~\ref{fig:domain_features}).
Each of those tools categorizes a domain based on its content. The number of categories ranges from 55 (Dr. Web) up to 451 (VirusTotal), and they include such classes as social networks, banking, ads, government and etc.

The domain-name classifier applies one-hot encoding schema to represent categorical data as a fixed length feature vectors.
Specifically, it creates a "zero" feature vector with the number of elements equal to the total number of categories (767-dimensional
feature vectors in our case) and sets "one" in the positions corresponding to the assigned categories, 
which are not necessarily mutually exclusive. 
%The classifier (XGBoost~\cite{xgboost}) 
%uses numeric fields as is (without additional encoding).
%The classifier also uses VirusTotal domain-related statistics such as the number of malware samples distributed 
%by a particular domain, the number of samples that refer to a domain and the number of malicious URLs belonging to a domain. 
%These numeric values lie in the range from 0 to 100 (they seem to be capped at 100 level). 
%The statistic is aggregated across the entire observation period, i.e. since the first time a domain name was submitted to VirusTotal service. 
%After extracting feature vectors from the raw data, we procced with training an xgboost binary classifier~\cite{xgboost} using 10-fold CV and AUC as an objective to address the class imbalance. In the training set each domain that has ever served at least one malware sample in the Wine dataset is labeled as malicious, otherwise it is considered as benign. 
%Domain name classifier achieves 83.80\% AUC and we choose the operating point of 19.03\% FP rate and 95.41\% TP rate.
%\begin{figure}[tbp]
%   \vspace{-0.0in}
%   \centering
%   \includegraphics[width=0.45\textwidth]{figs/symantec/domain_name_classifier/domain_features.pdf}\par
%   \caption{Domain name classifier's features.}
%   \vspace{-0.2in}
%\label{fig:domain_features}
%\end{figure}

%\mikhail{move somewhere}
\section{Vector-Histogram Implementation}
\label{vector_histogram_mplementation}
% implementation-level text
In the Symantec Wine case study, \sysname deals with two types of alert-FVs: file and domain alert-FVs.
Thus, it builds two separate vector histograms per neighborhood and then concatenates them into a single vector histogram.
The file-level vector histogram has dimensionality of 10x50, i.e. each file alert-FV is projected on 10-dimensional
basis and binned into 50 bins along each dimension.
Similarly, a domain vector-histogram has dimensionality of 100x5, i.e. each domain alert-FV is projected on 100-dimensional
basis and binned into 5 bins along each dimension.
Then, the algorithm concatenates two matrix-shaped vector-histograms. To do that, it represents them 
as two 500 dimensional vectors by using a row-major order and appends the second one to the first one, 
thus, the resulting vector has 1,000 dimensions.

In the waterhole experiment, \sysname projects alert-FVs on 10-dimensional basis, 
bins projections into 50 bins along each dimension. 
Thus, a vector-histogram has dimensionality of 10x50.
%and \sysname uses a specially designed ShapeScore function to detect malicious activity.

%To train the binary classifier in the Symantec Wine experiment, we can use plenty of benign and malicious files
%together with the neighborhood instantiation algorithm (Section~\ref{subsec:neighborhood_instantiation}) to
%generate benign/malicious neighborhoods given ground truth file-level data.
%In comparison to the binary classifier, training the threshold test 
%in the waterhole case study requires only benign data, which can be collected
%using test inputs on benign apps or using record-and-replay tools to re-run
%real user inputs in a malware-free system.  

%\section{Computing Shape GD's parameters} 
%\label{sec:deployment}
%\input{deployment}

\section{ShapeScore}
\label{sec:shape-score-function}
We developed the Wasserstein-based distance -- \textit{ShapeScore} function --
to detect neighborhoods with high malware concentration.
ShapeScore quantifies how much a current vector-histogram, $H_B$, differs from 
a {\em reference histogram}, $H_{{\rm ref}}$, which is generated during the training phase
using only the false positive FVs of the LDs 
by following the procedure for generating a vector histogram, 
which is outlined in Section~\ref{subsec:neighborhood_feature_extraction}.
ShapeScore is thus the distance of a neighborhood
from a benign reference histogram -- a high score indicates potential malware.

The ShapeScore of the accumulated set of FVs, $B$,
is given by the sum of the coordinate-wise Wasserstein distances
\cite{vallender1974calculation} between 
$$H_B =
(H_{B}(1) \ H_{B}(2) \ \ldots \ H_{B}(L))$$ 
and 
%$$\mathcal{H}_{{\rm ref}} =
%(\mathcal{H}_{{\rm ref}}(1) \ \mathcal{H}_{{\rm ref}}(2) \ \ldots \
%\mathcal{H}_{{\rm ref}}(L)).$$ 
$$H_{{\rm ref}} = (H_{{\rm ref}}(1) \ H_{{\rm ref}}(2) \ \ldots \
H_{{\rm ref}}(L)).$$ 
In other words,
%$$
%{\rm ShapeScore} = \sum_{l=1}^L d_W(\mathcal{H}_{\mathcal{B}}(l),\mathcal{H}_{{\rm ref}}(l)),
%$$
$$
{\rm ShapeScore} = \sum_{l=1}^L d_W(H_{B}(l),H_{{\rm ref}}(l)),
$$
where for two scalar distributions $p, q,$ the Wasserstein distance
\cite{vallender1974calculation,wasser-wiki} is given by
$$
d_W(p, q) = \sum_{i=1}^b \left |\sum_{j=1}^i \left(p(j) - q(j) \right) \right |.
$$

This Wasserstein distance serves as an efficiently computable one dimensional
projection, that gives us a discriminatively powerful metric of distance
\cite{vallender1974calculation,benbre00}. Because the Wasserstein distance
computes a metric between distributions -- for us, histograms normalized to
have total area equal to 1 -- it is invariant to the number of samples that
make up each histogram. Thus, unlike count-based algorithms, {\em it is robust
to estimation errors in community size}. %Figure~\ref{fig:windows-hist} verifies
%this intuition, and shows that true positives and false positive feature
%vectors separate well when viewed through the ShapeScore. 

Finally, to determine whether a neighborhood has malware present \sysname performs
hypothesis testing. If ShapeScore is greater than a threshold $\gamma$, \sysname
declares a global alert, i.e., the algorithm predicts that there is malware in
the neighborhood.  The robustness threshold $\gamma$ is computed
via standard confidence interval or cross-validation methods with multiple sets
of false-positive FVs.% (see Section \ref{sec:power-shape}).

\section{Symantec Wine: Real-Time Detection}
\label{app:real_detection_symantec}
Though we do not discuss real-time detection results in the main text,
we designed \sysname to act as a real-time detector. In this section we
do a deep dive into the real-time detection in terms of individual files and
compromised machines.
We compare \sysname against the local detectors~\cite{vt_report_classification} and
we also present neighborhood detector's results for completeness 
to better understand \sysname{}'s real-time detection.
Unfortunately, we have to exclude from the comparison the downloader detector~\cite{downloader_graphs}
because it is not designed to be a real-time detector and the authors did not share the source code with us.
We use the standard metrics for comparison: precision, recall, and F-1 score.

%Specifically, we focus our analysis on the following two characteristics -- how well
%\sysname detects malicious files and how well it detects infected machines, i.e. 
%machines that have downloaded at least one malicious file.
%
%To perform real-time analysis, we replay the 5-year long history of download events in the Wine dataset 
%(each event has a timestamp associated with it)
%and execute \sysname every 30 days. In our experiments we found that
%shorter period between consecutive runs of \sysname does not significantly affect
%results, it only improves time to detection and early detection parameters~\ref{sec:}. 
%We intentionally stick to a 30-day period between consecutive runs of \sysname{} 
%to keep execution time ($\sim$12 hours) and resource consumption manageable.
%
%We compare real-time behavior of three detectors -- local detectors~\cite{vt_report_classification}, 
%a neighborhood detector, which conservatively labels all the files within a malicious neighborhood as malicious, 
%and \sysname that comprises of the neighborhood classifier and local detectors 
%-- in terms of precision, recall, and F-1 score, which is a commonly used way to combine precision and recall.

\subsection{File-level real-time detection}

\begin{figure*}[t]
\begin{minipage}[tbp]{0.24\linewidth}
   \includegraphics[width=\textwidth]{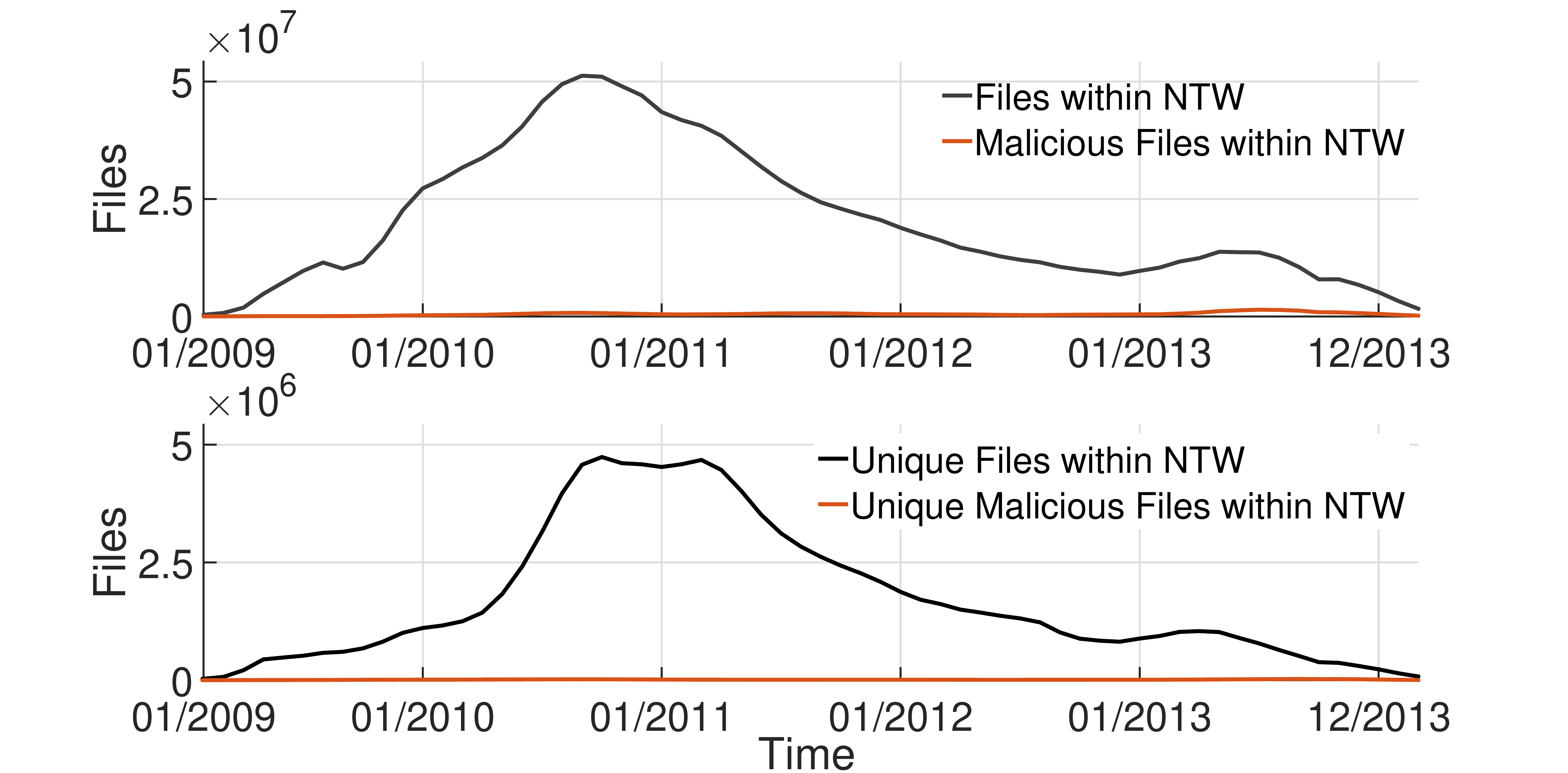}\par
   \label{fig:temporal-file-stat}
\end{minipage}
\begin{minipage}[tbp]{0.24\linewidth}
   \includegraphics[width=\textwidth]{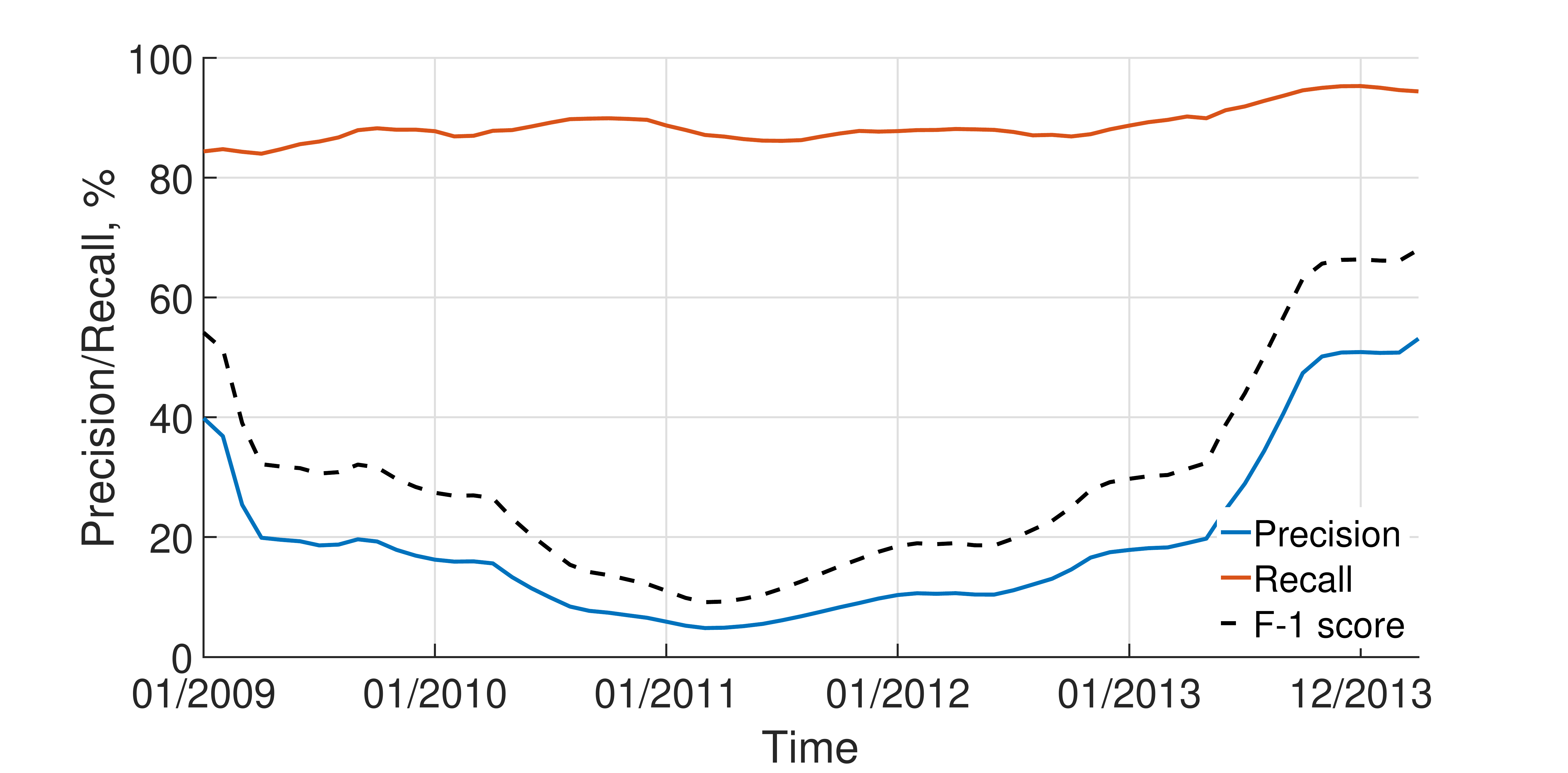}\par
   \label{fig:temporal-file-stat-LD}
\end{minipage}
\begin{minipage}[tbp]{0.24\linewidth}
   \includegraphics[width=\textwidth]{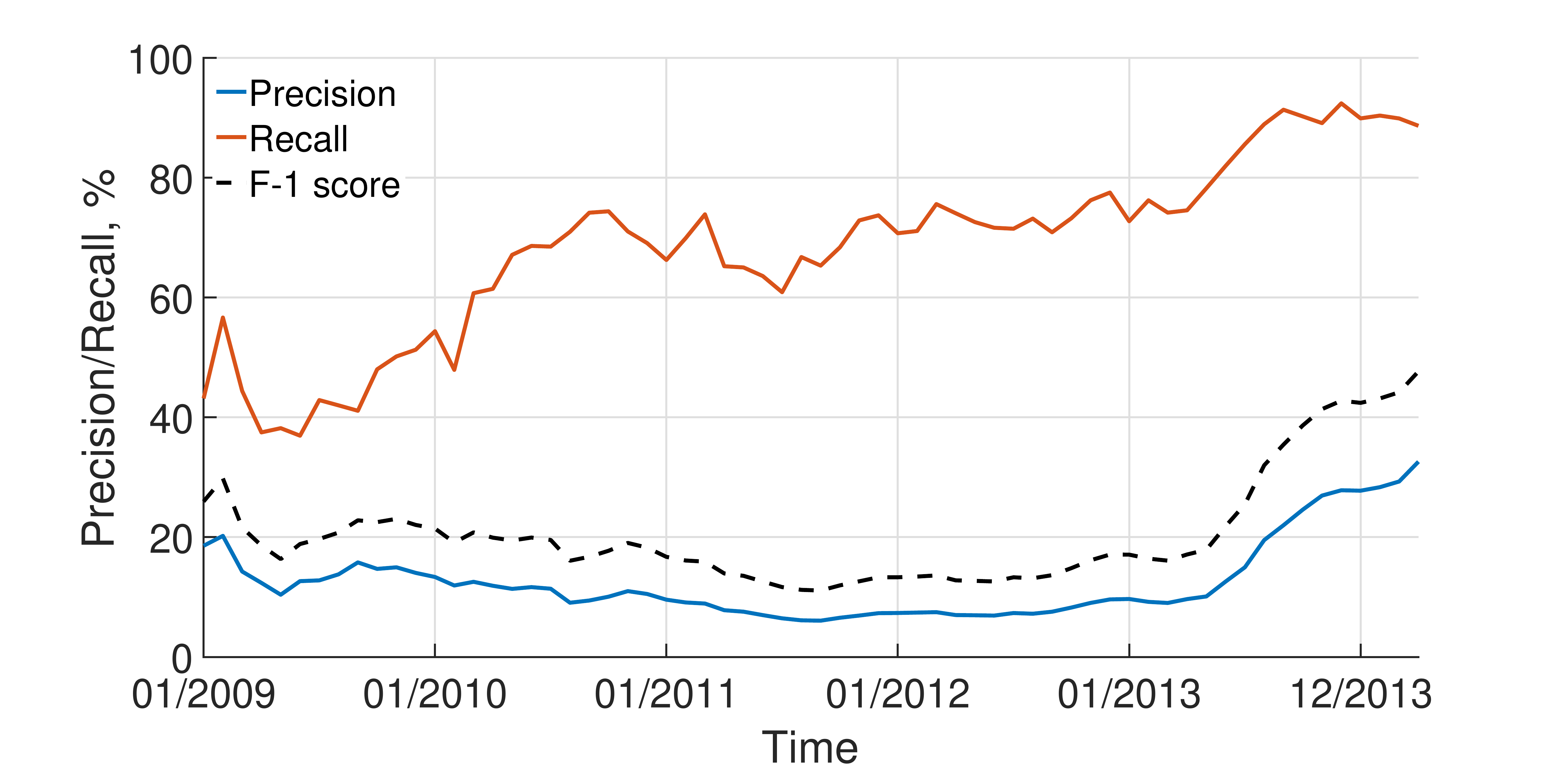}\par
   \label{fig:temporal-file-stat-NBD}
\end{minipage}
\begin{minipage}[tbp]{0.24\linewidth}
   \includegraphics[width=\textwidth]{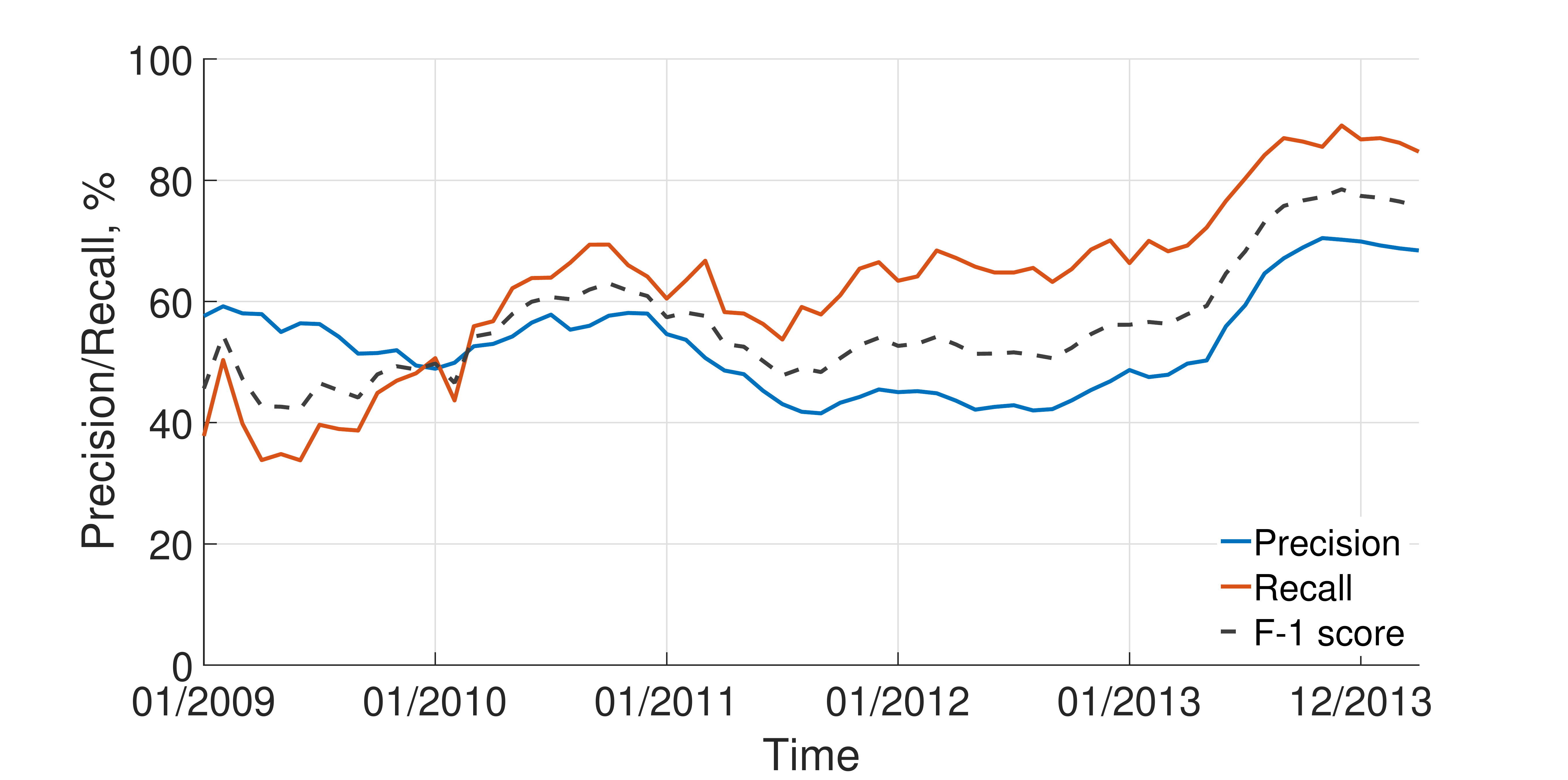}\par
   \label{fig:temporal-file-stat-Shape}
\end{minipage}
\caption{File-level dynamic behavior. Left to Right: Raw statistics, LD-level stat, NBD-level stat, Shape-LD - level stat}
\label{fig:temporal-file-stat-minipage}
\end{figure*}

%\textbf{File statistics.}
We start with the analysis of the temporal distribution 
of the download events (Figure~\ref{fig:temporal-file-stat}) to visualize file downloads over time.
Every time \sysname runs, it analyzes download events within a neighborhood time window (NTW), 
which is set to 150 days in our experiments.
Therefore, we represent the intensity of downloads over time as the number of downloads within each NTW.
Specifically, for each timestamp we compute the total number and the number of malicious file downloads 
within the previous NTW (the upper Figure~\ref{fig:temporal-file-stat}).
For example, the value on the Figure~\ref{fig:temporal-file-stat} labeled as 01/2011 includes file
downloads from 06/2010 till 01/2011.
We also visualize the total number of distinct downloads and
the number of distinct malicious downloads (the lower Figure~\ref{fig:temporal-file-stat}).

Every point on these curves characterizes the number of files \sysname has to deal with
when operating in a real-time detection mode. 
The large gap between the black and the red curves shows that only a small percentage of files
in the Symantec Wine dataset is malicious.
\sysname manages to filter out most benign files from further analysis to reduce the overall false positive rate.

When taking a deeper look at the plots, we notice that file downloads in the 
Wine dataset exhibit a nonuniform pattern over time. 
The total number of downloads increases from January 2008 and reaches its peak (51 million downloads per NTW)
within the NTW ending in October 2010, and after that it decreases over time.
The temporal pattern of distinct downloads slightly differs -- 
intensity of distinct downloads reaches a flat plato (4.74 million per NTW) in September 2010 and
remains on the approximately same level until April 2011.
However, malicious files are responsible for only
the small percentage of all downloads -- at most 1.43 million total malicious downloads and at most 27 thousand
unique malicious downloads.

Note that the low intensive ends of the distribution impose an obstacle for \sysname 
because of the insufficient number of correlated file downloads.
Due to this reason we discard file downloads before June 2008.
Therefore we run \sysname the first time on the neighborhood window spanning 
the interval from 06/2008 until 01/2009 and label the results with the `01/2009' timestamp.

\noindent{\bf Local detectors.}
When we analyze the temporal behavior of local detectors (Figure~\ref{fig:temporal-file-stat-LD}), we
notice anti-correlation between the total number of unique downloads and LDs' precision.
The peak of unique downloads corresponds to the large number of benign downloads.
Therefore, when LDs process them, they output a large number of false positive alerts, 
which results in a precision drop (it drops down to less than 5\% level). 
However, the recall stays in the range of 84\% -- 95\% 
because it depends only on LDs' ability to detect malicious files. 
F-1 score leans more towards precision than to recall, 
that is why LDs have mostly low F-1 score over the large period of time (between 9\% and 66\%).

\noindent{\bf Neighborhood detector.}
Before analyzing \sysname real-time detection, we briefly discuss the neighborhood detector's detection performance.
We assume that the neighborhood detector (Figure~\ref{fig:temporal-file-stat-NBD}) labels all the files within malicious
neighborhoods as malicious.
As local detectors, the neighborhood detector suffers from low precision as well, however,
the underlying cause is different. The neighborhood classifier is supposed to label
neighborhoods malicious if they contain more than 5\% of malicious files. Usually, most files
in a neighborhood are benign. Thus, when the neighborhood detector 
conservatively labels all the files malicious, it suffers from high false positive rate,
consequently, low precision.
Hence, the neighborhood detector is designed to be conservative.
Also the neighborhood detector inadvertently filters out some malicious files, which leads 
to lower than LDs' recall.

\noindent{\bf \sysname{}.}
Comparing to local detectors, \sysname boosts precision and inherits slightly 
lower recall from the neighborhood classifier because it aggregates LDs' predictions 
collected only across suspicious neighborhoods (Figure~\ref{fig:temporal-file-stat-Shape}). 
The reason why \sysname achieves high precision is because it has much lower false positive rate 
as many benign files are already filtered out by
the neighborhood detector. Thus LDs running within suspicious neighborhoods analyze fewer
benign files than LDs in the traditional deployment scenario.
At the same time, \sysname has slightly lower recall than both LDs and the neighborhood detector
because \sysname labels 
a file as malicious only if it is contained within a suspicious neighborhood and a local detector
raises a file-level alert. However, the neighborhood and domain name classifiers are imperfect --
they may fail to correctly label malicious neighborhoods and domains respectively. 
Therefore, \sysname does not aggregate LDs' output across
all malicious files, which results in a slightly lower recall. 
\sysname{}'s F-1 score is bounded by close values of precision and recall and it is much higher
than the analogous parameter of local detectors and the neighborhood detector.

To quantitatively compare \sysname with local detectors we compute the area under F-1 curve. 
In the case of file
detection, \sysname achieves 96.6\% higher area under F-1 curve than the local detector.
\sysname{}'s F-1 score is bounded by close values of precision and recall and
it is much higher than the analogous parameter of local detectors and the
neighborhood detector.

\noindent{\bf Machine-level real-time detection.}
\begin{figure*}[t]
\begin{minipage}[tbp]{0.24\linewidth}
   \includegraphics[width=\textwidth]{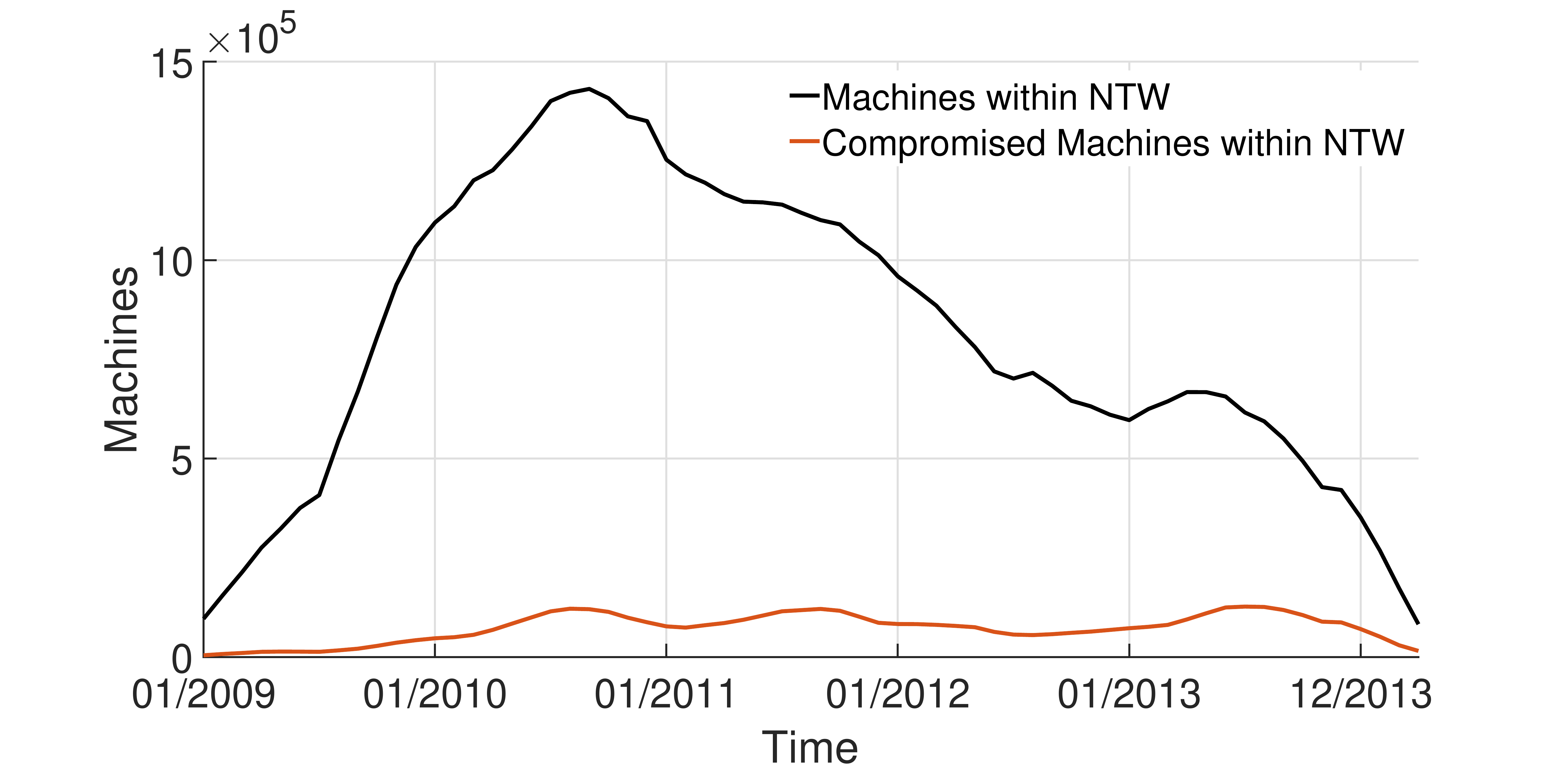}\par
   \label{fig:temporal-machine-stat}
\end{minipage}
\begin{minipage}[tbp]{0.24\linewidth}
   \includegraphics[width=\textwidth]{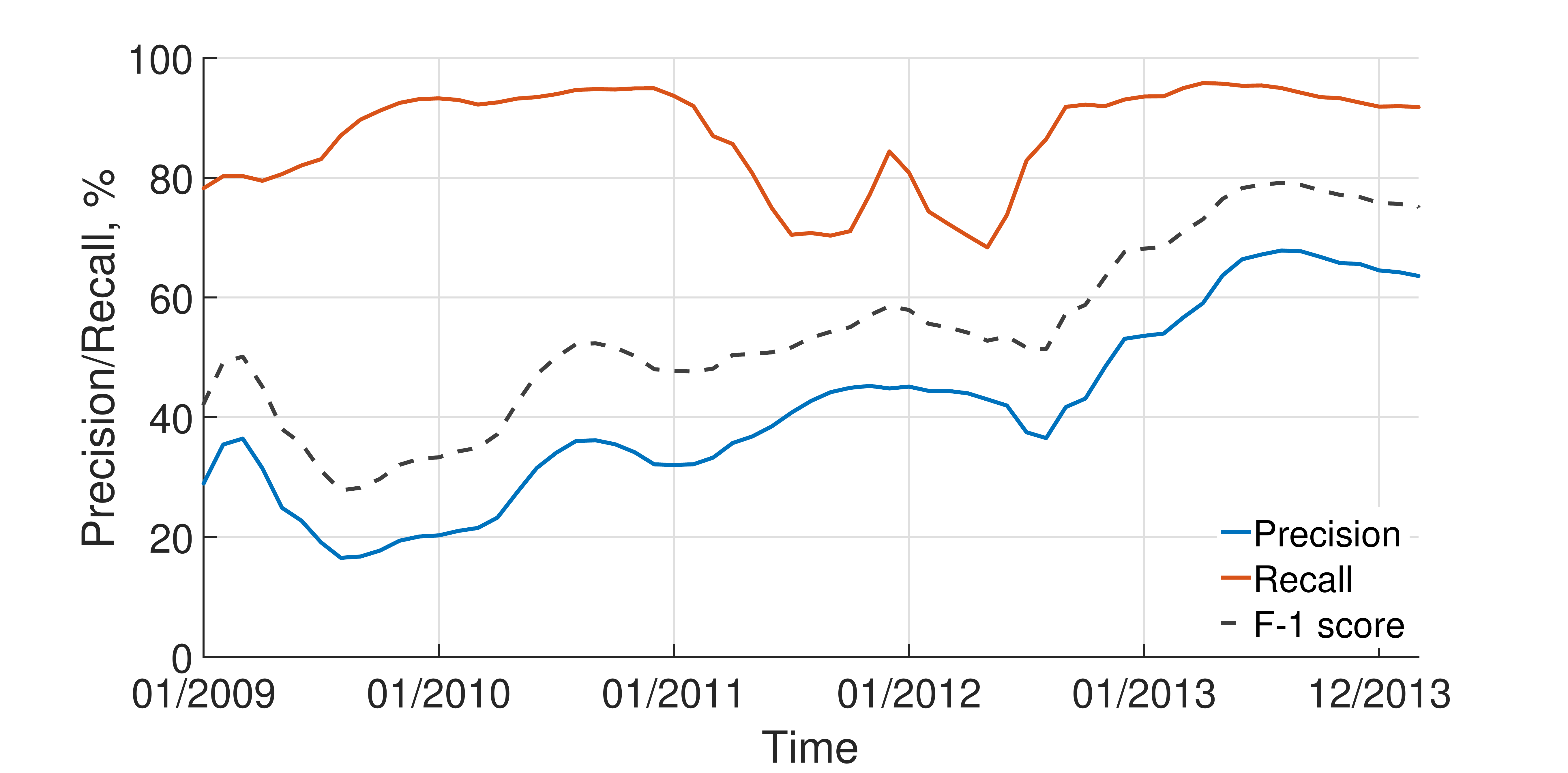}\par
   \label{fig:temporal-machine-stat-LD}
\end{minipage}
\begin{minipage}[tbp]{0.24\linewidth}
   \includegraphics[width=\textwidth]{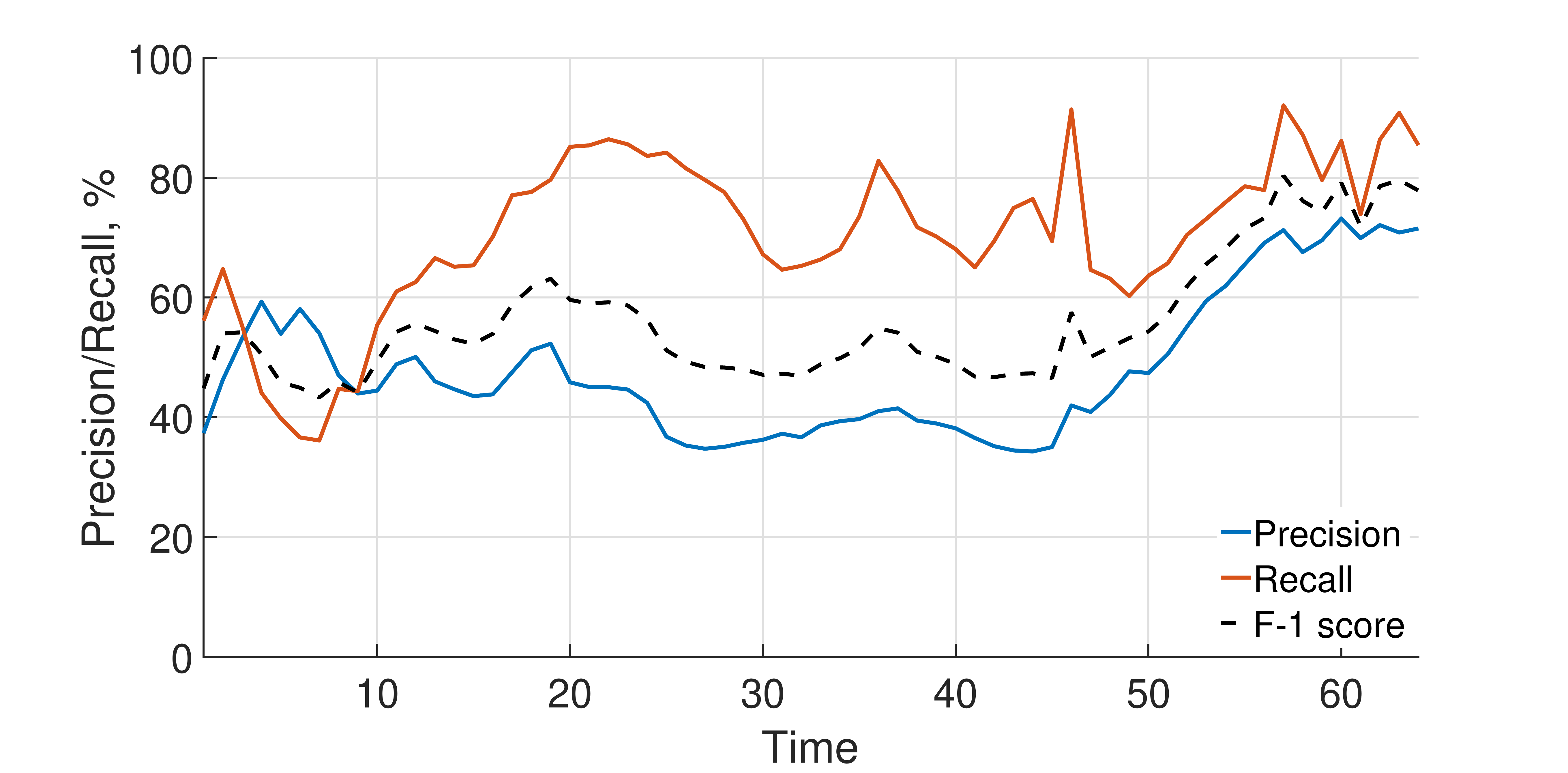}\par
   \label{fig:temporal-machine-stat-NBD}
\end{minipage}
\begin{minipage}[tbp]{0.24\linewidth}
   \includegraphics[width=\textwidth]{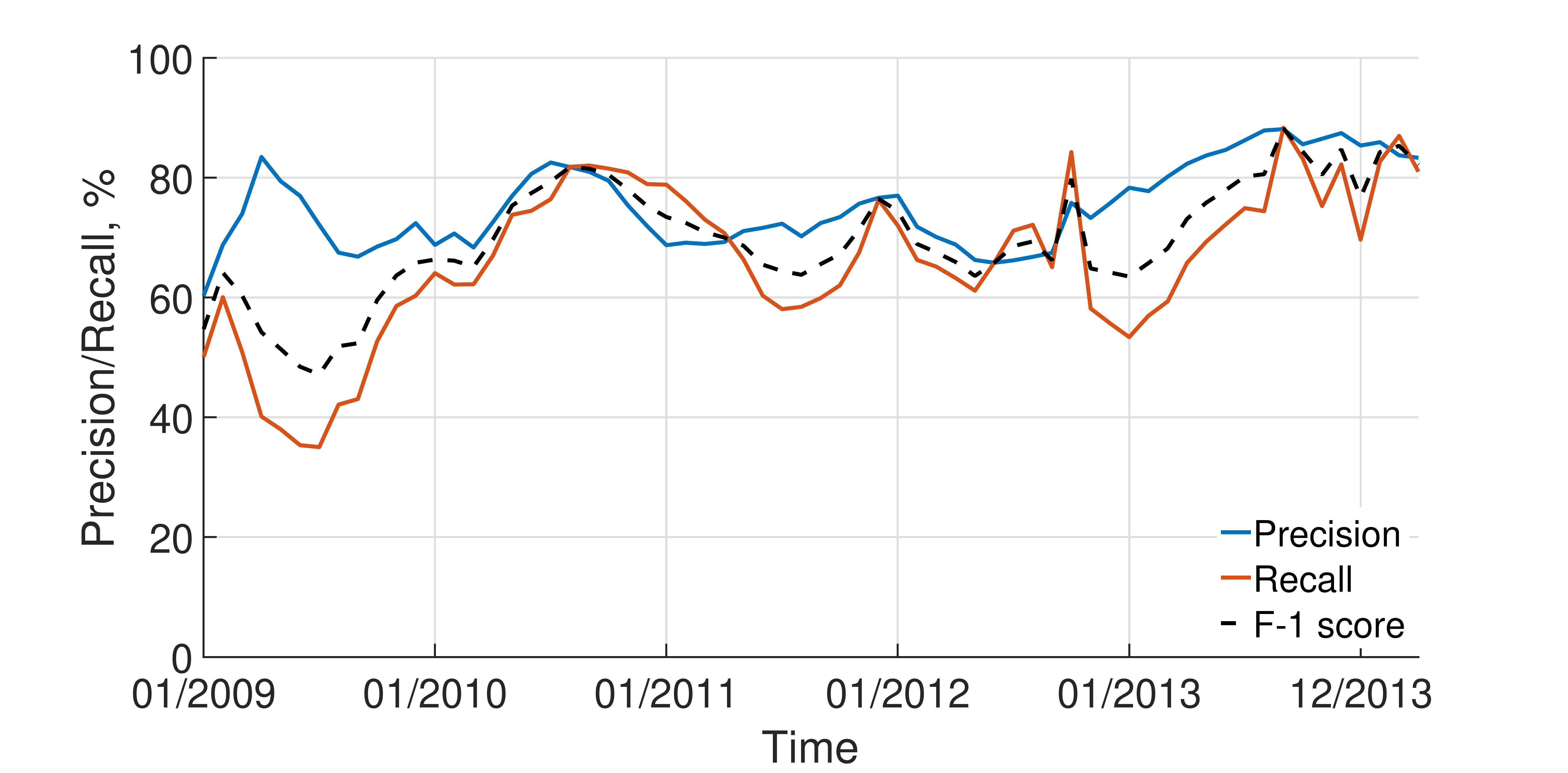}\par
   \label{fig:temporal-machine-stat-Shape}
\end{minipage}
\caption{Machine-level dynamic behavior. Left to Right: Raw statistics, LD-level stat, NBD-level stat, Shape-LD - level stat}
\label{fig:temporal-machine-stat-minipage}
\end{figure*}
%
%\textbf{Machine statistics.}
Machine-level statistics (Figure~\ref{fig:temporal-machine-stat}) is similar to the file-level statistics --
only a small percentage of machines is compromised within each NTW window.
The number of machines and compromised machines reach their peak values of 1.43 million and 126.9 thousand
respectively in October 2010, i.e. less than 8.9\% of compromised machines at the peak.

Overall, we observe higher values of precision and recall for all detectors (Figure~\ref{fig:temporal-machine-stat-minipage}) 
because, when interpreting detection results at the machine level, the detectors do not have to be very precise --
 they need to detect at least one malicious file on a machine, and 
file-level false positives on a particular machine do not count if that machine is infected.

Similar to file-level detection results, local detectors suffer from low precision because of the high
number of false positives. However, precision is significantly higher 
-- its mean value reaches 41\% as opposed to the mean value of 19\% for the file-level detection. 
Such dramatic difference is attributed to file-level false positives on compromised machines 
not affecting detectors' precision at a machine-level.
In both cases, recall curve exhibits similar behaviors.

We observe a similar trend for the neighborhood detector -- the mean precision value is 48\% versus
12.5\% in the case of file-level detection. The recall value remains
in the range of 36\% -- 92\%.
Finally, \sysname brings the precision curve up at the cost of slightly lowering the recall -- this is
exactly the same effect that we see in the case of file-level malware detection.

%reduces the FP rate even further and as a result the precision curve gets shifted 
%upward, but it looses some true positives. Interestingly, \sysname brings two curves --
%precision and recall -- together and achieves higher F-1 score than other detectors.
%\mikhail{add area under F-1 curve.}

Overall, \sysname achieves better results at the machine-level than at the file-level, which means that it can
identify infected machines earlier and more robustly than individual malware samples.
In the case of real-time detection, the main \sysname{}'s competitor is a local detector.
However, \sysname{}'s area under F-1 curve is 28.6\% higher than the analogous parameter for the local detector.

\section{Aggregate Machine-level Detection}
\label{agg_machine_detection}
\begin{figure}[tbp]
   \vspace{-0.0in}
   \centering
   \includegraphics[width=0.45\textwidth]{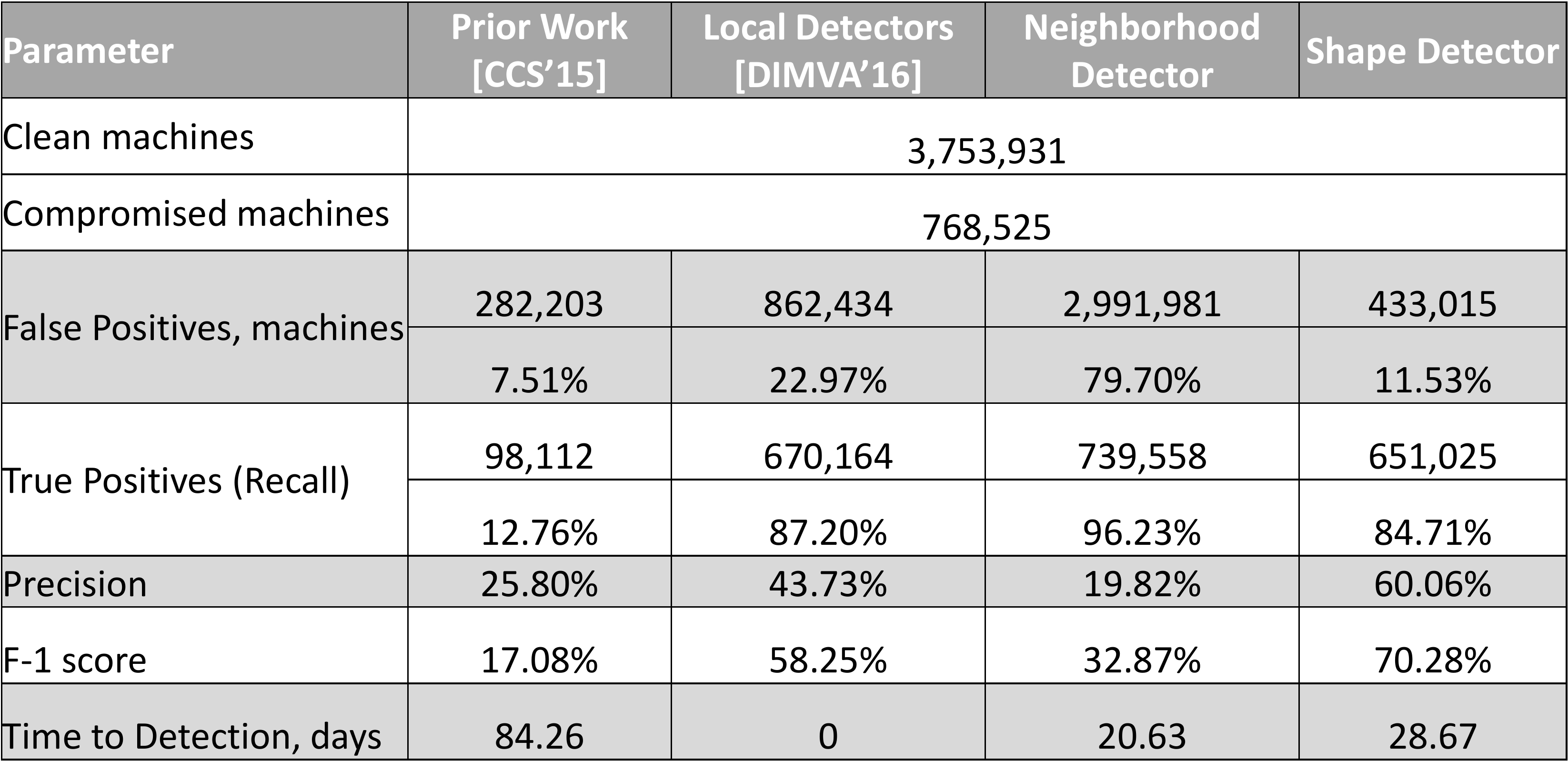}\par
   \caption{Machine-level aggregate analysis.}
   \vspace{-0.2in}
\label{fig:machine_agg_analysis}
\end{figure}
\begin{figure}[tbp]
   \vspace{-0.0in}
   \centering
   \includegraphics[width=0.45\textwidth]{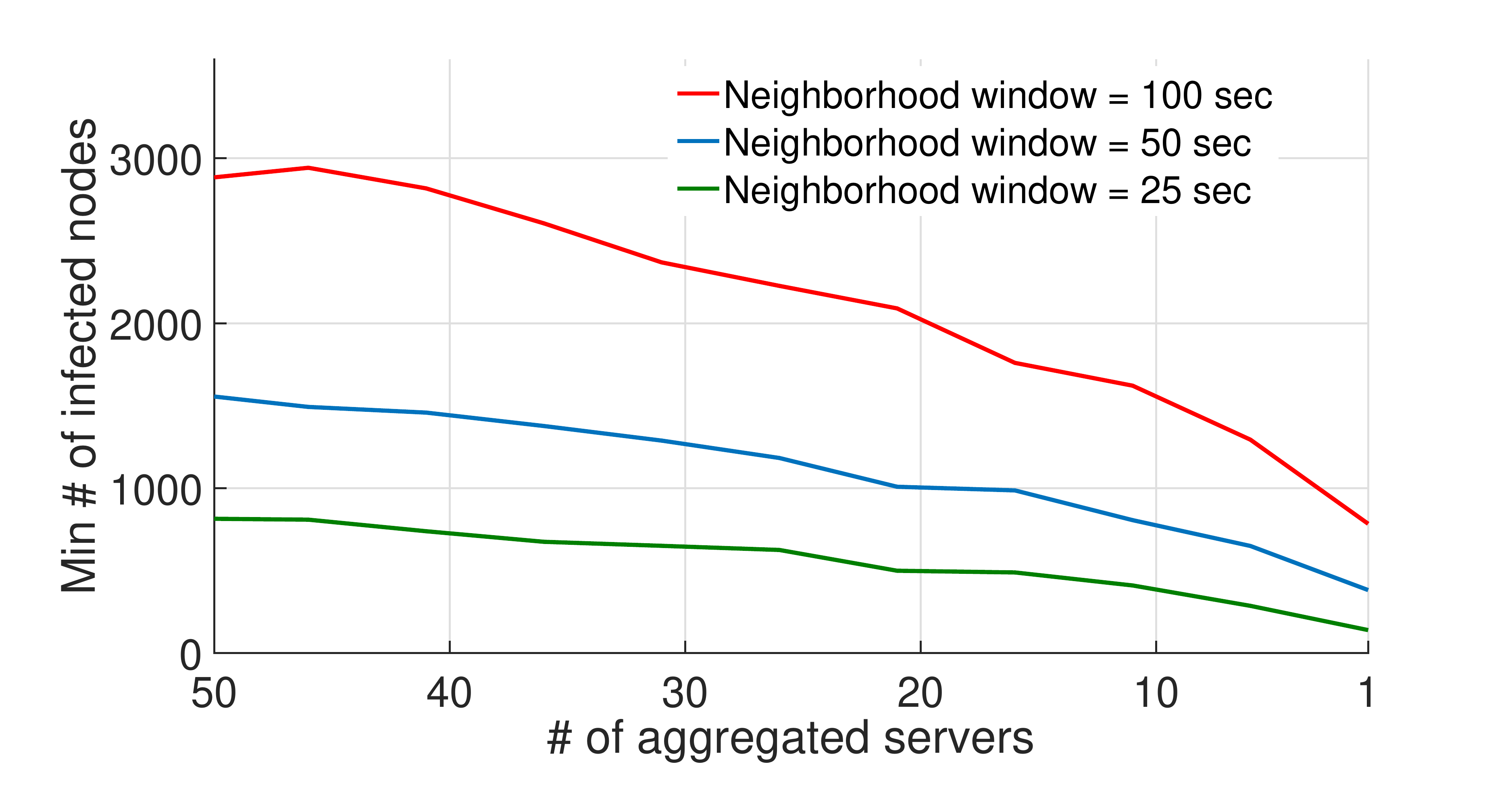}\par
   \caption{{\em(Waterhole attack) Comparing to pure time-based NF,
     structural filtering algorithm improves Shape GD's performance by
     $3.75\times$ -- $5.8\times$ by aggregating alerts on a server basis.  }}
   \vspace{-0.2in}     
\label{fig:waterhole-structural-NF}
\end{figure}
In the main text we present file-level aggregate results.
For completeness here we describe machine-level detection results as well, however, 
we mainly focus our attention on the new trends unobserved at the file-level.
As before, we consider a machine to be compromised (or infected) if it has downloaded at least one malicious file.
Note \sysname is meant to be a file detector, not the machine level detector.

\noindent{\bf False positive rate.}
We notice two opposite trends. First, the machine-level false positive rate is higher
in comparison to the file-level FP rate for all detectors because if detectors mislabel a single benign file,
this may dramatically affect false positive rate if the file has been downloaded on multiple clean machines, i.e.
those machines become false positives.
Second, if we do detectors' pairwise comparison in terms of false positives, we notice that their relative FP rates
becomes more different.
For example, the downloader detector has only 1.53 times lower FP rate than \sysname at the machine level
in comparison to 7.1 time difference at the file level.
The downloader detector's results worsen mainly because the detector often mislabels benign files 
that are frequently downloaded on multiple clean machines, so those machines are considered as false positives.
Surprisingly, the neighborhood detector's FP rate reaches almost 80\% and makes it completely
unusable -- due to this reason we exclude it from the further discussion.

\noindent{\bf True positive rate.}
In comparison to the FP rate, the TP rate does not exhibit a single trend -- the direction,
in which it moves, depends on a particular detector.
The downloader detector's TP rate drops down by almost 3 times because the majority of machines
in the Wine dataset is infected by non-downloaders (malware that does not download other files).
As a result, the downloader detector misses almost 87\% of infected machines.

\sysname{}'s TP rate demonstrates the opposite trend -- it increases in comparison to the file-level
detection by 6.7\% because \sysname searches for correlated malicious downloads and thus it is likely to 
detect similar malware that infects multiple machines. As a result, spatial correlations between 
malware downloads boost detection results -- they raise from 78.03\% up to 84.71\%.

\noindent{\bf F-1 score.}
Overall, \sysname achieves the best FP/TP trade-off -- the highest F-1 score (60.06\%). 
The downloader detector demonstrates the poorest detection results -- the lowest F-1 score (17.08\%) --
mainly due to low its low TP rate.

\noindent{\bf Time to detection.}
We observe that average time to detection slightly increases for \sysname (from 20.33 days up to 28.67 days),
but it is almost 3 times lower than the same parameter of the downloader detector because
\sysname makes a decision regarding a file without waiting until it downloads other files.

\section{Time to Detection Using Structural Information}
%Can neighborhood structural information improve Shape GD's
%  time to detection?}
\label{sec:time-struct}
%

%Structural filtering adds structure-based correlations on top of the time-based ones, whose concrete form highly depends on a particular use-case.
%In other words, structural filtering embeds features peculiar to a specific attack into the detection process.
%In the case of phishing, structural NF considers "sender--receiver" relation between entities exchanging emails, for example, it may aggregate data across recipients of a particular email.
%When detecting waterhole attacks, structural NF may collect suspicious data only from the clients that access a specific set of servers.

%\mohit{check: are we defining NF and Shape GD as two separate algos
% all the way?  Something else instead of 'structure'?}

Waterhole attack imposes a logical structure on nodes (beyond
their time of infection): it infects only the clients that
access a compromised server.  
This structure suggests that temporal neighborhoods can be further refined
based on the specific server accessed by a client (i.e., grouping clients that visit 
a server into one neighborhood).

To analyze the effect of such structural filtering on GD's performance, we vary
filtering from coarse- (no structural filtering, only
time-based filtering) to fine-grained (aggregating alerts 
across clients accessing each server separately)
(Figures~\ref{fig:waterhole-structural-NF}).  Specifically, the aggregation parameter
changes from 50 servers down to 1. \ignore{why 50? link back
to experimental setup.}
%\ignore{ It is worth noting that the maximum value of such parameter does not
%affect the results because nodes generate FPs irrespective of received benign
%emails or accessed legitimate servers.  And there is only one malicious email
%and only one compromised server in the simulated network.  Such value only
%defines the resolution of a plot.  For example, if it is 50, then we can
%compute at most 50 intermediate points.}
As before, we measure detection in terms of the minimum number of infected
nodes that lead to raising a global alert.  Also we consider three NTW values
-- 25-, 50-, and 100-sec long.

%\todo{conclusion? cannot make NTW
%arbitrarily small or large: shape GD won't work well. Both time and structure are imp.}
%This shows that structural filtering the infection-signal from alert-FVs   

%\ignore{The effect of large neighborhood windows is slightly
%more noticeable without structural filtering (aggregating alerts across all
%recipients' lists).  Comparing to 1-hour window, the largest window (3 hours)
%leads to $\sim 12\%$ decrease in the number of compromised nodes.}
%\begin{figure}[tbp]
%   \vspace{-0.0in}
%%\vspace{-0.2in}
%   \centering
%   \includegraphics[width=0.45\textwidth]{figs/windows/detection_threshold_vs_censoring_threshold/phishing_window_ALL_5_trial.eps}\par
%   \caption{{\em(Phishing attack) Comparing to pure time-based NF,
%     structural filtering algorithm improves Shape GD's performance by
%     $\sim 4\times$ by taking into consideration logical structure of
%     electronic communication (sender -- receiver relation).  }}
%   \vspace{-0.2in}     
%\label{fig:windows-detection_vs_censoring_threshold}
%\end{figure}

Structural filtering improves time to detection by
5.82x, 4.07x, and 3.75x for 25-, 50-, 100-sec long windows respectively.
Interestingly, structural filtering requires \sysname to use longer NTWs than
before -- small NTWs (such as 6 seconds from the last sub-section)
%limits the size of a neighborhood window from the bottom: we cannot anymore
%use a small 6-sec long window because it does not contain enough suspicious 
no longer supply a sufficient number of alert-FVs for \sysname to operate
robustly.  Even though structural filtering with a 25 second NTW improves detection
by 5.82x over temporal filtering with 25 second NTWs, the number of infected
nodes at detection time is 139.9 -- higher than the 107 infected nodes for
temporal filtering with a {\em 6 second} NTW (Figure
\ref{fig:waterhole_time_NF}). Temporal and structural filtering thus present
different trade-offs between detection time and work performed by \sysname{} -- 
their relative performance is affected by the rate at which true and false positive
FVs are generated.

%
%with an NTW of 6 seconds, the Shape GD performs more work 
%than using an NTW of 25 seconds, but can detect infections faster. Similarly, 
%a less frequently accessed waterhole server might require structural filtering
% 

%
%Waterhole attacks through popular servers
%thus present a  
%
%However, experiments show that
%structural filtering is important to prevent compromising a least frequently
%accessed server -- pure time-based NF may not work because malicious FVs will
%dissolve in a large volume of benign FVs.

%% file: clustering_results.tex
\ignore{
\begin{figure}[t]
\includegraphics[width=0.48\textwidth]{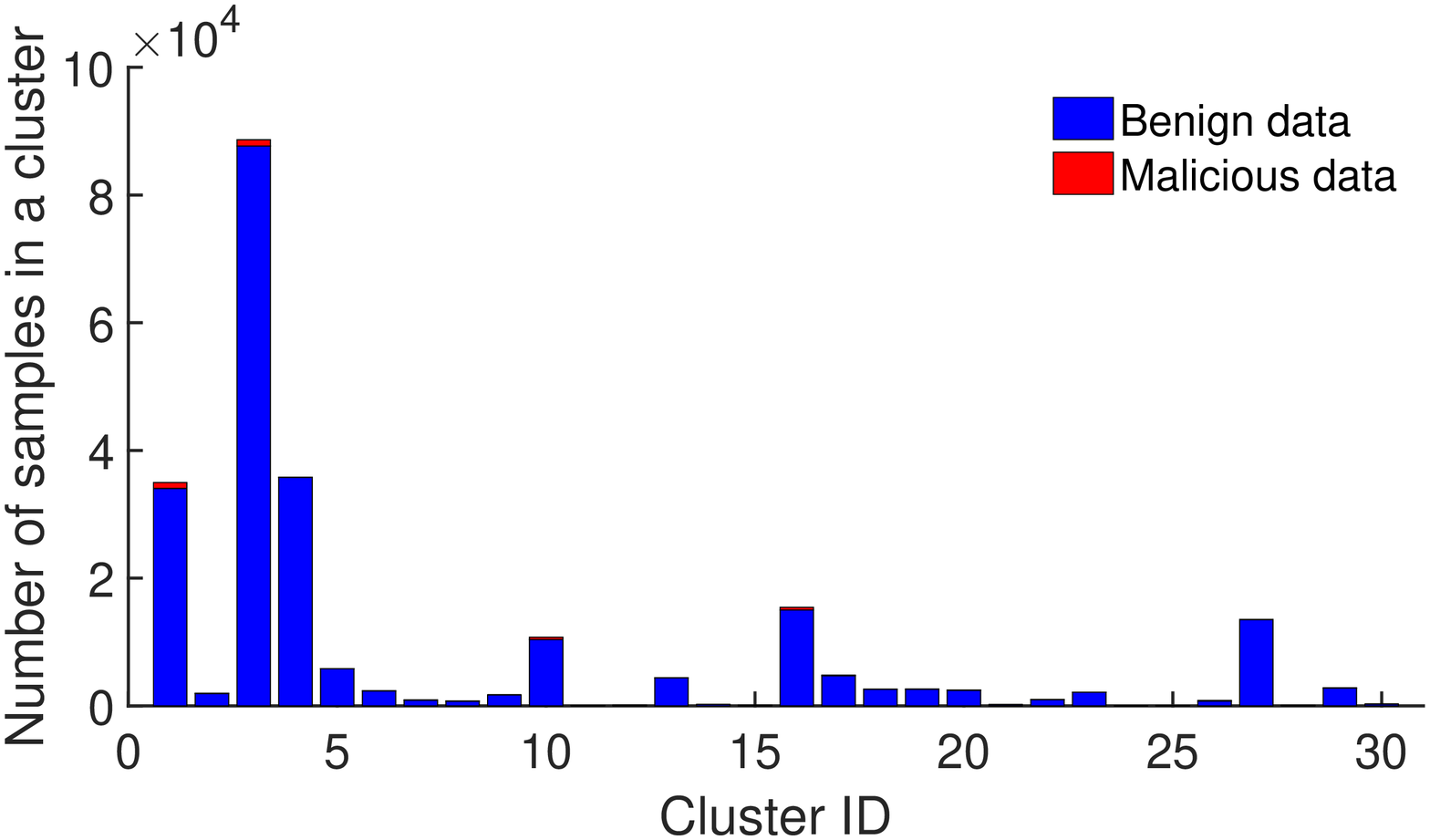}\par
\caption{(Waterhole attack) A common clustering approach \cite{beehive2013, Seurat2004, kaufman_clustering} fails to
  distinguish between malicious and benign feature vectors in our
  setting, i.e. each cluster includes significant number of both
  malicious and benign FVs.}
%  \vspace{-0.2in}  
\label{fig:clustering-bars-waterhole}
\end{figure}
}

%\mohit{Move to Intro: Clustering is another common approach to anomaly
%detection and it is widely used in security literature\cite{beehive2013,
%Seurat2004}.  Clustering methods rely on the following two assumptions: that
%malicious behavior deviates from benign w.r.t. a chosen metric (e.g. L1
%distance) and that most of the nodes in an enterprise network are uninfected.
%Thus, we expect to observe large aggregations of uninfected hosts and sporadic,
%low populated clusters mostly composed of misbehaving hosts.}

While Count-GD is fragile, clustering GDs are inaccurate in the early stages of
infection.  This is why prior work~\cite{beehive2013} uses clustering to
(offline) identify high-priority incidents from security logs for human
analysis (instead of as an alwayWs-on GD) -- this use case is complementary to
an always-on global detector. 
%-- instead of an accurate GD its goal is to opportunistically discover
%malware-related incidents. 
We quantify a recent clustering GD's~\cite{beehive2013} detection rate 
in the waterhole case study as well. 
We observed similar detection results (very low AUC metric) in the Symantec Wine case study.

%that this approach is not
%sufficiently accurate to be directly used as a global detector.  

First, we reduce dimensionality of 390-dimensional FVs by projecting them on
the top 10 PCA components, which retain 95.72\% of the data variance.  Second,
we use an adaptation of the K-means clustering algorithm that does not require
specifying the number of clusters in advance \cite{beehive2013, Seurat2004,
kaufman_clustering}. Specifically, the algorithm consists of the following
three steps: (1) select a vector at random as the first centroid and assign all
vectors to this cluster; (2) find a vector furthest away from its centroid 
(following Beehive~\cite{beehive2013}, we use L1 distance) and
make it a center of a new cluster, and reassign every vector to the cluster
with the closest centroid; and (3) repeat step 2 until no vector is further
away from its centroid than half of the average inter-cluster distance.

%In our implementation of the clustering algorithm, we use L1 distance, which
%is defined between arbitrary \textit{n}-dimensional vectors \textit{a} and \textit{b} as follows:
%$$
%L1(a, b) = \sum_{i=1}^n \left|a(i) - b(i) \right|.
%$$

The evaluation settings of the clustering algorithm match exactly the settings
where \sysname detects infected neighborhoods with 99\% confidence.
Specifically, the algorithm clusters the data that we collected in a
17,178-node neighborhood under a waterhole attack within 30 seconds. 
As we have already demonstrated
(Section~\ref{sec:time-nf}),\sysname starts detecting malware when 107
(waterhole attack) nodes get compromised (Figure~\ref{fig:windows-shape_vs_count-waterhole}).

%\begin{figure}[t]
%\includegraphics[width=0.48\textwidth]{figs/clustering/clusters_combined_roc.eps}\par
%\caption{(Waterhole case study) Receiver operating curve shows detection
%accuracy of the clustering-based malware detector
%\cite{beehive2013}. Its Area Under the Curve (AUC) parameter averaged for 10 runs reaches
%only 48.3\%; such low AUC value makes it unusable as a global detector.} 
%%\vspace{-0.3in}
%\label{fig:clustering-roc}
%\end{figure}

Clustering does not fare well.
It partitions the waterhole dataset into 30 clusters.  We
observe three large clusters that aggregate most of the benign FVs. However,
the algorithm fails to find small 'outlying' clusters consisting of
predominantly malicious data. Each cluster heavily mixes benign and malicious data, hence the
clustering approach suffers from poor discriminative ability, i.e. it is unable
to separate malicious and benign samples.

\begin{figure}[t]
\includegraphics[width=0.48\textwidth]{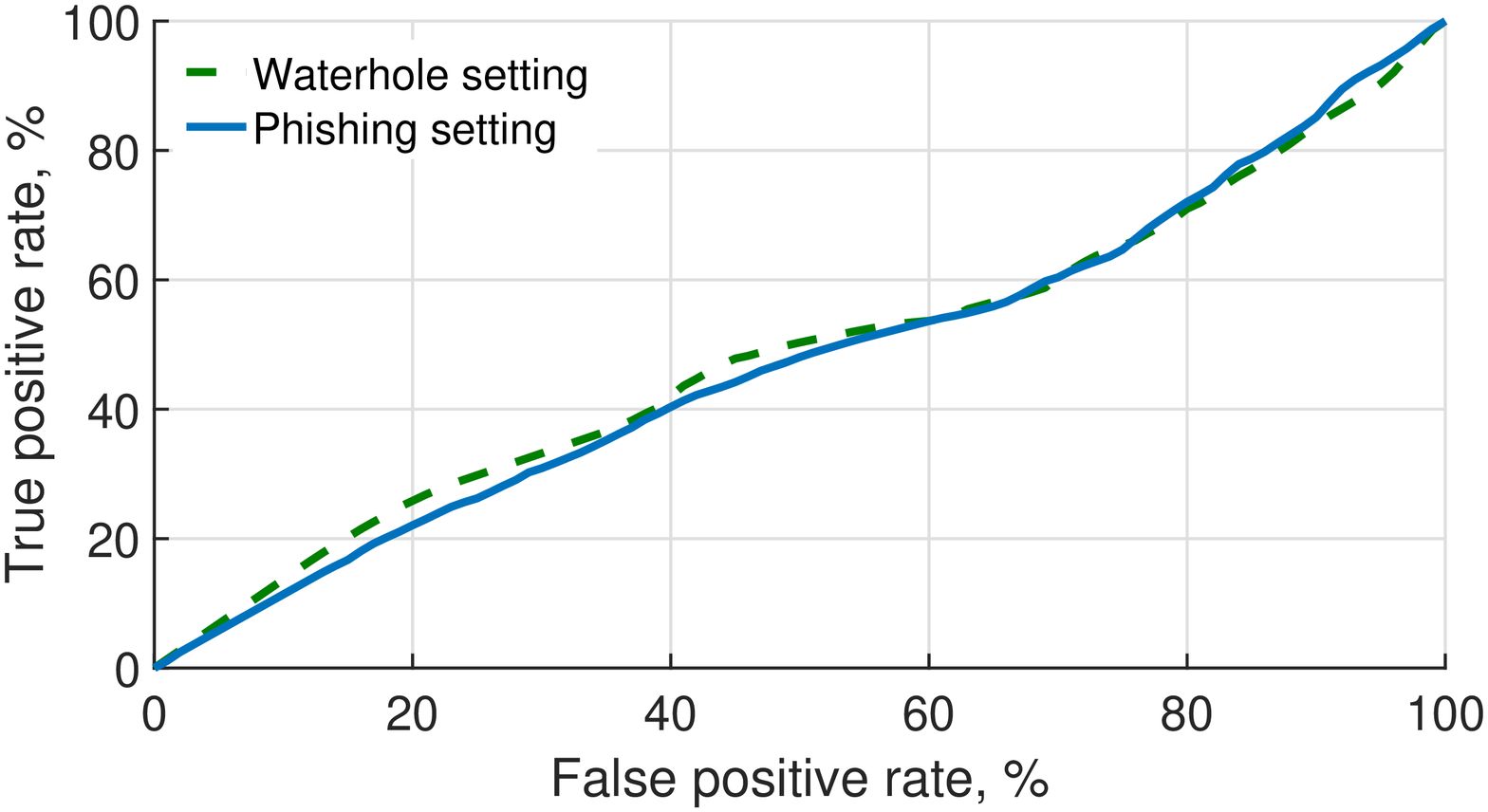}\par
\caption{(Waterhole case study) Receiver operating curve shows detection
accuracy of the clustering-based malware detector
\cite{beehive2013}. Its Area Under the Curve (AUC) parameter averaged for 10 runs reaches
only 48.3\%; such low AUC value makes it unusable as a global detector.} 
%\vspace{-0.3in}
\label{fig:clustering-roc}
\end{figure}

Note the clustering algorithm enforces explicit ordering across the clusters.
That is, the algorithm forms a new cluster around an FV that is furthest away
from its cluster centroid. Thus, earlier a cluster is created, the more
suspicious it is.  By design of the clustering algorithm, the clusters are
subject to a deeper analysis in order of their suspiciousness.  
Such an inherent ordering allows us to build a receiver operating curve (Figure
\ref{fig:clustering-roc}) and compute a typical metric for a binary classifier
-- Area Under the Curve (AUC) by averaging across 10 runs.  
%The ROC curve shows the true positive rate achieved by a binary classifier
%when operating at a fixed false positive rate.  And AUC characterizes the
%overall performance of a binary classifier: the higher AUC, the more accurate
%the classifier.  However, 
The AUC reaches only 48.3\% in the waterhole case study.

This experiment illustrates the failure of the traditional recipe of
dimensionality reduction plus clustering. There is a fundamental reason for
this -- the neighborhoods
we seek to detect are small compared to the total number of nodes in the
system.  %As has been observed and explained in various settings \cite{cite}, 
Optimization-based algorithms that exploit density, including
K-Means and related algorithms, fail to detect small clusters in high
dimensions, even under dimensionality reduction.  The reason is that the
dimensionality reduction is either explicitly random (e.g., as in
Johnson-Lindenstrauss type approaches), or, if data-dependent (like PCA), it is
effectively independent of small clusters, as these represent very little of
the energy (the variance) of the overall data. Spectral clustering style
algorithms~\cite{spectral_clustering_tutorial_2007,on_spectral_clustering_ng_01,evolutionary_spectral_clustering}
are also notoriously unable to deal with highly
unbalanced sized clusters, and in particular, are unable to find small
clusters.

\sysname also reduces dimensionality but does so after neighborhood
filtering.  This amplifies the impact of small neighborhoods. The combination
of dimensionality reduction, small-neighborhood-amplification, and then
aggregation represents a novel approach to this detection problem, and our
experiments validate this intuition.

%The low AUC metric demonstrates that clustering is ineffective without a 
%way of discovering the latent low-dimensional structure in data -- 
%several theoretical studies also show
%that classical clustering is ineffective in high dimensions~\cite{donoho1983notion,huber2011robust,xu2013outlier}.
%ShapeGD addresses this by using side-information about attack vectors
%to define neighborhoods followed by ShapeScore to robustly classify the
%(noisy) neighborhoods.

%reducing the dimensionality of data such as neighborhoods and
%ShapeScore.  Note that this is in concordance with theoretical studies showing
%the weakness of classical clustering in high dimensions
%\cite{donoho1983notion,huber2011robust,xu2013outlier}.

